\pdfoutput=1
\RequirePackage{amsmath}
\documentclass[graybox,envcountchap]{svmono}

\usepackage[british]{babel}
\usepackage{mathptmx}
\usepackage{helvet}
\usepackage{courier}

\usepackage{type1cm}         

\usepackage{makeidx}         %
\usepackage{graphicx}        %
\usepackage{multicol}        %
\usepackage[bottom]{footmisc}%

\usepackage{tikz}
\usetikzlibrary{decorations.pathmorphing,patterns}
\usepackage[colorlinks, citecolor=blue, linkcolor=blue, urlcolor=blue, filecolor=blue, linktocpage=true]{hyperref}
\usepackage{bbold,dsfont}
\usepackage[refpage]{nomencl}

\makenomenclature

\clearpage{}%
\usepackage{microtype}
\usepackage{amssymb}

\usepackage{amsthm}

\newtheorem{Th}{Theorem}[chapter]  
\newtheorem{Prop}[Th]{Proposition}  
\newtheorem{Cor}[Th]{Corollary}  
\newtheorem{Lem}[Th]{Lemma}  

\newtheorem{Def}[Th]{Definition}  
\newtheorem{Ex}[Th]{Example}

\theoremstyle{remark}

\newtheorem*{remark}{Remark}

\usepackage[noconfig]{refstyle}
\let\eqref=\relax
\newref{eq}{name={eq.~},Name={Eq.~},names={eqs.~},Names={Eqs.~},rngtxt={-},refcmd=(\ref{#1})}
\newref{f}{name={footnote~},Name={Footnote~},names={footnotes~},Names={Footnotes~}}
\newref{app}{name={Appendix~},Name={Appendix~},names={Appendixes~},Names={Appendixes~}}
\newref{tab}{name={table~},Name={Table~},names={tables~},Names={Tables~}}
\newref{ch}{name={Chapter~},Name={Chapter~},names={Chapters~},Names={Chapters~}}
\newref{sec}{name={Section~},Name={Section~},names={Sections~},Names={Sections~}}
\newref{fig}{name={figure~},Name={Figure~},names={figures~},Names={Figures~}}
\newref{th}{name={Theorem~},Name={Theorem~},names={Theorems~},Names={Theorems~}}
\newref{prop}{name={Proposition~},Name={Proposition~},names={Propositions~},Names={Propositions~}}
\newref{cor}{name={Corollary~},Name={Corollary~},names={Corollaries~},Names={Corollaries~}}
\newref{lem}{name={Lemma~},Name={Lemma~},names={Lemmas~},Names={Lemmas~}}
\newref{def}{name={Definition~},Name={Definition~},names={Definitions~},Names={Definitions~}}
\newref{ex}{name={Example~},Name={Example~},names={Examples~},Names={Examples~}}
\newref{qu}{name={Question~},Name={Question~},names={Questions~},Names={Questions~}}
\newref{exc}{name={Exercise~},Name={Exercise~},names={Exercises~},Names={Exercises~}}

\newcommand{\bt}{\begin{Th}\ \ }  
\newcommand{\et}{\end{Th}}  
\newcommand{\bp}{\begin{Prop}\ \ }  
\newcommand{\ep}{\end{Prop}}  
\newcommand{\bc}{\begin{Cor}\ \ }  
\newcommand{\ec}{\end{Cor}}  
\newcommand{\bl}{\begin{Lem}\ \ }  
\newcommand{\el}{\end{Lem}}  
\newcommand{\bd}{\begin{Def}\ \ }  
\newcommand{\ed}{\end{Def}}  
\newcommand{\bex}{\begin{Ex}\ \ }  
\newcommand{\eex}{\end{Ex}}  

\newcommand{\pf}{\begin{proof}}
\newcommand{\epf}{\end{proof}}

\newcommand{\be}{\begin{equation}}  
\newcommand{\ee}{\end{equation}}  

\newcommand\re[1]{(\ref{#1})}

\newcommand{\bC}{\mathbb{C}}  
\newcommand{\bR}{\mathbb{R}}

\newcommand{\bK}{\mathbb{K}}  

\renewcommand{\a}{\alpha}  
\renewcommand{\b}{\beta}  
  
\renewcommand{\d}{\delta}  
\newcommand{\e}{\epsilon}  
  
\newcommand{\g}{\gamma}

\renewcommand{\k}{\kappa}  
\renewcommand{\l}{\lambda}  
\renewcommand{\o}{\omega}

\newcommand{\s}{\sigma}

\newcommand{\G}{\Gamma}

\renewcommand{\O}{\Omega}

\newcommand{\ra}{\rightarrow}
\newcommand{\p}{\partial}
\newcommand{\n}{\nabla}  
\newcommand{\ot}{\otimes}  

\DeclareMathOperator\tr{tr}  
\DeclareMathOperator\End{End}

\newcommand\GL[1]{\mathrm{GL}(#1)}

\newcommand\SO[1]{\mathrm{SO}(#1)} 

\newcommand\U[1]{\mathrm{U}(#1)}

\renewcommand{\gg}{\mathfrak{g}}

\newcommand{\gu}{\mathfrak{u}}

\newcommand{\so}{\mathfrak{so}}  
\newcommand{\su}{\mathfrak{su}}

\newcommand{\ggl}{\mathfrak{gl}}

\DeclareMathAlphabet{\mathcurly}{OML}{cmm}{m}{it}
\newcommand{\vc}{\mathcurly{v}}\clearpage{}%

\allowdisplaybreaks

\makeindex             %

\begin{document}

\author{Vicente Cort\'es, Alexander S. Haupt}
\title{Mathematical Methods of Classical Physics}
\subtitle{ }
\hypersetup{pageanchor=false}
\maketitle
\hypersetup{pageanchor=true}

\frontmatter%

\clearpage{}%
\extrachap{Acknowledgements}

We are very grateful to David~Lindemann for careful proof-reading of earlier versions of the manuscript of this book and for numerous 
constructive comments which helped improve the presentation of this book. We thank Thomas~Leistner for drawing our attention to the reference~\cite{Eisenhart} and Thomas~Mohaupt for useful remarks concerning \chref{intro}.\clearpage{}%

\tableofcontents

\clearpage{}%

\renewcommand{\nomname}{List of Symbols}
\markboth{\nomname}{\nomname}
\printnomenclature[32mm]
\clearpage{}%

\mainmatter%
\clearpage{}%

\chapter{Introduction}
\chlabel{intro} 

\abstract*{We define the framework of classical physics as considered in the present book. We briefly point out its place in the history of physics and its relation to modern physics. As the prime example of a theory of classical physics we introduce Newtonian mechanics and discuss its limitations. This leads to and motivates the study of different formulations of classical mechanics, such as Lagrangian and Hamiltonian mechanics, which are the subjects of later chapters. Finally, we explain why in this book, we take a mathematical perspective on central topics of classical physics.}

\abstract{We define the framework of classical physics as considered in the present book. We briefly point out its place in the history of physics and its relation to modern physics. As the prime example of a theory of classical physics we introduce Newtonian mechanics and discuss its limitations. This leads to and motivates the study of different formulations of classical mechanics, such as Lagrangian and Hamiltonian mechanics, which are the subjects of later chapters. Finally, we explain why in this book, we take a mathematical perspective on central topics of classical physics.}

\section*{}

Classical physics refers to the collection of physical theories that do not use quantum theory and often predate modern quantum physics. They can be traced back to Newton (17th century) and in some sense even further all the way to Aristotle, Archimedes, and other Greek philosophers of antiquity (starting in the 4th century BC). However, this does by far not mean that theories of classical physics are exclusively a subject of the past. They continue to play important roles in modern physics, for example in the study of macroscopic systems, such as fluids and planetary motions, where the effects of the quantum behavior of the microscopic constituents are irrelevant.

Newtonian mechanics is arguably the first mathematically rigorous and self-contained theory of classical physics. In its traditional formulation, Newton's theory comprises three physical laws known as Newton's laws of motion, describing the relationship between a body (usually assumed to be a point particle of constant mass $m>0$) and forces acting upon it. They also quantify the resulting motion of the body in response to those forces and can be summarised, in an inertial reference frame, as follows.
\begin{enumerate}
 \item[\textbf{First law:}] A body at rest will stay at rest and a body in uniform motion will stay in motion at constant velocity, unless acted upon by a net force.
 \item[\textbf{Second law:}] The force $F$ acting on a body is equal to the mass of the body times the acceleration $a$ of the body:
 \be\eqlabel{2ndlaw} F = m\, a \; .\ee
 \item[\textbf{Third law:}] For every action force there is a corresponding reaction force which is equal in magnitude and opposite in direction. This is often abbreviated as \emph{actio} $=$ \emph{reactio}. 
\end{enumerate}
Considering a point particle moving in $\bR^3$, its acceleration at time $t$ is given by the second time derivate of its position vector, denoted by $a(t) = \ddot{x}$.\nomenclature[sdot]{$\dot{(\cdot)}$}{time derivate, $\dot{f}(t) = f'(t)$, $\ddot{f}(t) = f''(t)$} Throughout the entire book, we work in the $C^\infty$-category, unless stated otherwise, that is, manifolds and maps between them are usually assumed to be smooth. The force $F$ is generally allowed to depend on position $x$, velocity $\dot{x}$, and time $t$, that is $F = F(x,\dot{x},t)$. \Eqref{2ndlaw} then turns into 
\be\eqlabel{newtoneom}
 \ddot{x} = \frac{1}{m} F(x,\dot{x},t) \; . 
\ee
For $m$ and $F(x,\dot{x},t)$ given, this is a set of second-order ordinary differential equations known as \emph{Newton's equations of motion}. Note that $F=0$ if and only if the motion $t \to x(t)$ is linear and therefore Newton's first law is a special case of the second law. Also, the third law is a consequence of the second law in combination with conservation of momentum, which ultimately follows from translational invariance (see Corollary~\ref{momCor}). Despite this redundancy in Newton's laws, all three highlight different aspects of important concepts of modern physics, namely the notion of inertial frame in Einstein's theory of relativity and the relation between different concepts of mass (inertial versus gravitational) in gravitational theories (see, for example,~\cite{Einstein}).

We remark that \eqref{2ndlaw} is also sometimes written as $F = \dot{p}$, where $p=m\dot{x}$ is the \emph{momentum}\index{Momentum} of the particle. This allows for the consideration of bodies with non-constant mass, such as rockets consuming their fuel, where however the second law needs to be applied to the total system including the lost (or gained) mass.

In order to solve \eqref{newtoneom} for a given mechanical system, one first needs to determine an actual expression for the net force $F(x,\dot{x},t)$. However, this can be intricate in practice, especially when constraint forces are present. Typically for constraint forces it is easy to describe the constraints geometrically, while it is difficult to explicitly obtain the corresponding function $F(x,\dot{x},t)$. For example, consider a ball moving on the surface of a flat table. Geometrically, the constraint imposed by the presence of the table implies that the vertical component of $x$ (usually denoted by $x^3$) is held fixed. On the other hand, incorporating this constraint into \eqref{newtoneom} amounts to specifying an actual functional expression for the force exerted by the table on the ball. Another problem with Newton's formulation, partly related to the previous issue, is the intrinsic use of Cartesian coordinates. Indeed, changing to a different coordinate system is generally cumbersome. 

Given these limitations, one is led to consider more geometric formulations, such as Lagrangian and Hamiltonian mechanics, in which generalized coordinates are introduced. These two formulations were introduced by Lagrange in 1788 and by Hamilton in 1833, respectively. They can be adapted to the mechanical system at hand, for example in order to incorporate geometrically the presence of constraint forces. In addition, symmetries can be identified more straightforwardly and exploited more efficiently in these alternative formulations of classical mechanics. Historically, in particular Hamilton-Jacobi theory -- yet another formulation -- has also played an important role in the development of quantum mechanics~\cite{BW,Goldstein,Sakurai}. It has also paved the way for the development of classical field theory, which incorporates Einstein's theory of relativity and nowadays underlies many areas of modern physics.

For these reasons, a profound understanding of these central areas of classical physics is very important and hence, many textbooks on these subjects exist, such as for example refs.~\cite{AM,A,Goldstein,Goldstein2}. However, many of these textbooks require a strong background in physics and do not put strong emphasis on mathematical rigor. This presents difficulties for the more mathematically inclined reader. 

As a complementary contribution to the existing literature, the present book takes a different, namely more mathematical, perspective at these central topics of classical physics. It puts emphasis on a mathematically precise formulation of the topics while conveying the underlying geometrical ideas. For this purpose the mathematical presentation style (definition, theorem, proof) is used and all theorems are proved. In addition, the field theory part is formulated in terms of the theory of jet bundles, highlighting the relativistic covariance. Jet bundles do not receive widespread attention in the physics literature up to now. Each chapter of the book is accompanied by a number of exercises, which can be found collectively in \appref{exercises}. The interested reader is highly encouraged to try the exercises, as this will greatly help in gaining a deeper understanding of the subject.

This book grew out of a lecture course on ``mathematical methods of classical physics'' held in the winter semester of 2015/2016 as part of the master's program in mathematics and mathematical physics at the Department of Mathematics at the University of Hamburg. It is therefore primarily directed at readers with a background in mathematical physics and mathematics. Also, physicists with a strong interest in mathematics may find this text useful as a complementary resource.

\clearpage{}%
\clearpage{}%

\chapter{Lagrangian Mechanics}
\chlabel{lagmech} 

\abstract*{In this chapter, we lay out the foundations of Lagrangian Mechanics. We introduce the basic concepts of Lagrangian mechanical systems, namely the Lagrangian, the action, and the equations of motion, also known as the Euler-Lagrange equations. We also discuss important examples, such as the free particle, the harmonic oscillator, as well as motions in central force potentials, such as Newton's theory of gravity and Coulomb's electrostatic theory. Highlighting the importance of symmetries, we study integrals of motion and Noether's theorem. As an application, we consider motions in radial potentials and, further specializing to motions in Newton's gravitational potential, we conclude this section with a derivation of Kepler's laws of planetary motion.}

\abstract{In this chapter, we lay out the foundations of Lagrangian Mechanics. We introduce the basic concepts of Lagrangian mechanical systems, namely the Lagrangian, the action, and the equations of motion, also known as the Euler-Lagrange equations. We also discuss important examples, such as the free particle, the harmonic oscillator, as well as motions in central force potentials, such as Newton's theory of gravity and Coulomb's electrostatic theory. Highlighting the importance of symmetries, we study integrals of motion and Noether's theorem. As an application, we consider motions in radial potentials and, further specializing to motions in Newton's gravitational potential, we conclude this section with a derivation of Kepler's laws of planetary motion.}

\section{Lagrangian mechanical systems and their equations of motion}
\bd \label{firstDef} A \emph{Lagrangian mechanical system}\index{Lagrangian!mechanical system} is a pair $(M,\mathcal{L})$ consisting of a smooth manifold $M$\nomenclature[aM]{$M$}{smooth manifold (configuration space)} and a smooth function $\mathcal{L}$\nomenclature[aLcal]{$\mathcal{L}$}{Lagrangian (function)} on $TM$\nomenclature[aTM]{$TM$}{tangent bundle of $M$}. The manifold $M$ is called the \emph{configuration space}\index{Configuration space} and the function $\mathcal{L}$ is called the \emph{Lagrangian function} (or simply the \emph{Lagrangian})\index{Lagrangian} of the system. 
\ed 
More generally, one can allow for the Lagrangian to depend explicitly on an extra variable interpreted as time\index{Time}. 

\begin{Ex} \exlabel{NewtonEx} The Lagrangian of a point particle\index{Particle!Point} of \emph{mass}\index{Mass} $m>0$\nomenclature[am]{$m$}{mass of a point particle} moving in Euclidean space $\bR^n$ under the influence of a \emph{potential}\index{Potential} $V\in C^\infty (\bR^n)$\nomenclature[aV]{$V$}{potential} is 
\[ \mathcal{L}(v) = \frac12 m \langle v,v\rangle -V(x), \quad v\in T_x\bR^n,\quad x\in \bR^n, \]
where $\langle \cdot , \cdot \rangle$\nomenclature[s]{$\langle \cdot , \cdot \rangle$}{Euclidean scalar product on $\bR^n$} denotes the Euclidean scalar product.   Given a smooth curve 
$t\mapsto \gamma (t)$ in $\bR^n$, the quantities $E_{\textnormal{kin}}(t)=\frac12 m \langle \g'(t) ,\g'(t)\rangle$ and  $E_{\textnormal{pot}}(t)=V(\g(t))$ are called the 
\emph{kinetic energy}\index{Energy!Kinetic} and the \emph{potential energy}\index{Energy!Potential} at time $t$\nomenclature[at]{$t$}{time}, respectively. Their sum
\nomenclature[aE]{$E$}{(total) energy}\nomenclature[aEkin]{$E_{\textnormal{kin}}$}{kinetic energy}\nomenclature[aEpot]{$E_{\textnormal{pot}}$}{potential energy}
\[ 
 E(t) = E_{\textnormal{kin}}(t) + E_{\textnormal{pot}}(t)
\]
is the \emph{total energy}\index{Energy}\index{Energy!Total} at time $t$. \\
By specializing $V$ one obtains many important mechanical systems:
\begin{enumerate}
\item The \emph{free particle:}\index{Particle!Free} $V=0$.  
\item The \emph{harmonic oscillator:}\index{Harmonic!oscillator} $n=1$ and  $V = \frac12 kx^2$, where $k$ is a positive constant known as \emph{Hooke's constant}\index{Hooke's constant|see{Harmonic oscillator}}.  
\item \emph{Newton's theory of gravity:}\index{Newton!theory of gravity} $n=3$ and 
\[ V=-\frac{\k mM}{r},\] 
where $r=|x|$ denotes the Euclidean norm of the vector $x$, $M>0$ is the mass of the particle (placed at the origin $0\in \bR^3$) generating the gravitational potential\index{Gravitational!potential|see{Newton's theory of gravity}}\index{Potential!Gravitational|see{Newton's theory of gravity}}, 
and $\k$ is a positive constant known as 
\emph{Newton's constant}\index{Newton!constant|see{theory of gravity}}.
(Notice that in this case $V$ is not defined at the origin, so the configuration space is $\bR^3 \setminus \{ 0\}$ rather than $\bR^3$.)
\item \emph{Coulomb's electrostatic theory:}\index{Electrostatic theory} $n=3$ and 
\[ V=\frac{k_eq_1q_2}{r},\]
where  $q_1$ is the charge\index{Charge|see{Electrostatic theory}} of the particle of mass $m$ moving under the influence of the electric potential\index{Coulomb!potential|see{Electrostatic theory}}\index{Potential!Coulomb|see{Electrostatic theory}} generated by a particle of charge $q_2$ and 
$k_e$ is a positive constant known as \emph{Coulomb's constant}\index{Coulomb!constant|see{Electrostatic theory}}.  Notice that up to a constant factor Coulomb's potential is of the same
type as Newton's potential. However, contrary to Newton's potential, the Coulomb potential can have either sign. As we shall see, this 
corresponds to the fact that electric forces can be attractive or repulsive, depending on the sign of the product $q_1q_2$, whereas 
gravitational forces are always attractive.  Another important difference is that the gravitational potential contains the mass
$m$ as a factor whereas the electric potential does not. As we shall see, the former property implies that the acceleration $\g''$ of the particle
in Newton's theory of gravity is independent of its mass.   
\end{enumerate} 
\end{Ex}

\begin{Ex} \label{T-V:Ex}
More generally, given a smooth function $V$ on a pseudo-Riemannian manifold $(M,g)$, one can consider 
the Lagrangian
\nomenclature[ag]{$g$}{metric on $M$ (Riemannian or pseudo-Riemannian)}
\[ 
 \mathcal{L}(v) = \frac12 g(v,v)-V(x), \quad v\in T_xM,\quad x\in M.
\]
\end{Ex}

\bd \label{actionDef} Let $(M,\mathcal{L})$ be a Lagrangian mechanical system. 
The \emph{action}\index{Action} of a smooth curve $\g : [a,b] \ra M$\nomenclature[ggamma]{$\g$}{smooth curve in $M$} is defined as
\nomenclature[aS]{$S$}{action}
\[ 
 S(\g ) := \int_a^b \mathcal{L}(\g ' (t))dt.
\]
A \emph{motion}\index{Motion} of the system is a critical point of $S$ under smooth variations with 
fixed endpoints. (This statement is a mathematical formulation of \emph{Hamilton's principle of least action}, which should 
better be called principle of stationary action.)
\ed 
Next we will derive the \emph{equations of motion}\index{Equations of motion|see{Euler-Lagrange equations}}, which are the differential equations describing critical points of the action functional. 
These are known as the \emph{Euler-Lagrange equations}\index{Euler-Lagrange!equations} and are given in  \re{EL:eq}. For this we introduce coordinates $(x^1,\ldots ,x^n)$\nomenclature[ax1]{$(x^1,\ldots ,x^n)$}{local coordinates on an open subset $U\subset M$} on an open subset $U\subset M$. 
They induce a system of coordinates $(q,\hat{q})=(q^1,\ldots ,q^n, \hat{q}^1,\ldots ,\hat{q}^n)$\nomenclature[aq1]{$(q^1,\ldots ,q^n, \hat{q}^1,\ldots ,\hat{q}^n)$}{induced/canonical local coordinates on $TM$ associated with local coordinates $(x^1,\ldots ,x^n)$ on $M$} on the open subset $TU = \pi^{-1}(U)\subset TM$, 
where $\pi : TM \ra M$\nomenclature[gpi]{$\pi$}{canonical projection from $TM$ to $M$} denotes the canonical projection. The \emph{induced coordinates}\index{Induced coordinates}\index{Coordinate!Induced} are 
defined by
\[ q^i:= x^i\circ \pi,\quad \hat{q}^i (v) := dx^i(v), \quad v\in TU.\]
Let $\g_s : [a,b] \ra M$, $-\e < s<\e$, $\e>0$, be a smooth variation of $\g : [a,b] \ra M$ with fixed endpoints. Then in local coordinates as above 
we compute 
\begin{eqnarray*} \left. \frac{d}{ds}\right|_{s=0} \mathcal{L}(\g_s') &=& \sum \left( \frac{\p \mathcal{L}}{\p q^i}(\g' )\left. \frac{d}{ds}\right|_{s=0}q^i(\g_s') + \frac{\p \mathcal{L}}{\p \hat{q}^i}(\g')\left. \frac{d}{ds}\right|_{s=0}\hat{q}^i(\g_s') \right)\\
&=& \sum \left( \frac{\p \mathcal{L}}{\p q^i}(\g' )\left. \frac{d}{ds}\right|_{s=0}\g_s^i + \frac{\p \mathcal{L}}{\p \hat{q}^i}(\g')\left. \frac{d}{ds}\right|_{s=0}\dot{\g}_s^i \right),
\end{eqnarray*}
where $\g^i := x^i \circ \g$ and $\dot{\g}^i := (\g^i)'$. Note that here and throughout the rest of the book, we use an adapted version of \emph{Einstein's summation convention}\index{Einstein!summation convention}, where repeated upper and lower indices are to be summed over. This is indicated by a plain sum symbol (see also page~\pageref{page_los_einstein_sum_conv}).\nomenclature[ssum]{$\sum$}{We use an adapted version of \emph{Einstein's summation convention} throughout the book: Upper and lower indices appearing with the same symbol within a term are to be summed over. We write the symbol $\sum$ to indicate whenever this convention is employed. Owing to the aforementioned convention, the summation indices can be (and are usually) omitted below the symbol $\sum$\label{page_los_einstein_sum_conv}}
If we denote by $\mathcal{V} := \left. \frac{d}{ds}\right|_{s=0}\g_s$ the variation vector field along $\g$, then 
we can rewrite this as
\begin{eqnarray*}   \left. \frac{d}{ds}\right|_{s=0} \mathcal{L}(\g_s') &=&  \sum \left( \frac{\p \mathcal{L}}{\p q^i}(\g' )\mathcal{V}^i+ \frac{\p \mathcal{L}}{\p \hat{q}^i}(\g')\dot{\mathcal{V}}^i \right) \\ &=&
\sum \left( \frac{\p \mathcal{L}}{\p q^i}(\g' )-\frac{d}{dt}\frac{\p \mathcal{L}}{\p \hat{q}^i}(\g') \right)\mathcal{V}^i + f' ,\end{eqnarray*}
where $\mathcal{V}^i := dx^i(\mathcal{V})=\hat{q}^i(\mathcal{V})$  
are the components of $\mathcal{V}$ and 
\[ f:= \sum  \frac{\p \mathcal{L}}{\p \hat{q}^i}(\g')\mathcal{V}^i.\] 
Notice that the local vector field $\mathcal{V}^{\textnormal{ver}} = \sum \mathcal{V}^i\frac{\p}{\p \hat{q}^i}$ along $\g'$ is independent of the choice of 
local coordinates and, hence, defines a global vector field along $\g'$. This follows from the fact that it corresponds to $\mathcal{V}$ under the 
canonical identification of $T_{\g (t)}M$ with the vertical tangent space 
\[ T^{\textnormal{ver}}_{\g'(t)}(TM):= \ker d\pi|_{\g'(t)}\subset T_{\g'(t)}(TM).\]   
Using this vector field we can rewrite the function $f$ as
\[ f=\mathcal{V}^{\textnormal{ver}} (\mathcal{L})|_{\g'}.\]
This shows that also $f$ is globally defined on the interval $[a,b]$ and vanishes at the endpoints.  Therefore we have  
\begin{eqnarray} \left. \frac{d}{ds}\right|_{s=0} S(\g_s ) &=& \left. \frac{d}{ds}\right|_{s=0} \int_a^b\mathcal{L}(\g_s'(t))dt =  \int_a^b \left. \frac{d}{ds}\right|_{s=0} \mathcal{L}(\g_s'(t))dt\\ 
&=&   \int_a^b \left( \left. \frac{d}{ds}\right|_{s=0} \mathcal{L}(\g_s'(t))-f'(t)\right)dt ,\label{S':eq}\end{eqnarray} 
where the integrand is a globally defined function on the interval $[a,b]$ with the following local expression:
\[ \left. \frac{d}{ds}\right|_{s=0} \mathcal{L}(\g_s'(t))-f'(t) = \sum \left( \frac{\p \mathcal{L}}{\p q^i}(\g'(t) )-\frac{d}{dt}\frac{\p \mathcal{L}}{\p \hat{q}^i}(\g'(t)) \right)\mathcal{V}^i(t) .\] 
\bt Let $(M,\mathcal{L})$ be a Lagrangian mechanical system, $n=\dim M$\nomenclature[an]{$n$}{dimension of $M$}. The motions\index{Motion} of the system are the 
solutions $\g : [a,b] \ra M$ of the following system of ordinary differential equations:
\be \label{EL:eq}  \a_i:=  \frac{\p \mathcal{L}}{\p q^i}(\g' )-\frac{d}{dt}\frac{\p \mathcal{L}}{\p \hat{q}^i}(\g') =0, \quad i=1,\ldots n.\ee\nomenclature[galphai]{$\a_i$}{component of the Euler-Lagrange one-form in some local coordinate system}
\et

\pf Let us first remark that for every vector field $\mathcal{V}$ along $\g$  the function $\sum \a_i\mathcal{V}^i=  \left. \frac{d}{ds}\right|_{s=0} \mathcal{L}(\g_s')-f'$ is coordinate independent and thus globally 
well-defined. From this we can deduce that $\a = \sum \a_idx^i|_{\g}$ is a well-defined one-form along $\g$ (cf. \appref{exercises}, \excref{1}). 
In virtue of \re{S':eq}, we see that $\g$ is a motion if and only if
\[ \int_a^b \a (\mathcal{V})|_tdt =0.\]
for all vector fields $\mathcal{V}$ along $\g$ vanishing at the endpoints. To see that this implies that $\a (t_0)=0$ for all 
$t_0\in (a,b)$ we  take $\d>0$ such that $\g ([t_0-\d,t_0+\d])$ is contained in a coordinate domain and consider $\mathcal{V}^i = h\alpha_i$, where $h\ge 0$ is a smooth 
function on $[a,b]$ with support contained in  $(t_0-\d,t_0+\d )$ such that $h(t_0)>0$. This defines a smooth vector field
$\mathcal{V}$ along $\g$ vanishing at the endpoints and $0=\int_a^b \a (\mathcal{V})|_tdt= \int_{t_0-\d}^{t_0+\d}h\sum \a_i^2dt$ now implies 
that $\a (t_0) =0$. This proves that $\a|_{(a,b)}=0$ and, by continuity, $\a=0$. 
\epf 

\bd The one-form $\a$\nomenclature[galpha]{$\a$}{Euler-Lagrange one-form} along $\g$ will be called the \emph{Euler-Lagrange one-form}\index{Euler-Lagrange!one-form}. 
\ed 
\begin{Ex}\label{ex_geodesic} Let $(M,g)$ be a pseudo-Riemannian metric and consider the Lagrangian $\mathcal{L}(v) = \frac12 g(v,v)$, $v\in TM$. 
In canonical local coordinates $(q^1,\ldots ,q^n,$ $\hat{q}^1,\ldots, \hat{q}^n)$ on $TM$ associated with local coordinates $(x^1,\ldots ,x^n)$ on $M$ it is given by 
\[ \mathcal{L}=\frac12 \sum \tilde{g}_{ij}\hat{q}^i\hat{q}^j,\] 
where $g=\sum g_{ij}dx^idx^j$ and $\tilde{g}_{ij} = g_{ij}\circ \pi$. We compute 
\begin{eqnarray*}  \frac{\p \mathcal{L}}{\p q^i} &=& \frac12 \sum \frac{\p \tilde{g}_{kj}}{\p q^i}\hat{q}^k\hat{q}^j= \frac12 \sum \left(\frac{\p g_{kj}}{\p x^i}\circ \pi \right)\hat{q}^k\hat{q}^j,\\
 \frac{\p \mathcal{L}}{\p \hat{q}^i} &=& \sum \tilde{g}_{ij}\hat{q}^j,\\
   \frac{\p \mathcal{L}}{\p q^i}(\g') &=& \frac12 \sum \frac{\p g_{kj}}{\p x^i}(\g )\dot{\g}^k\dot{\g}^j,\\
    \frac{\p \mathcal{L}}{\p \hat{q}^i}(\g') &=& \sum g_{ij}(\g)\dot{\g}^j,\\
    \frac{d}{dt}\frac{\p \mathcal{L}}{\p \hat{q}^i}(\g') &=& \sum \frac{\p g_{ij}}{\p x^k}(\g)\dot{\g}^k\dot{\g}^j + \sum g_{ij}(\g)\ddot{\g}^j,\\
    -\a_i &=& \sum g_{ij}(\g)\ddot{\g}^j + \frac12 \sum \left(\frac{\p g_{ij}}{\p x^k}(\g)+  \frac{\p g_{ik}}{\p x^j}(\g) -\frac{\p g_{kj}}{\p x^i}(\g )\right)\dot{\g}^k\dot{\g}^j\\
    &=& \sum_\ell g_{i\ell }(\g)\left(\ddot{\g}^\ell + \sum_{j,k}\G_{jk}^\ell (\g ) \dot{\g}^j\dot{\g}^k\right).
 \end{eqnarray*} 
 This shows that the Euler-Lagrange equations\index{Euler-Lagrange!equations} are equivalent to the geodesic equations\index{Geodesic!equations} 
 \[ \ddot{\g}^\ell + \sum \G_{jk}^\ell  (\g )\dot{\g}^j\dot{\g}^k=0,\quad  \ell =1,\ldots ,n.\] 
 Using the covariant derivative\index{Covariant derivative}\index{Derivative!Covariant} $\n$\nomenclature[snabla]{$\n$}{covariant derivative or connection} they can be written as $\n_{\g'}\g'=0$ or $\frac{\n}{dt}\g'=0$.  
\end{Ex}
\bp Let  $V$ be a smooth function on a pseudo-Riemannian manifold $(M,g)$ and consider the Lagrangian
\be\eqlabel{cm_std_lagr} \mathcal{L}(v) = \frac12 g(v,v) -V(\pi (v)),\quad v\in TM, \ee 
of Example \ref{T-V:Ex}. Then a curve $\g : I \ra M$ is a motion if and only if it satisfies the following equation
\be \label{geo+pot:Eq} \frac{\n}{dt}\g'+ \mathrm{grad}\, V|_{\g}=0\ee 
\ep 

\pf We have already shown that the Euler-Lagrange one-form in the case $V=0$ is given by  
$\a = -g\left(\frac{\n}{dt}\g',\cdot \right)$. It remains to compute the contribution of the potential $V$ to the Euler-Lagrange one-form, 
which is 
\[ -\sum \frac{\p (V\circ \pi) }{\p q^i}(\g')dx^i= -\sum \frac{\p V }{\p x^i}(\g)dx^i =-dV_\g.\]
So in total we obtain 
\[ -\a = g\left( \frac{\n}{dt}\g',\cdot \right) +dV|_\g\]
and, hence, 
\[ -g^{-1}\a =  \frac{\n}{dt}\g'+ \mathrm{grad}\, V|_{\g},\]
where $g^{-1} : T^*M \ra TM$ denotes the inverse of the 
map 
\[ g : TM \ra T^*M, \quad v\mapsto g(v,\cdot ).\] 
\epf

\section{Integrals of motion}
\bd Let  $(M,\mathcal{L})$ be a Lagrangian mechanical system. A  smooth function
$f$ on $TM$ is called an \emph{integral of motion}\index{Integral of motion}\index{Motion!Integral of} if it is constant along
every motion of the system, that is for every motion $\g : I \ra M$ the function
$f(\g' )$ is constant.  
\ed 

Obviously, the integrals of motion are completely determined by the equations of motion and do
not depend on the precise Lagrangian.   Therefore, the notion of an integral of motion  
is meaningful if we are just given a system of second order differential equations\footnote{Such a system will be usually given by 
a consistent specification of a system of second order differential equations for the components of the curve in each local coordinate system. 
A typical example is \re{EL:eq}.}  for a curve $\g : I \ra M$ in a smooth manifold $M$. 
 
 \begin{Prop}[Conservation of energy]  \label{EnergyProp}\index{Energy!Conservation of}\index{Conservation!of energy} Let  $V$ be a smooth function on a pseudo-Riemannian manifold $(M,g)$ and consider the Lagrangian
\be \mathcal{L}(v) = \frac12 g(v,v) -V(\pi (v)),\quad v\in TM, \ee 
of Example \ref{T-V:Ex}. Then the \emph{total energy}  
\[ v\mapsto E(v) := \frac12 g(v,v) +V(\pi (v))\]
is \emph{conserved}, that is it is an integral of motion.
If $V$ is constant then the motion is geodesic. 
 \end{Prop}
 
 \pf We compute the derivative of $t\mapsto E(\g'(t))$ along a motion $\g$: 
 \[ \frac{d}{dt} \left( \frac12 g(\g',\g') +V(\g )\right) = g\left(\frac{\n}{dt}\g',\g'\right) + dV|_\g \g'\stackrel{\re{geo+pot:Eq}}{=}0.\]
If $V$ is constant then the equation of motion \re{geo+pot:Eq} reduces to the geodesic equation. 
 \epf
 
\begin{remark}
The last statement of the proposition is related to conservation of momentum in Newtonian mechanics. 
Recall from \chref{intro} that the momentum of a particle $\g : I \ra \bR^n$ of mass $m$ moving in Euclidean space is 
$\vec{p}= m\g'$. Newton's second law of mechanics 
has the form $\vec{p}' =F$, where $F$ is the force acting on the particle, which may depend on the position 
$\g$ and momentum $\vec{p}$ of the particle. Obviously, $F=0$ implies that the momentum is constant along every motion.  
The conservation of energy holds in Newtonian mechanics if the force field $F$ depends only on the position and is a gradient vector field. 
In fact, then we can write $F= -\mathrm{grad}\, V$ for some function $V$ on $\bR^n$ and hence, Newton's equation is a special case of \re{geo+pot:Eq}. 
\end{remark} 
\bd Let $(M,\mathcal{L})$ be a Lagrangian mechanical system. A diffeomorphism $\varphi : M \ra M$ is called an \emph{automorphism}\index{Automorphism}\index{Symmetry|see{Automorphism}} of
the system if
\[ \mathcal{L}(d\varphi v) = \mathcal{L}(v),\]
 for all $v\in TM$. A vector field $X\in \mathfrak{X}(M)$\nomenclature[aXfrakM]{$\mathfrak{X}(M)$}{set of all smooth vector fields on $M$} is called an \emph{infinitesimal automorphism}\index{Automorphism!Infinitesimal} if its flow consists
 of local automorphisms of $(M,\mathcal{L})$.    
\ed 

 \begin{Th}[Noether's theorem] \label{NoetherThmMech}\index{Noether!theorem} With every infinitesimal automorphism $X$ of a Lagrangian mechanical system
$(M,\mathcal{L})$ we can associate an integral of motion $f=d\mathcal{L}X^{\textnormal{ver}}$, where $X^{\textnormal{ver}}\in \mathfrak{X}(TM)$\nomenclature[aXver]{$X^{\textnormal{ver}}$}{vertical lift of $X$} denotes the vertical lift of  $X$. 
(In local coordinates we have $X^{\textnormal{ver}} = \sum (X^i\circ \pi )\frac{\p}{\p \hat{q}^i}$ if $X= \sum X^i\frac{\p}{\p x^i}$.) \end{Th} 
 
 \pf Let us denote by $\varphi_s$ the flow of $X$. For every motion $\g : I \ra M$ of the system with values in the domain 
 of definition of $\varphi_s$ the curve $\g_s=\varphi_s \circ \g$ is again a motion and 
 \[ \g_s' = d\varphi_s\g',\quad  \left. \frac{\p}{\p s}\right|_{s=0} \g_s' = \tilde{X}|_{\g'},\]
 where 
 \[ TM\ni v\mapsto \tilde{X}(v)=\left.\frac{\p}{\p s}\right|_{s=0}d\varphi_s(v)\] 
 denotes the vector field on $TM$ the flow of which is $d\varphi_s$.  
Differentiating the equation  $\pi \circ d\varphi_s =  \varphi_s \circ \pi$  with respect to $s$, we obtain
$d\pi \circ \tilde{X} = X \circ \pi$.  In local coordinates $(x^i)$ on $M$ and corresponding local coordinates $(q^i,\hat{q}^i)$ on $TM$ 
this means that 
 \[ \tilde{X}= \sum \left( (X^i\circ \pi) \frac{\p}{\p q^i}+ Y^i \frac{\p}{\p \hat{q}^i}\right) , \quad X=\sum X^i\frac{\p}{\p x^i}.\]  
 The locally defined functions $Y^i$ on $TM$ can be computed at $v\in TM$ as follows:
 \begin{eqnarray*} Y^i(v) &=& d\hat{q}^i\tilde{X}(v) = \left. \frac{\p}{\p s}\right|_{s=0}\hat{q}^i(d\varphi_sv) = 
 \left. \frac{\p}{\p s}\right|_{s=0}\hat{q}^i\left(\sum_{j,k} v^j\frac{\p \varphi_s^k}{\p x^j}\frac{\p}{\p x^k}\right)\\
 &=& \left. \frac{\p}{\p s}\right|_{s=0}\sum_{j} v^j\frac{\p \varphi_s^i}{\p x^j} 
 =dX^i(v), 
 \end{eqnarray*}
 where $\varphi_s^i := x^i\circ \varphi_s$ and we have used that $\left. \frac{\p}{\p s}\right|_{s=0}\varphi_s^i =X^i$. 
 The above implies that 
 \be  \label{zero:Equ} 0=\left. \frac{\p}{\p s}\right|_{s=0} \mathcal{L}(\g_s') =d\mathcal{L}\tilde{X}|_{\g'} = \sum  \left( \left. \frac{\p \mathcal{L}}{\p q^i}\right|_{\g'}X^i(\g )   + 
 \left.\frac{\p \mathcal{L}}{\p \hat{q}^i}\right|_{\g'}dX^i (\g')\right).\ee
 Using the Euler-Lagrange equations\index{Euler-Lagrange!equations} we can now compute 
  \begin{eqnarray*}  \frac{d}{d t}f(\g') &=& \frac{d }{d t}\sum \left. \frac{\p \mathcal{L}}{\p \hat{q}^i}\right|_{\g'} X^i(\g)\\
  &=& \sum \underbrace{\left( \frac{d }{d t} \left.\frac{\p \mathcal{L}}{\p \hat{q}^i}\right|_{\g'}\right)}_{=\left. \frac{\p \mathcal{L}}{\p q^i}\right|_{\g'}}X^i(\g)  + \sum \left.\frac{\p \mathcal{L}}{\p \hat{q}^i}\right|_{\g'}dX^i(\g')
  \stackrel{\re{zero:Equ}}{=} 0\end{eqnarray*}  
 \epf 
 \bp The group of automorphisms of the Lagrangian system of Example \ref{T-V:Ex} is given by the Lie subgroup 
 \[ \mathrm{Aut}(M,\mathcal{L}) = \{ \varphi \in \mathrm{Isom}(M,g)| V\circ \varphi = V\}\subset \mathrm{Isom}(M,g).\]
 Its Lie algebra consists of all Killing vector fields $X$ such that  $X(V)=0$.  
 \ep 
 \pf See \appref{exercises}, \excref{6}.
 \epf 
\bc Let  $V$ be a smooth function on a pseudo-Riemannian manifold $(M,g)$ and consider the Lagrangian $\mathcal{L}(v) = \frac12 g(v,v) -V(\pi (v))$, $v\in TM$,
of Example \ref{T-V:Ex}. Then every Killing vector field $X$ such that $X(V)=0$ gives rise to an integral of motion 
$f(v)= g(v,X(\pi (v)))$, $v\in TM$. 
\ec 
\bc \label{momCor} (Conservation of momentum)\index{Conservation!of momentum}\index{Momentum!Conservation of} Consider, as in the previous corollary, the Lagrangian of Example \ref{T-V:Ex}. Assume that 
with respect to some coordinate system on $M$ the metric $g$ and the potential $V$ are both invariant under translations
in one of the coordinates $x^i$. Then the function 
\[ p_i := \sum \tilde{g}_{ij}\hat{q}^j\]
is an integral of motion on the coordinate domain.  
\ec 
Since the Euclidian metric is translational invariant we have the following special case of Corollary \ref{momCor}.
\bc Let $V=V(x^2,\ldots, x^n)$ be a smooth function on Euclidean space $\bR^n$ which does not depend on the first coordinate and consider the Lagrangian $\mathcal{L}(v)=\frac12 m\langle v,v\rangle -V(\pi (v))$, 
$v\in T\bR^n$, of \exref{NewtonEx}. 
Then the first component $p_1$ of the momentum vector $\vec{p}=\sum p_ie_i$\nomenclature[apvec]{$\vec{p}$}{momentum vector} is an integral of motion, where 
$p_i(v) := m\langle v,e_i\rangle$, $v\in T\bR^n$. 
\ec
Note that the result still holds if we replace the Euclidean scalar product by a pseudo-Euclidean scalar product, such as the Minkowski scalar product. 
\begin{Cor}[Conservation of angular momentum] \label{LCor}\index{Conservation!of angular momentum}\index{Angular momentum}\index{Angular momentum!Conservation of}\index{Momentum!Angular} Let $V$ be a smooth \emph{radial}  function on $\bR^3 \setminus \{ 0\}$, that is $V$ depends only on the radial coordinate\index{Radial!coordinate}\index{Coordinate!Radial} $r$, and consider the Lagrangian
$\mathcal{L}(v) = \frac12 m\langle v,v\rangle - V(\pi (v))$, $v\in T(\bR^3 \setminus \{ 0\})$. Then the components of the angular momentum vector\nomenclature[aLvec]{$\vec{L}$}{angular momentum vector}
\[ \vec{L}(v) = \pi (v) \times \vec{p}(v) = m \left( \begin{array}{c} x^1\\
x^2\\
x^3 
\end{array}\right) \times  \left( \begin{array}{c} v^1\\
v^2\\
v^3 
\end{array}\right) \]
are integrals of motion. Here, $x^1, x^2, x^3$ denote the components of $x=\pi (v)$. 
\end{Cor}

\pf This can be proven either directly from the equations of motion or by applying Noether's theorem (see \appref{exercises}, \excref{9}). The first proof uses the fact that the moment of force $\g \times F(\g )$, $F=-\mathrm{grad}\, V$, is zero for every curve 
$\g : I \ra \bR^3 \setminus \{ 0\}$ and yields also the following proposition. 
\epf 
\bd A vector field $F$ on $\bR^3 \setminus \{ 0\}$ is called \emph{radial}\index{Radial!vector field} if there exists a radial function\index{Radial!function} 
$f$ on $\bR^3 \setminus \{ 0\}$ such that $F(x)=f(x)\frac{x}{|x|}$ for all $x\in \bR^3 \setminus \{ 0\}$.
\ed 
\bp Let $F$ be a smooth radial vector field on $\bR^3 \setminus \{ 0\}$. 
Then the angular momentum is constant for every solution of Newton's equation $\frac{d}{dt}\vec{p}(t)=F(t)$, where we are using the 
usual notation $\vec{p}(t) = \vec{p}(\g' (t))$ and $F(t) = F(\g (t))$. 
\ep 
\begin{remark} A function or vector field on $\bR^3\setminus \{ 0 \}$ is radial if and only if it is \emph{spherically symmetric}\index{Symmetry!Spherical}, that is invariant under $\mathrm{SO}(3)$ (or, equivalently, $\mathrm{O}(3)$), see \appref{exercises}, \excref{11,12}. The name is due to the fact that $\mathrm{SO}(3)$ is the group of orientation preserving isometries of the sphere $S^2$ and $\mathrm{O}(3)= \mathrm{Isom}(S^2)$.  
\end{remark}
In the next section we will see how to use the conservation of energy and angular momentum to analyze the motion in a 
radial potential.

\section{Motion in a radial potential}\index{Radial!potential}\index{Potential!Radial}
We consider the Lagrangian $\mathcal{L}(v) = \frac12 m \langle v,v\rangle -V(r)$ on $\bR^3 \setminus \{ 0 \}$, where $V$ is a smooth function of the radial coordinate 
$r=|x|$ alone. Since the mass can be absorbed into the definitions of $V$ and $\mathcal{L}$ (denoting $\mathcal{L}/m$ and $V/m$ again by $\mathcal{L}$ and $V$) we may as well put $m=1$.  So from now on we consider a particle of unit mass. 

We know by the results of the previous section that the energy $E(t) = \frac12 \langle \g '(t) ,\g '(t)\rangle$ $+ V(|\g (t)|)$ and 
angular momentum $\vec{L}(t)=\g (t) \times \g'(t)$ are constant for every motion $\g : I \ra \bR^3 \setminus \{ 0 \}$ of the system. 
  
The first observation is that the conservation of angular momentum implies that the motion is \emph{planar}\index{Motion!Planar}, that is contained in a plane. 
\bp If the vector $\vec{L}= \g \times \g'$ is constant along the curve $\g : I \ra \bR^3\setminus \{ 0\}$ then $\g$ is a planar curve. 
\ep 

\pf If $\vec{L}=0$ then $\g$ is a radial curve (see \appref{exercises}, \excref{7}) and thus planar. Therefore we can assume that $\vec{L}$ is a nonzero
constant vector. Since the cross product of two vectors is always perpendicular to both of them, we know that $\g (t) \in \vec{L}^\perp$ for all $t$. 
So $\g$ is contained in the plane $\vec{L}^\perp$. 
\epf 

By a change of coordinates we can assume that the motion is restricted to the plane $e_3^\perp$ and use polar coordinates 
$(r,\varphi)$  in that plane 
rather than Cartesian coordinates $(x^1,x^2)$ to describe the motion. In these coordinates, 
the Euclidean metric is $dr^2 +r^2d\varphi^2$ and, hence,  
\be \label{EEqu} E=\frac12  \left(\dot{r}^2 + r^2\dot{\varphi}^2\right) + V(r).\ee
To compute the cross product $\g \times \g'$ in polar coordinates we recall that
\[ \p_r=\cos \varphi \p_1 + \sin \varphi \p_2,\quad \p_\varphi = -r\sin \varphi \p_1 + r\cos \varphi \p_2,\]
where we have abbreviated $\p_r = \p /\p r$, $\p_\varphi = \p/\p \varphi$ and $\p_i = \p/\p x^i$. From these 
relations we easily obtain 
\[ \p_r \times \p_\varphi = r e_3\]
and hence
\[ \vec{L} = \g \times \g' = r\p_r  \times \left(\dot{r}\p_r + \dot{\varphi} \p_\varphi\right) = 
r^2\dot{\varphi}e_3.\]
Changing the time parameter $t$ to $-t$, if necessary, we can assume that the constant $r^2\dot{\varphi}\ge 0$ 
and, hence, 
\be \label{LEqu} L := | \vec{L}| = r^2\dot{\varphi } .\ee\nomenclature[aL]{$L$}{length of the angular momentum vector $\vec{L}$}
Substituting $\dot{\varphi}=L/r^2$ into the energy equation \re{EEqu} we arrive at
\be \label{EffEEqu} E= \frac12 \dot{r}^2 + V_{\textnormal{eff}}(r),\quad V_{\textnormal{eff}}(r) :=  V(r) + \frac{L^2}{2r^2},\ee
where $V_{\textnormal{eff}}$\nomenclature[aVeff]{$V_{\textnormal{eff}}$}{effective potential} is called the \emph{effective potential}\index{Potential!Effective}. 
As a consequence, we obtain the following result. 
\bt Consider a particle of unit mass moving in a radial potential $V(r)$ in $\bR^3$. Then the radial coordinate obeys the equation of motion 
$\ddot{r}=-V_{\textnormal{eff}}'(r)$ 
of  a particle $t\mapsto r(t)>0$  moving in the Euclidean line under the potential
\be \label{VeffEq} V_{\textnormal{eff}}(r) :=  V(r) + \frac{L^2}{2r^2},\ee
which reduces to the first order equation 
\be \label{dotrEqu} \dot{r} =\pm  \sqrt{2\left(E-V_{\textnormal{eff}}(r)\right)}\ee
solvable by separation of variables. 
The angular coordinate is then given by 
\[ \varphi (t)  = L\int_{0}^t\frac{ds}{{r(s)^2}} + \varphi (0),\]
where the initial value $\varphi (0)$ is freely specifiable or, as a function of $r$, by
\[ \varphi (r) = \pm L\int_{r_0}^r\frac{ds}{s^2\sqrt{2\left(E-V_{\textnormal{eff}}(s)\right)}}+ \varphi (r_0).\]
\et

\pf The energy equation  \re{EffEEqu} immediately implies \re{dotrEqu}, which can be solved by separation of variables:  
\[ \pm \frac{dr}{\sqrt{2\left(E-V_{\textnormal{eff}}(r)\right)}} =dt.\] 
Integration yields $t$ as a function of $r$. Then $r$ as a function of $t$ is the inverse of that function. 
Differentiating \re{dotrEqu} gives Newton's equation  $\ddot{r}=-V_{\textnormal{eff}}'(r)$.  
Given $r$ as a function of $t$, the angular coordinate is now determined from $\dot{\varphi}=L/r^2$ by integration as claimed.
Finally, to express $\varphi$ as a function of $r$ one does not need to know the inverse of the function $r\mapsto t(r)$ but can proceed as follows: 
\[ \frac{d}{dr}\varphi = \dot{\varphi}\frac{dt}{dr}=  \pm \frac{L}{r^2\sqrt{2\left(E-V_{\textnormal{eff}}(r)\right)}}  .\]
\epf 
Next we specialize the discussion to Newton's potential $V(r) = -\frac{M}{r}$, from which we will derive, in particular, 
Kepler's laws of planetary motion.

\subsection{Motion in Newton's gravitational potential}\index{Newton!theory of gravity}
By \re{VeffEq} we know that the motion in Newton's gravitational potential is governed by the following
effective potential\index{Potential!Effective}  
\be \label{fEq}  f(r):=V_{\textnormal{eff}}(r) = -\frac{M}{r}+ \frac{L^2}{2r^2}.\ee
We will assume that $L>0$; the case $L=0$ is treated in \appref{exercises}, \excref{8}. 
The qualitative behavior of the effective potential is essential for the qualitative analysis of the possible orbits. The  proof of the next proposition
is elementary. 
\bp \label{fProp} The function $f: \bR^{>0} \ra \bR$ defined in \re{fEq} has the following properties:
\begin{enumerate}
\item $f$ has a unique zero, at $r_0:=\frac{L^2}{2M}$. 
\item It is positive for $r<r_0$  and negative for $r>r_0$. 
\item $\lim_{r\ra 0}f(r) = \infty$, $\lim_{r\ra \infty}f(r) =0$.
\item $f$ has a unique critical point, at $r_{\textnormal{min}} := \frac{L^2}{M}>r_0$, where $f$ attains its global minimum $f(r_{\textnormal{min}})=-\frac{M^2}{2L^2}$.
\item $f$ is strictly decreasing for $r<r_{\textnormal{min}}$ and strictly increasing for $r>r_{\textnormal{min}}$. 
\end{enumerate}
\ep
\bp \begin{enumerate}
\item[(i)] The range of possible energies for motions with nonzero angular momentum in Newton's gravitational potential is given by the 
interval $\left[-\frac{M^2}{2L^2},\infty\right)$. The motions  are unbounded if $E\ge 0$ and bounded 
if $E<0$. The orbit of minimal energy $E_{\textnormal{min}}=-\frac{M^2}{2L^2}$ is a circle of radius $\frac{L^2}{M}$.  
\item[(ii)] The perihelion distance $r_{\textnormal{per}}$, that is the minimal distance to the origin, for a given energy $E\ge 0$ (and $L>0$) 
is given by the unique positive solution of the equation $r^2+\frac{M}{E}r-\frac{L^2}{2E}=0$, which is $r_{\textnormal{per}}=\frac{M}{2E} \left(-1+\sqrt{1+2\frac{L^2E}{M^2}}\right)$. For $E_{\textnormal{min}}<E<0$ (and $L>0$)  the distance to the origin varies between the two positive solutions of the 
equation $r^2+\frac{M}{E}r-\frac{L^2}{2E}=0$, which are 
$r_{\textnormal{per}}=\frac{M}{2|E|} \left(1-\sqrt{1+2\frac{L^2E}{M^2}}\right)$ and $r_{\textnormal{aph}}=\frac{M}{2|E|} \left(1+\sqrt{1+2\frac{L^2E}{M^2}}\right)$
(the aphelion distance, that is the maximal distance to the origin). 
\end{enumerate}
\ep 

\pf The first part follows immediately from Proposition \ref{fProp}. For the second part note that at the points of minimal or 
maximal distance to the origin we have $\dot{r}=0$ and, hence, $E=V_{\textnormal{eff}}=f$. The last equation is equivalent 
to the quadratic equation  $r^2+\frac{M}{E}r-\frac{L^2}{2E}=0$, which has a unique solution for $E=E_{\textnormal{min}}$
(the radius of the circular orbit), two positive solutions  for $E_{\textnormal{min}} < E<0$, and only one positive solution
for $E\ge 0$. 
\epf 
Next we will solve the equations of motion. The function $t\mapsto r(t)$ is determined by
\[ \ddot{r}= -f'(r)= -\frac{M}{r^2}+ \frac{L^2}{r^3}.\]
To solve this equation we make the substitution $u=1/r$ and compute 
\[ \frac{du}{d\varphi} = \left. \frac{du}{dt}\right|_{t(\varphi )}\frac{dt}{d\varphi} = -\left. \frac{\dot{r}}{r^2\dot{\varphi}}\right|_{t(\varphi )} = -\left.\frac{\dot{r}}{L}\right|_{t(\varphi )},\] 
which implies 
\[ \frac{d^2u}{d\varphi^2}= -\frac{\ddot{r}}{L}\frac{dt}{d\varphi}=-\frac{\ddot{r}}{L\dot{\varphi}}=-\frac{\ddot{r}r^2}{L^2}=\frac{f'(r)r^2}{L^2}= \frac{M}{L^2}- u.\]
So we obtain the equation 
\[ \frac{d^2u}{d\varphi^2} +u = \frac{M}{L^2},\]
the general solution of which is
\[ u = k \cos (\varphi -\varphi_0)+ \frac{M}{L^2},\]
where $k\ge 0$ and $\varphi_0$ are constants. 
By choosing $\varphi_0=0$, this allows us to write 
\be \label{rpeEqu} r = \frac{p}{1+\varepsilon \cos \varphi},\quad p:= \frac{L^2}{M},\quad \varepsilon := \frac{kL^2}{M}.\ee 
For $0\le \varepsilon< 1$ this is an ellipse of eccentricity $\varepsilon$. 

For $\varepsilon = 1$ it is a parabola and for $\varepsilon >1$ a component of a hyperbola. (Observe that for the parabola and hyperbola the angle $\varphi$ is constrained by
the condition that $\varepsilon \cos \varphi > -1$.)  Recall that for an ellipse with major half-axis $a$ and minor half-axis $b$ 
the distance between the two focal points is $2c$, where $c^2 = a^2-b^2$, and the eccentricity and parameter are given by 
\[ \varepsilon = \frac{c}{a}, \quad p= \frac{b^2}{a} \; , \]
see e.g.\ \cite[VIII.43]{K} for a detailed discussion of conic sections. 
 
Summarizing we obtain:
\bt The motions  of a particle (with nonzero angular momentum)  in Newton's gravitational potential are along conic sections with a focus at the origin. 
The bounded motions are ellipses.
\et 
The latter statement is the content of \emph{Kepler's first law}\index{Kepler's Laws} (of planetary motion)\index{Planetary motion}\index{Motion!Planetary}: The planets move along ellipses with the sun at a focal point.  
 \emph{Kepler's second law}\index{Kepler's Laws} asserts that the area swept out  during a time
 interval $[t,t+s]$ by the line segment connecting the sun and the planet  depends only on its duration $s$ and not on the initial time $t$.  
 This is a simple consequence of the conservation of angular momentum as shown in \appref{exercises}, \excref{13}:
 \bp \label{AreaProp} Let $\g : I \ra \bR^3\setminus \{ 0\}$ be the motion of a particle in a radial force field $F$ according to Newton's law $m\g'' = F(\g )$. 
Then the angular momentum\index{Angular momentum}\index{Momentum!Angular} vector $\vec{L}$ is constant, the motion is planar and the area $A(t_0,t_1)$ swept out by the vector $\g$ during a time interval $[t_0,t_1]$ 
is given by
\[ A(t_0,t_1) = \frac{L}{2}(t_1-t_0),\]
where $L$ is the length of the angular momentum vector. 
 \ep 
As a corollary we obtain \emph{Kepler's third law}\index{Kepler's Laws}: 
\bc The elliptic orbits in Newton's gravitational potential are periodic with period
\[ T =  \frac{2\pi}{\sqrt{M}}a^{3/2},\]
where $a$ is the major half-axis of the ellipse.  
\ec
\pf 
For a given initial time $t_0\in \bR$ let us denote by $T=T(t_0)$ the smallest positive real number such that  $(r,\varphi)|_{t_0+T}=(r,\varphi)_{t_0}$, 
where $t\mapsto (r(t),\varphi (t))$ is an elliptic  motion of the system. Since $A(t_0,t_0+T)=\frac{L}{2}T$ is the area of the ellipse, $T$ 
is clearly independent of $t_0$, which shows that the motion is periodic. 
The area of the ellipse is $A=\pi ab$ and, hence,  
\[ T=\frac{2}{L}A = \frac{2\pi ab}{L}\stackrel{\re{rpeEqu}}{=} \frac{2\pi ab}{\sqrt{pM}}= \frac{2\pi a^{3/2}}{\sqrt{M}}.\]  
\epf\clearpage{}%
\clearpage{}%

\chapter{Hamiltonian Mechanics}
\chlabel{Hamiltonian_systems} 

\abstract*{We present Hamilton's formulation of classical mechanics. In this formulation, the $n$ second-order equations of motion of an $n$-dimensional mechanical system are replaced by an equivalent set of $2n$ first-order equations, known as Hamilton's equations. There are problems where it is favorable to work with the $2n$ first-order equations instead of the corresponding $n$ second-order equations. After introducing basic concepts from symplectic geometry, we consider the phase space of a mechanical system as a symplectic manifold. We then discuss the relation between Lagrangian and Hamiltonian systems. We show that, with appropriate assumptions, the Euler-Lagrange equations of a Lagrangian mechanical system are equivalent to Hamilton's equations for a Hamiltonian, which can be obtained from the Lagrangian by a Legendre transformation. In the last part, we consider the linearization of mechanical systems as a way of obtaining approximate solutions in cases where the full non-linear equations of motion are too complicated to solve exactly. This is an important tool for analyzing physically realistic theories as these are often inherently non-linear.}

\abstract{We present Hamilton's formulation of classical mechanics. In this formulation, the $n$ second-order equations of motion of an $n$-dimensional mechanical system are replaced by an equivalent set of $2n$ first-order equations, known as Hamilton's equations. There are problems where it is favorable to work with the $2n$ first-order equations instead of the corresponding $n$ second-order equations. After introducing basic concepts from symplectic geometry, we consider the phase space of a mechanical system as a symplectic manifold. We then discuss the relation between Lagrangian and Hamiltonian systems. We show that, with appropriate assumptions, the Euler-Lagrange equations of a Lagrangian mechanical system are equivalent to Hamilton's equations for a Hamiltonian, which can be obtained from the Lagrangian by a Legendre transformation. In the last part, we consider the linearization of mechanical systems as a way of obtaining approximate solutions in cases where the full non-linear equations of motion are too complicated to solve exactly. This is an important tool for analyzing physically realistic theories as these are often inherently non-linear.}

\section{Symplectic geometry and Hamiltonian systems}

\bd A \emph{symplectic manifold}\index{Symplectic!manifold}\index{Manifold!Symplectic} $(M,\o )$ is a smooth manifold $M$ endowed with a \emph{symplectic form}\index{Symplectic!form} $\o$, that is a non-degenerate
closed $2$-form $\o$. 
\ed 
\begin{Ex}[Symplectic vector space] \label{ocan:ex}Let $\o$ be a non-degenerate skew-symmetric bilinear form on a finite dimensional real vector space $V$. 
Then $(V,\o )$ is called a \emph{(real) symplectic vector space}\index{Symplectic!vector space}. Every symplectic vector space is of even dimension and there exists a linear
isomorphism $V \ra \bR^{2n}$, $2n=\dim V$, which maps $\o$ to the \emph{canonical symplectic form} 
\be \label{ocan:eq} \o_{\textnormal{can}} = \sum_{i=1}^n dx^i\wedge dx^{n+i},\ee
where $(x^1,\ldots ,x^{2n})$ are the standard coordinates on $\bR^{2n}$. 
\end{Ex}
It is a basic result in symplectic geometry, known as Darboux's theorem\index{Darboux!theorem|see{coordinates}}\index{Darboux!coordinates}\index{Coordinate!Darboux}, see e.g.\ \cite[Thm.\ 3.2.2]{AM},  that for 
every point in a symplectic manifold $(M,\o )$ there exists a local 
coordinate system $(x^1,\ldots ,x^{2n})$ defined in a neighborhood $U$ of that point, such that 
\[ \o|_U = \sum_{i=1}^n dx^i\wedge dx^{n+i}.\] 
So $\o|_U$ looks like the canonical symplectic form on $\bR^{2n}$. 
\begin{Ex}[Cotangent bundle] Let $\pi : N=T^*M\ra M$ be the cotangent bundle of a manifold $M$. We define a $1$-form
$\lambda$ on $N$ by
\[ \lambda_\xi (v) := \xi (d\pi (v)),\]
for all $\xi \in N$, $v\in T_\xi N$. The $1$-form is called the  \emph{Liouville form}\index{Liouville form}. Its differential
$\o =d\l$\nomenclature[gomega]{$\o$}{canonical symplectic form}\nomenclature[glambda]{$\l$}{Liouville form} is a symplectic form on $N$, which is called the \emph{canonical symplectic form}\index{Canonical symplectic form} of the cotangent bundle $N$. 
To check that $\o$ is indeed non-degenerate, let us compute the Liouville form in coordinates 
$(q^1,\ldots ,q^n,p_1,\ldots ,p_n)$ on $\pi^{-1}(U)=T^*U \subset T^*M$ associated with coordinates $(x^1,\ldots ,x^n)$ on some 
open set $U\subset M$:
\[ q^i = x^i \circ \pi ,\quad p_i (\xi ) = \xi \left(\frac{\p}{\p x^i}\right), \quad \xi \in T^*U.\]
Since under the projection $d\pi : TN\ra TM$ the vector fields $\p/\p q^i$ and $\p/\p p_i$ are mapped to $\p/\p x^i$ and zero, respectively, at every point $\xi = \sum \xi_jdx^j|_{\pi (\xi)}\in T^*U$ we have 
\[ \l_\xi \left(\frac{\p}{\p q^i}\right) = \xi \left( \frac{\p}{\p x^i}\right)=\xi_i = p_i(\xi ),\quad\l_\xi \left(\frac{\p}{\p p_i}\right)=0.\] 
This shows that 
\[ \l|_U = \sum p_idq^i,\quad \o|_U = \sum dp_i\wedge dq^i,\]
proving that $\o$ is a symplectic form. 
\end{Ex}

Given a smooth function $f$ on a symplectic manifold $(M,\o )$, there is a unique vector field $X_f$ 
such that 
\[ df = -\o (X_f,\cdot ).\]
\bd The vector field $X_f$\nomenclature[aXf]{$X_f$}{Hamiltonian vector field associated with a smooth function $f$} is called the \emph{Hamiltonian vector field}\index{Hamiltonian!vector field} associated with $f$. 
\ed 

According to Darboux's theorem,  we can locally write $\o = \sum dp_i\wedge dq^i$ for some local coordinate
system $(q^1,\ldots ,q^n,p_1,\ldots ,p_n)$\nomenclature[aq1p1]{$(q^1,\ldots ,q^n,p_1,\ldots ,p_n)$}{Darboux coordinates} on $M$. In such coordinates, which will be called \emph{Darboux coordinates}\index{Darboux!coordinates}\index{Coordinate!Darboux}, 
we can easily compute 
\be \label{HamvfEq} X_f= \sum\left(  \frac{\p f}{\p p_i}\frac{\p}{\p q^i} -\frac{\p f}{\p q^i}\frac{\p}{\p p_i} \right), \ee
see \appref{exercises}, \excref{19}. 
\bd A \emph{Hamiltonian system}\index{Hamiltonian!system} $(M,\o , H)$ is a symplectic manifold $(M,\o )$ endowed 
with a function $H\in C^\infty (M)$\nomenclature[aH]{$H$}{Hamiltonian}, called the \emph{Hamiltonian}\index{Hamiltonian}. An integral curve  $\g : I \ra M$ 
of the Hamiltonian vector field $X_H$, defined on an open interval $I\subset \bR$, is called a \emph{motion}\index{Motion} of the Hamiltonian system. The corresponding system of ordinary differential equations
\be\eqlabel{Hamilton_eq} \g' = X_H(\g)\ee
is called \emph{Hamilton's equation}\index{Hamilton's equation}. An \emph{integral of motion}\index{Integral of motion}\index{Motion!Integral of}
of the Hamiltonian system is a function $f\in C^\infty (M)$ which is constant along every motion. The symplectic manifold
$(M,\o )$ is called the \emph{phase space}\index{Phase space} of the Hamiltonian system. 
\ed 
\bp Let $(M,\o , H)$ be a Hamiltonian system. Then $H$ is an integral of motion. 
\ep 

\pf Let $\g : I \ra M$ be a motion of the system. Then 
\[ \frac{d}{dt}H(\g (t))= dH\g'(t) = -\o (X_H(\g (t)),\g'(t))= -\o (X_H,X_H)|_{\g(t)}=0.\]
\epf 
\bp \label{HamEqu}In Darboux coordinates\index{Darboux!coordinates}\index{Coordinate!Darboux}, Hamilton's equation\index{Hamilton's equation} for a curve $\g : I \ra M$ takes the form
\be\eqlabel{Hamilton_eqs} \dot{q}^i(t) = \frac{\p H}{\p p_i}(\g (t)),\quad \dot{p}_i(t) = -\frac{\p H}{\p q^i}(\g (t)),\ee
for all $i=1,\ldots,n$ (here, $n=\frac{\dim M}{2}$), where $q^i(t) = q^i(\g (t))$, $p_i(t) = p_i (\g (t))$. 
\ep 
\pf This follows immediately from \re{HamvfEq} by comparing $\g'= \sum \left( \dot{q}^i\frac{\p}{\p q^i} + \dot{p}_i \frac{\p}{\p p_i}\right)$ with 
$X_H(\g )$.  
\epf 
We refer henceforth to \eqref{Hamilton_eq} as Hamilton's equation and to \eqref{Hamilton_eqs} as Hamilton's equation\emph{s}.

\section{Relation between Lagrangian and Hamiltonian systems} 

In this section, we clarify the relation between Lagrangian and Hamiltonian systems. We will show, under certain assumptions, that the Euler-Lagrange equations\index{Euler-Lagrange!equations} of a Lagrangian mechanical system $(M,\mathcal{L})$ are equivalent to Hamilton's equation\index{Hamilton's equation} for a certain Hamiltonian $H$ on $T^*M$ endowed with the canonical 
symplectic form $\o$. If this is true, then the Lagrangian mechanical system $(M,\mathcal{L})$ should possess an integral of motion,
corresponding to the integral of motion $H$ of the Hamiltonian system $(T^*M,\o , H)$.

\subsection{Hamiltonian formulation for the Lagrangian systems of Example \ref{T-V:Ex}}
In the case of Lagrangians 
of the form $\mathcal{L}(v) = \frac12 g(v,v) -V(\pi (v))$, $v\in TM$, considered in Example \ref{T-V:Ex}, we showed in Proposition \ref{EnergyProp}
that the energy $E(v) = \frac12 g(v,v) +V(\pi (v))$, $v\in TM$, is indeed an integral of motion. Moreover, we have a natural identification
of $TM$ with $T^*M$ by means of the pseudo-Riemannian metric:
\[ \phi =\phi_g : TM \ra T^*M,\quad v\mapsto g(v,\cdot ).\]
\bp \label{Legendre1Prop} Let $(M,g)$ be a pseudo-Riemannian manifold and denote by $\phi :TM\ra T^*M$ the isomorphism of vector bundles induced by $g$. Let $V$ be a smooth function on  
$M$ and consider the Lagrangian $\mathcal{L}$ of
Example \ref{T-V:Ex}. 
\begin{enumerate}
\item[(i)] 
Then a smooth curve $\g: I \ra M$ is a solution of the  Euler-Lagrange equations  if and only if the curve $\phi \circ \g' : I \ra T^*M$ is a solution of Hamilton's equation for the Hamiltonian $H = E\circ \phi^{-1}$. 
\item[(ii)] 
Conversely, if a smooth curve $\tilde{\g} : I \ra T^*M$ is a motion of the Hamiltonian system $(T^*M,\o ,H)$ then 
the curve $\pi \circ \tilde{\g} : I \ra M$ is a motion of the Lagrangian system $(M,\mathcal{L})$, where $\pi : T^*M \ra M$ denotes the projection. The maps $\g \mapsto \phi \circ \g'$ and $\tilde{\g} \mapsto \pi \circ \tilde{\g}$
are inverse to each other when restricted to solutions of the Euler-Lagrange equations and Hamilton's equations, respectively. 
\end{enumerate}
\ep 
\pf We prove (i). Part (ii) is similar and part of \appref{exercises}, \excref{22}. Let $(x^1,\ldots ,x^n)$ be coordinates defined on some open set $U\subset M$. They induce 
coordinates $(q^1,\ldots ,q^n,$ $\hat{q}^1,\ldots , \hat{q}^n)$ on $TU \subset TM$ and 
$(q^1,\ldots ,q^n$, $p_1,\ldots ,p_n)$ on $T^*U\subset T^*M$. In terms of these coordinates
$\phi$ is given by 
\[ q^i\circ \phi = q^i,\quad p_i \circ \phi = \sum (g_{ij}\circ \pi )\hat{q}^j =\frac{\p \mathcal{L}}{\p \hat{q}^i},\]
where $\pi : TM \ra M$ denotes the projection. The inverse $\psi : T^*M\ra TM$ is given by
\[ q^i\circ \psi = q^i,\quad \hat{q}^i \circ \psi = \sum (g^{ij}\circ \pi ) p_j,\]
where $(g^{ij})$ is the matrix inverse to $(g_{ij})$ and $\pi : T^*M \ra M$ is the projection. 
Therefore we obtain
\[ H= E \circ \psi = \frac12 \sum (g^{ij}\circ \pi )p_ip_j + V\circ \pi,\quad \pi : T^*M \ra M,\]
and by Proposition \ref{HamEqu}, Hamilton's equation takes the form
\begin{eqnarray*} \dot{q}^i &=& \sum (g^{ij}(\g ) )p_j=\hat{q}^i,\\
\dot{p}_i &=& - \frac12 \sum \frac{\p g^{k\ell}}{\p x^i}(\g ) p_kp_\ell - \frac{\p V}{\p x^i}(\g )\\
&=&  \frac12 \sum \frac{\p g_{k\ell}}{\p x^i}(\g )\hat{q}^k\hat{q}^\ell - \frac{\p V}{\p x^i}(\g )= \frac{\p \mathcal{L}}{\p q^i}(\g' ),\end{eqnarray*}
where $\g : I \ra M$, $q^i(t) = x^i(\g (t))$, $\hat{q}^i(t) = \hat{q}^i(\g'(t))$ and $p_i(t) = p_i(\phi (\g' (t)))$.   
So we see that $\phi \circ \g' : I \ra T^*M$ is a motion of the Hamiltonian system if and only if $\g' : I \ra TM$ satisfies 
\[ \dot{q}^i = \hat{q}^i,\quad \dot{p}_i = \frac{\p \mathcal{L}}{\p q^i}(\g' ),\]
where $p_i(t) = \frac{\p \mathcal{L}}{\p \hat{q}^i}(\g'(t))$. Substituting the expression for $p_i$ into the second equation one obtains precisely the Euler-Lagrange equations, whereas the first equation holds for every curve $\g$. 
\epf 
Next we will generalize the above constructions to any Lagrangian satisfying an appropriate non-degeneracy  assumption analogous to the 
non-degeneracy of the pseudo-Riemannian metric.

\subsection{The Legendre transform}\seclabel{legendre_trafo}
As a first step we will show that for every Lagrangian mechanical system in the sense of Definition \ref{firstDef}  (that is for which the Lagrangian does not explicitly depend 
on time) we can define an integral of motion which generalizes the energy defined for Example \ref{T-V:Ex}. For this we remark that for the Lagrangian
of Example \ref{T-V:Ex} the energy can be written in the form
\[ E = \sum \frac{\p \mathcal{L}}{\p \hat{q}^i}\hat{q}^i-\mathcal{L}.\]
We use this formula to define the \emph{energy}\index{Energy} $E$ for any Lagrangian $\mathcal{L}$. To show that 
the definition is coordinate independent it is sufficient to remark that the vector field
\[ \xi = \sum \hat{q}^i\frac{\p}{\p \hat{q}^i}\]
is coordinate independent. It is in fact the vector field generated by the one-parameter group of dilatations 
\[ \varphi_t : TM \ra TM,\quad v\mapsto e^tv.\]
Its value at $v\in TM$ is simply 
\[ \xi (v) = \left. \frac{d}{dt}\right|_{t=0} \varphi_t(v)=v,\]
where on the right-hand side $v\in T_xM$, $x=\pi (v)$, is interpreted as vertical vector by means of the canonical 
identification $T_v^{\textnormal{ver}}TM=T_xM$.  
The following result is a generalization of Proposition \ref{EnergyProp}.  
 \begin{Prop}[Conservation of energy]\index{Energy!Conservation of}\index{Conservation!of energy}  \label{conserEProp} Let $(M,\mathcal{L})$ be a Lagrangian mechanical system. Then the 
 energy $E=\xi (\mathcal{L})-\mathcal{L}$ is an integral of motion\index{Integral of motion}\index{Motion!Integral of}. 
\end{Prop}
\pf It suffices to differentiate $t\mapsto E(\g'(t))$ along a motion  $\g : I \ra M$. We obtain
\begin{eqnarray*}  \frac{d}{dt} E(\g') &=&  \frac{d}{dt}\left( \sum \frac{\p \mathcal{L}}{\p \hat{q}^i}\hat{q}^i-\mathcal{L}\right)\\ 
&=& 
\sum \left( \left(\frac{d}{dt}\frac{\p \mathcal{L}}{\p \hat{q}^i}\right)\hat{q}^i  +  \frac{\p \mathcal{L}}{\p \hat{q}^i}\dot{\hat{q}}^i  - \frac{\p \mathcal{L}}{\p q^i}\dot{q}^i-
\frac{\p \mathcal{L}}{\p \hat{q}^i}\dot{\hat{q}}^i \right) ,\end{eqnarray*}
which vanishes by the Euler-Lagrange equations. 
\epf 
Next we generalize the isomorphism $\phi=\phi_g : TM \ra T^*M$ defined by a pseudo-Riemannian metric $g$.
As we have shown, it can be written in terms of the Lagrangian as
\[ q^i\circ \phi = q^i,\quad p_i \circ \phi =\frac{\p \mathcal{L}}{\p \hat{q}^i}.\]
We claim that these formulas define a smooth map $\phi=\phi_\mathcal{L} : TM \ra T^*M$ for any Lagrangian $\mathcal{L}$. 
To see this let us define $\phi_\mathcal{L}$ in a coordinate independent way. For $v\in T_xM$, $x\in M$, we define 
\be \label{phiLEqu} \phi_\mathcal{L} (v) = d\left( \mathcal{L}|_{T_xM}\right)|_v \in (T_xM)^*=T^*_xM.\ee 
What we obtain is a smooth map that maps any vector $v\in TM$ to a covector $\phi_\mathcal{L}(v)\in T^*_xM$ at the point 
$x=\pi (v)$. In other words, $\phi_\mathcal{L}$ is a one-form along the projection map $\pi : TM \ra M$, or, equivalently,  a section of the 
vector bundle $\pi^*T^*M $ over $TM$. Notice that as a map from $TM$ to $T^*M$ it is fiber-preserving but (contrary to $\phi_g$) in general not linear on fibers. 
Also, in general, it is not even a local diffeomorphism. In local coordinates we have 
\[ \phi_\mathcal{L}(v) = \sum \frac{\p \mathcal{L}(v)}{\p \hat{q}^i}dx^i|_x.\]
\bd Let $(M,\mathcal{L})$ be a Lagrangian mechanical system, $n=\dim M$. Then $\mathcal{L} $ is called 
\emph{non-degenerate}\index{Lagrangian!Non-degenerate} if $\phi_\mathcal{L}$ is of maximal rank, that is if 
$d\phi_\mathcal{L}$ has rank~$2n$ everywhere. The Lagrangian is called \emph{nice}\index{Lagrangian!Nice} if it is non-degenerate
and $\phi_\mathcal{L}$ is a bijection. 
\ed
\bp Let $(M,\mathcal{L})$ be a Lagrangian mechanical system. Then the following conditions are equivalent:
\begin{enumerate}
\item $\mathcal{L} $ is non-degenerate.
\item $\phi_\mathcal{L}: TM \ra T^*M$ is a local diffeomorphism. 
\item For all $x\in M$, $\phi_{\mathcal{L}}|_{T_xM} : T_xM \ra T^*_xM$ is of maximal rank.
\item For all $v\in TM$, there exists a coordinate system $(x^i)$ defined on an open neighborhood $U$ of $\pi (v)$ such that the matrix $\left( \frac{\p^2 \mathcal{L}(v)}{\p \hat{q}^i\p \hat{q}^j}\right)$ is  invertible, where $(q^1,\ldots ,q^n,\hat{q}^1,\ldots ,\hat{q}^n)$ are the corresponding coordinates on $TU$. 
\item For all $v\in TM$ and every coordinate system $(x^i)$ defined on an open neighborhood $U$ of $\pi (v)$, the matrix $\left( \frac{\p^2 \mathcal{L}(v)}{\p \hat{q}^i\p \hat{q}^j}\right)$ is  invertible. 
\end{enumerate}
\ep 
\pf See \appref{exercises}, \excref{20}.
\epf 
\bp Let $(M,\mathcal{L})$ be a Lagrangian mechanical system. Then the following conditions are equivalent:
\begin{enumerate}
\item $\mathcal{L} $ is nice.
\item $\phi_\mathcal{L}: TM \ra T^*M$ is a diffeomorphism. 
\item $\mathcal{L} $ is non-degenerate and for all $x\in M$, $\phi_{\mathcal{L}}|_{T_xM} : T_xM \ra T^*_xM$ is a bijection. 
\item For all $x\in M$, $\phi_{\mathcal{L}}|_{T_xM} : T_xM \ra T^*_xM$ is a diffeomorphism. 
\end{enumerate}
\ep 
\pf See \appref{exercises}, \excref{21}.
\epf 
We can now generalize Proposition \ref{Legendre1Prop} to the case of non-degenerate Lagrangians. For simplicity we assume that 
$\mathcal{L}$ is nice such that $\phi_\mathcal{L} : TM \ra T^*M$ is not only  a local but a global diffeomorphism. The general (local) result 
for non-degenerate Lagrangians is left as an exercise (see \appref{exercises}, \excref{23}). 
\bt Let $(M,\mathcal{L})$ be a Lagrangian mechanical system with nice Lagrangian. Then $\phi=\phi_\mathcal{L}: TM \ra T^*M$
is a diffeomorphism and the following hold:
\begin{enumerate}
\item[(i)] A smooth curve $\g: I \ra M$ is a solution of the  Euler-Lagrange equations  if and only if the curve $\phi \circ \g' : I \ra T^*M$ is a solution of Hamilton's equation for the Hamiltonian $H = E\circ \phi^{-1}$, where $E$ is the energy, defined in Proposition \ref{conserEProp}. 
\item[(ii)] 
Conversely, if a curve $\tilde{\g} : I \ra T^*M$ is a motion of the Hamiltonian system $(T^*M,\o ,H)$, then 
the curve $\pi \circ \tilde{\g} : I \ra M$ is a motion of the Lagrangian system $(M,\mathcal{L})$, where $\pi : T^*M \ra M$ denotes the projection. 
\item[(iii)] The maps $\g \mapsto \phi \circ \g'$ and $\tilde{\g} \mapsto \pi \circ \tilde{\g}$
are inverse to each other when restricted to solutions of the Euler-Lagrange equations and Hamilton's equations, respectively. 
\end{enumerate}
\et    
\pf Let us first remark that the function $H$ on $T^*M$ gives rise to a smooth map $\psi = \psi_H : T^*M \ra TM$. For $\a \in T_x^*M$, $x\in M$, we define 
\be \label{psiHEq} \psi_H (\a ) = d\left( H|_{T^*_xM}\right)|_\a \in (T_x^*M)^*=T_xM.\ee
In local coordinates 
$(q^i,p_i)$ on $T^*U\subset T^*M$ and $(q^i,\hat{q}^i)$ on $TU\subset TM$ associated with local coordinates $(x^i)$ on $U\subset M$, it is 
given by 
\[  q^i \circ \psi = q^i,\quad \hat{q}^i\circ \psi = \frac{\p H}{\p p_i}.\] 
We claim that $\phi$ and $\psi$ are inverse to each other. In particular, $\psi$ is a diffeomorphism. 
Since we already know that $\phi$ is bijective, we only need to check that $\psi \circ \phi = \mathrm{Id}_{TM}$. 
Obviously $q^i  \circ \psi \circ \phi =q^i$. Thus it is sufficient to check that $\hat{q}^i \circ \psi \circ \phi =\hat{q}^i$. 
Let us denote by $(a^{ij})$ the $n\times n$-matrix inverse to the matrix $(a_{ij})$ with matrix coefficients $a_{ij} := \frac{\p^2\mathcal{L}}{\p \hat{q}^i \p \hat{q}^j}$.  
Then we have  
\begin{eqnarray*} 
\frac{\p H}{\p p_i} &=&  dE \circ d \left(\phi^{-1}\right) \frac{\p}{\p p_i}=dE \circ \left(d\phi \right)^{-1} \frac{\p}{\p p_i} \\ &=& \left( dE \sum a^{ij} \frac{\p}{\p \hat{q}^j}\right) \circ \phi^{-1}=
\left( \sum a^{ij} \frac{\p E}{\p \hat{q}^j}\right) \circ \phi^{-1}
\end{eqnarray*}
and 
\[  \frac{\p E}{\p \hat{q}^j} = \frac{\p }{\p \hat{q}^j} \left( \sum \hat{q}^k\frac{\p \mathcal{L}}{\p \hat{q}^k} -\mathcal{L} \right) =  \sum \hat{q}^k\frac{\p^2 \mathcal{L}}{\p \hat{q}^j \p 
\hat{q}^k} = \sum \hat{q}^k a_{jk}.\]
Substituting the second equation into the first equation, we obtain
\be \label{HpEqu} \frac{\p H}{\p p_i} =  \hat{q}^i\circ \phi^{-1}\ee
and, hence, 
\[ \hat{q}^i \circ \psi \circ \phi =  \frac{\p H}{\p p_i} \circ \phi =  \hat{q}^i.\]
This proves that $\psi=\phi^{-1}$ and, in particular, that 
$p_i \circ \phi \circ \psi = p_i$, that is 
\be \label{LpEqu} \frac{\p \mathcal{L}}{\p \hat{q}^i}\circ \psi = p_i.\ee
To relate Hamilton's equation to the Euler-Lagrange equations, we compute with the help of  
\re{HpEqu},\re{LpEqu}:  
\begin{eqnarray}\label{HqEqu} \frac{\p H}{\p q^i} &=&  \frac{\p }{\p q^i} \left( E \circ \psi \right) = \frac{\p }{\p q^i} \left( \left(\sum \hat{q}^j\frac{\p \mathcal{L}}{\p \hat{q}^j} -\mathcal{L}\right)\circ \psi \right) =\frac{\p }{\p q^i} \left( \sum \frac{\p H}{\p p_j}  p_j-\mathcal{L}\circ  \psi  \right) \nonumber \\ \nonumber
&=& \sum \frac{\p^2 H}{\p q^i \p p_j}  p_j-\frac{\p }{\p q^i} (\mathcal{L}\circ  \psi )\\ 
&=&  \sum_j \frac{\p^2 H}{\p q^i \p p_j}  p_j-\frac{\p \mathcal{L}}{\p q^i}\circ  \psi - \sum_j \underbrace{\left( \frac{\p \mathcal{L}}{\p \hat{q}^j}\circ \psi \right)}_{=p_j}\underbrace{\frac{\p (\hat{q}^j\circ \psi )}{\p q^i}}_{=\frac{\p^2 H}{\p q^i \p p_j}} 
= -\frac{\p \mathcal{L}}{\p q^i}\circ  \psi .
\end{eqnarray}
Now we see from  \re{HpEqu}  and \re{HqEqu} that Hamilton's equations for a curve $\tilde{\g} : I \ra T^*M$ in canonical coordinates $(q^i,p_i)$  take the form 
\begin{eqnarray*} \frac{d}{dt} q^i(\tilde{\g} (t)) &=& \frac{\p H}{\p p_i}(\tilde{\g}(t))= \hat{q}^i (\psi (\tilde{\g}(t)),\\
 \frac{d}{dt}  p_i(\tilde{\g} (t)) &=& -\frac{\p H}{\p qi}(\tilde{\g}(t)) = \frac{\p \mathcal{L}}{\p q^i}(\psi (\tilde{\g}(t)).
\end{eqnarray*}
Notice that the first equation is satisfied if and only if the curve $\psi \circ \tilde{\g} : I \ra TM$ is the velocity vector field $\g'$ of the curve 
$\g := \pi \circ \tilde{\g} : I \ra M$. Using $\psi \circ \tilde{\g}=\g'$, the second equation can be written in the form 
\[ \frac{d}{dt}  (p_i\circ \phi ) (\g' (t)) = \frac{\p \mathcal{L}}{\p q^i}(\g'(t)) .\]
In view of \re{LpEqu} this is equivalent to the Euler-Lagrange equations for the curve $\g$. So we have proven (ii). 
This also proves (i) by considering the curve $\tilde{\g} = \phi \circ \g'$, which projects onto $\g$.  

In order to prove (iii), let us first remark that for every smooth curve $\g$ in $M$ we have $\pi \circ \phi \circ \g'=\g$, simply 
because $\phi$ maps $T_xM$ to $T_x^*M$ for all $x\in M$. Now let 
$\tilde{\g}$ be a solution of Hamilton's equation. Then, as shown above,  
$\psi \circ \tilde{\g} = \g'$ is the velocity vector of $\g = \pi \circ \tilde{\g}$ and, hence, 
$\phi \circ \g' = \tilde{\g}$. This proves (iii). 
\epf
Let us summarize for completeness some interesting facts, which we have established in the course of the proof. 
\bp Let $(M,\mathcal{L})$ be a Lagrangian mechanical system with nice Lagrangian. Then the inverse of the 
diffeomorphism $\phi=\phi_\mathcal{L}: TM \ra T^*M$ is given by the map $\psi = \psi_H: T^*M \ra TM$ defined by $H=E\circ \phi^{-1}$ in equation \re{psiHEq}.  
Under these diffeomorphisms the following functions on $TM$ and $T^*M$ are mapped to each other:
\[ \begin{array}{|c|c|c|c|c|}\hline TM&q^i&\hat{q}^i& \p \mathcal{L}/\p \hat{q}^i&E\\
\hline
T^*M&q^i&\p H/\p p_i&p_i&H\\
\hline
\end{array}
\]
\ep
Next we will explain the relation of the previous constructions with the notion of Legendre transform of a smooth function
$f:V \ra \bR$ on a finite-dimensional real vector space $V$.  To define the Legendre transform we  consider the smooth map 
\be \phi_f : V \ra V^*, \quad x\mapsto df_x.\ee 
For simplicity we will assume that $\phi_f$ is a diffeomorphism. (More generally, we could consider the case when $\phi_f$ is only locally
a diffeomorphism.) Then we can define a new function
$\tilde{f} : V^* \ra \bR$ by 
\be \label{tildefEq} \tilde{f} := (\xi (f) - f) \circ \phi_f^{-1},\ee
where $\xi$ is the position vector field in $V$, that is $\xi_x = x$ for all $x\in V$.
Evaluating this function at $y=\phi_f(x)$, we obtain
\[ \tilde{f}(y) = \xi_x(f) -f(x) = df_xx-f(x)=\langle y,x\rangle -f(x),\]
where $\langle y,x\rangle = yx=y(x)$ is the duality pairing.
This shows that \re{tildefEq} can be equivalently written as
\be \tilde{f}(y) = \left. \left( \langle y,x\rangle -f(x)\right)\right|_{x=\phi_f^{-1}(y)}.\ee
\bd The function $\tilde{f} : V^* \ra \bR$\nomenclature[aftilde]{$\tilde{f}$}{Legendre transform of a smooth function $f$} is called the \emph{Legendre transform}\index{Legendre transform} of 
$f: V \ra \bR$. 
\ed 
\bp \label{involProp} Let $f$ be a smooth function on a finite-dimensional real vector space $V$ such that 
$\phi_f : V \ra V^*$ is a diffeomorphism and consider its Legendre transform $\tilde{f}\in C^\infty (V^*)$. 
Then $\phi_{\tilde{f}} : V^* \ra V$ is a diffeomorphism and the Legendre transform
of $\tilde{f}$ is $f$. 
\ep 
\pf See \appref{exercises}, \excref{24}. \epf 
Notice that, with the above notations, for every Lagrangian $\mathcal{L}\in C^\infty (TM)$ the restriction 
of $\phi_\mathcal{L} : TM \ra T^*M$, defined in \re{phiLEqu}, to $T_xM$, $x\in M$, is given by 
\[ \phi_{\mathcal{L}}|_{T_xM} = \phi_{\mathcal{L}_x},\quad \mathcal{L}_x := \mathcal{L}|_{T_xM} : T_xM \ra \bR.\]
Similarly, for every function $H\in C^\infty (T^*M)$ 
\[ \psi_H|_{T_x^*M} =  \phi_{H_x},\quad H_x := H|_{T_x^*M}: T^*_xM\ra \bR.\]
In view of this relation, we will now unify the notation and
define $\phi_H := \psi_H$. 
\bp Let $\mathcal{L}\in C^\infty (TM)$ be a nice Lagrangian and $H\in C^\infty (T^*M)$ the corresponding Hamiltonian. 
Then $H$ is the fiber-wise Legendre transform of $\mathcal{L}$ and vice versa, that is 
$H_x$ is the Legendre transform of $\mathcal{L}_x$  and $\mathcal{L}_x$ is the Legendre
transform of $H_x$ for all $x\in M$.
\ep 
\pf By comparing the definition of the energy $E=\xi (\mathcal{L})-\mathcal{L}$ with \re{tildefEq} we see that 
$H_x = E_x \circ \phi^{-1}_{\mathcal{L}_x}= \tilde{\mathcal{L}}_x$ and this implies that 
$\tilde{H}_x=\mathcal{L}_x$ by Proposition \ref{involProp}.  
\epf

\section{Linearization and stability}

The Euler-Lagrange equations and Hamilton's equations are typically non-linear. We refer to Example~\ref{ex_geodesic} for an illustration of this observation. Non-linearity is also a key feature in many interesting physical applications. Examples from classical mechanics are the aerodynamic drag, where the drag force is proportional to the square of the velocity, and the Navier-Stokes equations describing the motion of viscous fluids. Important examples from modern physics are Einstein's theory of general relativity (see \secref{EHlagr}) as well as the Standard Model of particle physics.

However, non-linear equations are notoriously difficult to solve. A way out is to consider a linear approximation, instead. This often yields valuable insights into the true behavior of the underlying non-linear problem and serves as a starting point for more thorough studies (such as perturbation theory).

This section is based on Ref.~\cite[Ch. 5]{A}. In this section we denote by $I \subset \bR$ an open interval.

\bd
 A point $x_0 \in \bR^n$ is called an \emph{equilibrium position}\index{Equilibrium position} of the system of ordinary differential equations
 \be\eqlabel{eqilibpts}
 \frac{dx}{dt} = f(x) \; , \quad x: I \to \bR^n \; ,
 \ee
 if the constant curve $x(t)=x_0$ is a solution.
\ed

Notice that the equilibrium positions of the system~\eqref*{eqilibpts} are precisely the zeroes of $f$. 

For the rest of this section, we consider as the prime example a classical mechanical system on $M=\bR^n$ with canonical local coordinates $(q^1,\ldots ,q^n,$ $\hat{q}^1,\ldots, \hat{q}^n)$ on $TM$ and Lagrangian
\be\eqlabel{equilib_ex_lagr}
 \mathcal{L} = \frac12 \sum a_{ij} (q) \hat{q}^i \hat{q}^j - V(q) \; .
\ee
The functions $a_{ij} (q)$ are chosen such that $E_{\textnormal{kin}} = \frac12 \sum a_{ij} (q) \hat{q}^i \hat{q}^j > 0$ for all $\hat{q} \neq 0$. The motion is governed by the the Euler-Lagrange equations
\be\eqlabel{equilib_ex_Eleqs}
 \frac{d}{dt} \frac{\partial \mathcal{L}}{\partial \hat{q}^i} - \frac{\partial \mathcal{L}}{\partial q^i} = 0 \; , \qquad i=1,\ldots,n \; .
\ee

\bp
 The point $q=q_0$, $\hat{q}=\hat{q}_0$ is an equilibrium position of the Lagrangian mechanical system~\eqref*{equilib_ex_lagr} if and only if $\hat{q}_0 = 0$ and $q_0$ is a critical point of $V(q)$, that is
 \be\eqlabel{equilib_critpt}
  \left. \frac{\partial V}{\partial q^i} \right|_{q=q_0} = 0 \qquad \forall i=1,\ldots,n \; . 
 \ee
\ep
\pf
From Proposition \ref{HamEqu} we know that the Euler-Lagrange equations~\eqref*{equilib_ex_Eleqs} can be transformed into a system of $2n$ first-order equations of the form
\[
\dot{q}^i = \sum a^{ij} p_j=\hat{q}^i \; , \qquad\quad
\dot{p}_i = - \frac12 \sum \frac{\p a^{k\ell}}{\p q^i} p_k p_\ell - \frac{\p V}{\p q^i} \; ,
\]
where $(q^1,\ldots ,q^n,$ $p_1,\ldots, p_n)$ are local coordinates on $T^\ast M$ and $(a^{ij})$ denotes the inverse matrix of $(a_{ij})$. At an equilibrium position, we have $\dot{q} = \dot{p} = 0$ and hence from the first equation $\hat{q} = 0$, $p = 0$. With $p=0$ the first term in the second equation vanishes and hence we conclude that $q = q_0$ is an equilibrium position if~\eqref*{equilib_critpt} holds and only in that case.
\epf

Returning to the general case, $\frac{dx}{dt} = f(x)$, we may Taylor-expand $f(x)$ close to an equilibrium position $x_0$. For convenience, we may assume without loss of generality $x_0 = 0$ (by a translation of the coordinate system). The Taylor expansion close to $x_0 = 0$ then becomes
\be\eqlabel{taylor_exp}
 f(y) = Ay + \mathcal{O}(|y|^2) \; ,
\ee
where $A=\frac{\p f}{\p x}|_{x=0}$.

\bd
 The passage from the system
 \[
 \frac{dx}{dt} = f(x) \; , \qquad x: I \to \bR^n \; , 
 \]
 to the linear system
 \[
  \frac{dy}{dt} = Ay \; , \qquad  y: I \to T_0 \bR^n \cong \bR^n \; , 
 \]
 is called \emph{linearization}\index{Linearization}\index{Lagrangian!Linearized} around the equilibrium position $x_0 =0$.
\ed

The linearized system can be easily solved by
\[
 y(t) = e^{A t} y(0) \; ,
\]
where $e^{A t} = \mathds{1}_n + At + \frac12 A^2 t^2 + \ldots$ is the matrix exponential series. For small enough $y$, the higher-order corrections $\mathcal{O}(|y|^2)$ in~\eqref*{taylor_exp} are small compared to $y$ itself. Thus, for a long time, the solutions $y(t)$ of the linear system and $x(t)$ of the full system remain close to each other, provided that the initial conditions $y(0)=x(0)$ are chosen sufficiently close to the equilibrium position $x_0$. More precisely, for a given time $T>0$ and $\epsilon>0$, there exists a $\delta>0$ such that for any $x_0'\in \bR^n$ with $|x_0'-x_0|<\delta$ 
the solution $x(t)$ of~\eqref*{eqilibpts} with initial condition $x(0)=x_0'$ exists (at least) for all $t\in [0,T]$ and compares to the solution $y:\bR \ra \bR^n$ of the linearized system with 
$y(0)=x_0'$ by 
\[ \max_{[0,T]} |x-y| <\epsilon \; .\] 

Consider the Lagrangian mechanical system~\eqref*{equilib_ex_lagr} near the equilibrium position $q=q_0$ and choose coordinates such that $q_0=0$.

\bp
 The linearization\nomenclature[aLsub2]{$\mathcal{L}_2$}{linearized Lagrangian} of~\eqref*{equilib_ex_lagr} near $q=q_0$ is given by
 \be\eqlabel{defL2}
  \mathcal{L}_2 = \frac12 \sum a_{ij} (0) \hat{q}^i \hat{q}^j - \frac12 \left. \sum \frac{\p^2 V}{\p q^i \p q^j} \right|_{q=0} q^i q^j \; .
 \ee
 This is also known as \emph{quadratic approximation}\index{Quadratic approximation}.
\ep
\pf
The Hamiltonian corresponding to~\eqref*{equilib_ex_lagr} is given by
\[
 H = \frac12 \sum a^{ij} (q) p_i p_j + V(q) \; .
\]
Hamilton's equations can be written as
\[
 \dot{p}_i = - \frac{\p H}{\p q^i} =: f_i (p,q) \; , \qquad\quad 
 \dot{q}^i = \frac{\p H}{\p p_i} =: g^i (p,q) \; ,
\]
The linearization of this system is obtained by Taylor expanding $f$ and $g$ around $q=p=0$ keeping only terms that are at most linear in $p$ and $q$:
\begin{eqnarray*}
 \dot{p}_i &=& \left. \sum \frac{\p f_i}{\p q^j} \right| q^j + \left. \sum \frac{\p f_i}{\p p_j} \right| p_j 
            =  - \left. \sum \frac{\p^2 H}{\p q^i \p q^j} \right| q^j - \left. \sum \frac{\p^2 H}{\p q^i \p p_j} \right| p_j \\
           &=& - \left. \sum \frac{\p^2 V}{\p q^i \p q^j} \right| q^j - \left. \frac12 \sum \left( \frac{\p^2 a^{k\ell}}{\p q^i \p q^j} p_k p_\ell \right) \hspace*{-0.2em}\right| q^j - \left. \sum \left( \frac{\p a^{jk}}{\p q^i} p_k \right) \hspace*{-0.2em}\right| p_j \\
           &=& - \left. \sum \frac{\p^2 V}{\p q^i \p q^j} \right| q^j \; ,
\end{eqnarray*}
\begin{eqnarray*}
 \dot{q}^i &=& \left. \sum \frac{\p g^i}{\p q^j} \right| q^j + \left. \sum \frac{\p g^i}{\p p_j} \right| p_j 
            =  \left. \sum \frac{\p^2 H}{\p p_i \p q^j} \right| q^j + \left. \sum \frac{\p^2 H}{\p p_i \p p_j} \right| p_j \\
           &=& \left. \sum \left( \frac{\p a^{ik}}{\p q^j} p_k \right) \hspace*{-0.2em}\right| q^j + \sum a^{ij} (0) p_j
            =  \sum a^{ij} (0) p_j \; .
\end{eqnarray*}
In the above equations a vertical line is used as a shorthand symbol to denote the evaluation of the preceding expression at $q=p=0$. Differentiating and combining the two equations yields
\[
 \ddot{q}^i = \sum a^{ij} (0) \dot{p}_j = - \left. \sum a^{ij} (0) \frac{\p^2 V}{\p q^j \p q^k} \right|_{q=0} q^k \; .
\]
These are precisely the Euler-Lagrange equations obtained from $\mathcal{L}_2$.
\epf

\begin{Ex}\label{ex1d}
 Consider the case $n=1\!\! :$
 \[
  \mathcal{L} = \frac12 a (q) \dot{q}^2 - V(q) \; .
 \]
 Let $q(t) = q_0$ be an equilibrium position, that is
 \[
  \left. \frac{\p V}{\p q} \right|_{q=q_0} = 0 \; .
 \]
 Assuming without loss of generality $q_0=0$, the linearized Lagrangian becomes
 \[
  \mathcal{L}_2 = \frac12 \a \dot{q}^2 - \frac12 \b q^2 \; ,
 \]
 where $\a := a(0)$ and $\b := \left. \frac{\p^2 V}{\p q^2} \right|_{q=q_0}$. Note that $\a>0$ by assumption (cf. below \eqref{equilib_ex_lagr}). The Euler-Lagrange equation corresponding to $\mathcal{L}_2$ is given by
 \be\eqlabel{L2_1dof_ex_eom}
  \ddot{q} = - \o_0^2 q \; , \qquad\quad \o_0^2 := \frac{\b}{\a} \; .
 \ee
 The motion crucially depends on the sign of $\b$ or, in other words, on whether the potential $V(q)$ attains a local minimum or maximum at the equilibrium position $q_0$. Indeed, we find as solution of~\eqref*{L2_1dof_ex_eom} with integration constants $c_1, c_2 \in\bR$,
 \[
  q(t) = \begin{cases} c_1 \cos(\o_0 t) + c_2 \sin(\o_0 t) \; , &\qquad\text{for $\b>0$ (``small oscillations'')} \; , \\
                       c_1 \cosh(|\o_0| t) + c_2 \sinh(|\o_0| t) \; , &\qquad\text{for $\b<0$ (``runaway behavior'')} \; , \\
                       c_1 t + c_2 \; , &\qquad\text{for $\b=0$ (``uniform motion'')} \; .
         \end{cases}
 \]
 This observation leads to the notion of \emph{stability}.
\end{Ex}

Now, we consider the case of higher-dimensional motion on $\bR^n$, $n>1$. 
\bd\label{defnstable}
 Consider a Lagrangian mechanical system of the form~\eqref*{equilib_ex_lagr} with equilibrium position $q=q_0$ and set\nomenclature[gOmega]{$\O$}{Hessian matrix of the potential $V$}
 \[
  (\O)_{ij} := \left. \frac{\p^2 V}{\p q^i \p q^j} \right|_{q=q_0} \; , \qquad\quad i,j=1,\ldots,n \; .
 \]
 The equilibrium position $q=q_0$ is called
 \begin{enumerate}
  \item[(i)]   \emph{unstable}\index{Equilibrium position!Unstable}, if $\O$ is negative definite,
  \item[(ii)]  \emph{stable}\index{Equilibrium position!Stable}\index{Stability}, if $\O$ is positive definite,
  \item[(iii)] \emph{degenerate}\index{Equilibrium position!Neutral}, if $\O$ is degenerate,
  \item[(iv)]  a \emph{saddle point}\index{Saddle point}, if $\O$ is nondegenerate but indefinite.
 \end{enumerate}
\ed

This can be understood as follows. Condition (i) implies that the potential $V(q)$ attains a local maximum at the critical point $q_0$. Hence, any small displacement away from $q_0$ causes the system to deviate further from it, since the net-force is directed away from $q_0$. An example of such a situation is depicted in the following figure:
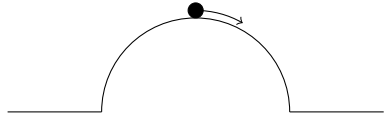
\begin{center}
\begin{tikzpicture}[scale=5]
\draw (0,1)--(0.25,1);
\draw (0.75,1)--(1,1);
\draw (0.75,1) arc [radius=0.25, start angle=0, end angle= 180];
\draw [fill] (0.5,1.27) circle [radius=0.02];
\draw [->] (0.5,1.27) arc [radius=0.25, start angle=90, end angle= 60];
\end{tikzpicture}
\end{center}
In case (ii), $V(q)$ attains a local minimum at $q=q_0$. For small perturbations, the system stays close to $q_0$ for all times, since the forces are ``restoring forces.'' This can be exemplified as follows:
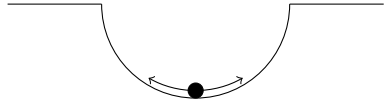
\begin{center}
\begin{tikzpicture}[scale=5]
\draw (0,1)--(0.25,1);
\draw (0.75,1)--(1,1);
\draw (0.25,1) arc [radius=0.25, start angle=180, end angle=360];
\draw [fill] (0.5,0.77) circle [radius=0.02];
\draw [->] (0.5,0.77) arc [radius=0.25, start angle=270, end angle=300];
\draw [->] (0.5,0.77) arc [radius=0.25, start angle=270, end angle=240];
\end{tikzpicture}
\end{center}
A typical example of case (iii) is a region where the potential $V(q)$ is flat:
\begin{center}
\begin{tikzpicture}[scale=5]
\draw (0,1)--(1,1);
\draw [fill] (0.5,1.02) circle [radius=0.02];
\draw [<->] (0.35,1.02)--(0.65,1.02);
\end{tikzpicture}
\end{center}
In that case, higher-order derivatives are needed to decide stability. Finally, condition (iv) can be understood as a situation where some directions are stable and some are unstable.

For the rest of this section, to streamline the presentation we only consider stable equilibrium positions. The general case can be analyzed similarly. For one-dimensional motion we saw in Example~\ref{ex1d} that the linearized motion around a stable equilibrium position is oscillatory in nature,
\[
  q(t) = c_1 \cos(\o_0 t) + c_2 \sin(\o_0 t) \; .
\]
Here, $\o_0=\sqrt{\b/\a}$\nomenclature[gomega0]{$\o_0$}{frequency of a small oscillation} is the \emph{frequency}\index{Frequency}, and $\tau_0=2\pi/\o_0$\nomenclature[gtau0]{$\tau_0$}{period of a small oscillation} the \emph{period}\index{Period} of the oscillation. This leads us to the next concept, namely \emph{small oscillations}.

\bd
Motions in a linearized system $\mathcal{L}_2$, as defined in~\eqref*{defL2}, are called \emph{small oscillations}\index{Oscillation!Small} near an equilibrium position $q=q_0$.
\ed

Recall that the linearized Lagrangian is given by
\[
  \mathcal{L}_2 = \frac12 \sum \a_{ij} \hat{q}^i \hat{q}^j - \frac12 \sum \O_{ij} q^i q^j \; .
\]
where $\a_{ij} := a_{ij} (0)$ and $\O_{ij} := \left. \frac{\p^2 V}{\p q^i \p q^j} \right|_{q=0}$. Note that $(\a_{ij})$ and $\O:=(\O_{ij})$ are symmetric, real matrices and $(\a_{ij})$ is positive definite.
It is useful to work in coordinates where $\a_{ij}=\d_{ij}$, where $\d_{ij}$ is the \emph{Kronecker delta}\nomenclature[gdeltaij]{$\d_{ij}$}{Kronecker delta. Its value is defined to be 1 if the indices are equal, and 0 otherwise}\index{Kronecker delta}.

The equations of motion following from $\mathcal{L}_2$ are a set of a priori coupled linear ordinary differential equations,
\[
 \ddot{q}^i = - \sum \d^{ij} \O_{jk} q^k \; .
\]
Here, the Kronecker delta $\d^{ij}$ (note that $\d^{ij}=\d_{ij}$) is written explicitly in order to make it manifest that the index positions on both sides of the equation match.
By a suitable choice of coordinates, we can \emph{de-couple} the ordinary differential equations. That is, we \emph{diagonalize}\index{Diagonalization} the real symmetric matrix $\O$ using an orthogonal transformation 
$M \in O(n)$, so that
\[
\qquad M^\intercal \O M = \mathrm{diag}(\l_1, \ldots, \l_n) \; ,
\]
where $\l_1, \ldots, \l_n \in \bR$ are the eigenvalues of $\O$ and $(\cdot)^\intercal$\nomenclature[stranspose]{$(\cdot)^\intercal$}{matrix transposition} denotes matrix transposition. Now, we define new coordinates $Q$ on $\bR^n$ as
\[
 Q = M^\intercal q \; .
\]
In terms of the new coordinates, the equations of motion become
\[
 \ddot{Q}^i = - \l_i Q^i \qquad\quad\text{(no sum over $i$)} \; .
\]
The corresponding Lagrangian is given by
\[
 \mathcal{L}_2 = \frac12 \sum_i \left(\dot{Q}^i\right)^2 - \frac12 \sum_i \l_i \left(Q^i\right)^2 \; ,
\]
up to an additive constant. The de-coupled system of equations can be solved straightforwardly. One finds $n$ independent harmonic oscillators of the form
\[
 Q^i (t) = c^i_1 \cos \left(\sqrt{\l_i}\, t\right) + c^i_2 \sin \left(\sqrt{\l_i}\, t\right) \qquad\quad\text{(no sum over $i$)}  \; ,
\]
with real integration constants $(c^i_{1,2})_{i=1,\ldots,n}$.
Recall that the stability criterion (see Definition~\ref{defnstable}) guarantees $\l_i>0$ for all $i$. Otherwise the linearized motion will in general not be periodic, but solutions can be described in a similar way using hyperbolic and linear functions for the cases $\lambda_i < 0$ and $\lambda_i = 0$, respectively.

What we have just learned is that a system performing small oscillations decomposes into a direct product of $n$ one-dimensional systems performing small oscillations. In particular, the system can perform small oscillations of the form
\be\eqlabel{ev_osc}
 q(t) = \left[c_1 \cos (\o t) + c_2 \sin (\o t)\right] \xi \; ,
\ee
where $\o = \sqrt{\l}$ and $\xi$\nomenclature[gxi]{$\xi$}{eigenvector of $\O$} is an eigenvector of $\O$ corresponding to $\l$, that is
\[
 \O\xi = \l\xi \; .
\]
This oscillation can be regarded as a product of $Q^i = c^i_1 \cos (\o_i t) + c^i_2 \sin (\o_i t)$ and $Q^j=0$, $j\neq i$, for some $i\in\{1,\ldots,n\}$.
The two-parameter family of solutions~\eqref*{ev_osc} is called a \emph{characteristic oscillation}\index{Oscillation!Characteristic} or \emph{eigen-oscillation}\index{Eigen-oscillation|see {Characteristic oscillation}} or \emph{normal mode}\index{Normal mode|see {Characteristic oscillation}} and $\o = \sqrt{\l}$ is called \emph{characteristic frequency}\index{Frequency!Characteristic} or \emph{eigen-frequency}\index{Eigen-frequency|see {Characteristic frequency}} (or sometimes also \emph{resonance frequency}\index{Frequency!Resonance|see {Characteristic frequency}}). Sometimes also an element of that family is called a characteristic oscillation. The vector $\xi$ is called the \emph{eigenvector} corresponding to the characteristic oscillation and a system of characteristic oscillation is called \emph{independent} if the corresponding eigenvectors are linearly independent.

The above results are summarized in the following theorem.
\bt
The linearized Lagrangian mechanical system of the form~\eqref*{defL2} near a stable equilibrium position $q=q_0$ performs small oscillations given by a sum of characteristic oscillations. 
The system has $n$ independent characteristic oscillations and the characteristic frequencies are the square roots of the eigenvalues of the Hessian matrix of the potential $V(q)$ at $q_0$ (assuming that $\a_{ij}=\d_{ij}$, without loss of generality).
\et
Note that a \emph{sum} of characteristic oscillations is generally \emph{not periodic}.\footnote{For example, consider a case with two eigenvalues $\l_1 = 1$, $\l_2 = 2$, and the solution $q(t) = \sin(t) \xi_1 + \sin(\sqrt{2}\, t) \xi_2$.}

\medskip
The linearized Lagrangian mechanical system $\mathcal{L}_2$ can now be solved in the following way:
\begin{enumerate}
 \item[(i)] Find the complex characteristic oscillations of the form $q(t) = e^{i\o t} \xi$ by substituting into the equations of motion $\ddot{q} = - \O q$. This yields a characteristic equation, $\O\xi = \o^2 \xi$. Solving this equation produces $n$ pairwise orthogonal eigenvectors $\xi_k$ with corresponding real eigenvalues $\l_k = \o_k^2$.
 \item[(ii)] The general real-valued solution is a linear combination of (i). That is,
 \[
  q(t) = \mathrm{Re} \sum_{k=1}^n c_k e^{i(\o_k t + \d_k)} \xi_k \; ,
 \]
 with $c_k$ and $\d_k$ real parameters.
\end{enumerate}
\begin{remark}
 This result is valid irrespective of the multiplicities of the eigenvalues $\l_k$.
\end{remark}
\begin{Ex}[See~{\cite[Ch.~5]{A}} for this and further examples]
 Consider two identical mathematical pendula of unit mass connected by a weightless spring as depicted in the following figure:
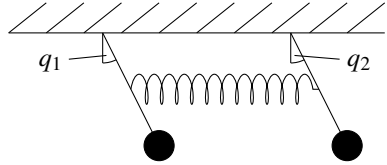
\begin{center}
\begin{tikzpicture}[scale=5]
\draw (0,0)--(1,0);
\draw (0,0)--(0.1,0.08);
\draw (0.1,0)--(0.2,0.08);
\draw (0.2,0)--(0.3,0.08);
\draw (0.3,0)--(0.4,0.08);
\draw (0.4,0)--(0.5,0.08);
\draw (0.5,0)--(0.6,0.08);
\draw (0.6,0)--(0.7,0.08);
\draw (0.7,0)--(0.8,0.08);
\draw (0.8,0)--(0.9,0.08);
\draw (0.9,0)--(1.0,0.08);
\draw (0.25,0)--(0.25,-0.08);
\draw (0.75,0)--(0.75,-0.08);
\draw (0.25,0)--(0.4,-0.3);
\draw (0.75,0)--(0.9,-0.3);
\draw [fill] (0.4,-0.3) circle [radius=0.04];
\draw [fill] (0.9,-0.3) circle [radius=0.04];
\draw[decoration={aspect=0.3, segment length=2mm, amplitude=2mm,coil},decorate] (0.325,-0.15)--(0.825,-0.15); 
\draw (0.25,-0.08) arc [radius=0.08, start angle=270, end angle=296.565];
\draw (0.75,-0.08) arc [radius=0.08, start angle=270, end angle=296.565];
\node[text width=1cm] at (0.18,-0.08) {$q_1$};
\draw (0.16,-0.07)--(0.26,-0.06);
\node[text width=1cm] at (1,-0.08) {$q_2$};
\draw (0.88,-0.07)--(0.76,-0.06);
\end{tikzpicture}
\end{center}
For small oscillations, we have $E_{\textnormal{kin}} = \frac12 \dot{q}_1^2 + \frac12 \dot{q}_2^2$ and $E_{\textnormal{pot}} = V = \frac12 q_1^2 + \frac12 q_2^2 + \frac{k}{2} (q_1 - q_2)^2$, where the last term in $V$ is due to the spring. We choose new diagonalizing coordinates
\[
 Q_1 = \frac{q_1 + q_2}{\sqrt{2}} \; , \qquad\qquad Q_2 = \frac{q_1 - q_2}{\sqrt{2}} \; .
\]
In terms of the new coordinates, the linearized Lagrangian becomes
\[
 \mathcal{L}_2 = E_{\textnormal{kin}} - E_{\textnormal{pot}} = \frac12 \dot{Q}_1^2 + \frac12 \dot{Q}_2^2 - \frac12 \o_1^2 Q_1^2 - \frac12 \o_2^2 Q_2^2 \; ,
\]
with $\o_1 = 1$ and $\o_2 = \sqrt{1+2k}$. There are two characteristic oscillations, namely
\begin{enumerate}
 \item[(a)] $Q_2=0$, that is $q_1 = q_2$.
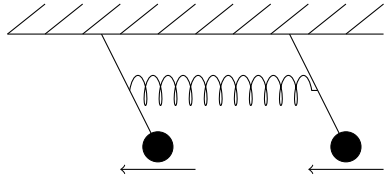
\begin{center}
\begin{tikzpicture}[scale=5]
\draw (0,0)--(1,0);
\draw (0,0)--(0.1,0.08);
\draw (0.1,0)--(0.2,0.08);
\draw (0.2,0)--(0.3,0.08);
\draw (0.3,0)--(0.4,0.08);
\draw (0.4,0)--(0.5,0.08);
\draw (0.5,0)--(0.6,0.08);
\draw (0.6,0)--(0.7,0.08);
\draw (0.7,0)--(0.8,0.08);
\draw (0.8,0)--(0.9,0.08);
\draw (0.9,0)--(1.0,0.08);
\draw (0.25,0)--(0.4,-0.3);
\draw (0.75,0)--(0.9,-0.3);
\draw [fill] (0.4,-0.3) circle [radius=0.04];
\draw [fill] (0.9,-0.3) circle [radius=0.04];
\draw[decoration={aspect=0.3, segment length=2mm, amplitude=2mm,coil},decorate] (0.325,-0.15)--(0.825,-0.15); 
\draw [->] (0.5,-0.36)--(0.3,-0.36);
\draw [->] (1.0,-0.36)--(0.8,-0.36);
\end{tikzpicture}
\end{center}
 This is the case where both pendula move in phase and the spring has no effect.
 \item[(b)] $Q_1=0$, that is $q_1 = -q_2$.
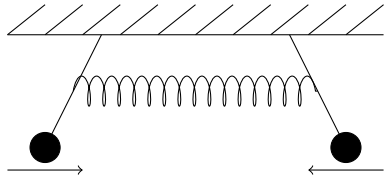
\begin{center}
\begin{tikzpicture}[scale=5]
\draw (0,0)--(1,0);
\draw (0,0)--(0.1,0.08);
\draw (0.1,0)--(0.2,0.08);
\draw (0.2,0)--(0.3,0.08);
\draw (0.3,0)--(0.4,0.08);
\draw (0.4,0)--(0.5,0.08);
\draw (0.5,0)--(0.6,0.08);
\draw (0.6,0)--(0.7,0.08);
\draw (0.7,0)--(0.8,0.08);
\draw (0.8,0)--(0.9,0.08);
\draw (0.9,0)--(1.0,0.08);
\draw (0.25,0)--(0.1,-0.3);
\draw (0.75,0)--(0.9,-0.3);
\draw [fill] (0.1,-0.3) circle [radius=0.04];
\draw [fill] (0.9,-0.3) circle [radius=0.04];
\draw[decoration={aspect=0.3, segment length=2mm, amplitude=2mm,coil},decorate] (0.175,-0.15)--(0.825,-0.15); 
\draw [->] (0.0,-0.36)--(0.2,-0.36);
\draw [->] (1.0,-0.36)--(0.8,-0.36);
\end{tikzpicture}
\end{center}
 In this case, the pendula move in opposite phase with increased frequency $\o_2 > 1$ due to the presence of the spring.
\end{enumerate}
\end{Ex}
%

%
\clearpage{}%

\chapter{Hamilton-Jacobi theory}
\chlabel{hjthy} 

\abstract*{Besides the Newtonian, Lagrangian, and Hamiltonian formulations of classical mechanics, there is yet a fourth approach, known as Hamilton-Jacobi theory, which is part of Hamiltonian mechanics. This approach is the subject of the present chapter. In Hamilton-Jacobi theory, the central equation capturing the dynamics of the mechanical system is the Hamilton-Jacobi equation which is a first-order, non-linear partial differential equation. Remarkably and contrary to the other formulations of classical mechanics, the entire multi-dimensional dynamics is described by a single equation. Even for relatively simple mechanical systems, the corresponding Hamilton-Jacobi equation can be hard or even impossible to solve analytically. However, its virtue lies in the fact that it offers a useful, alternative way of identifying conserved quantities even in cases where the Hamilton-Jacobi equation itself cannot be solved directly. In addition, Hamilton-Jacobi theory has played an important historical role in the development of quantum mechanics, since the Hamilton-Jacobi equation can be viewed as a precursor to the Schr\"odinger equation~\cite{BW,Goldstein,Sakurai}.}

\abstract{Besides the Newtonian, Lagrangian, and Hamiltonian formulations of classical mechanics, there is yet a fourth approach, known as Hamilton-Jacobi theory, which is part of Hamiltonian mechanics. This approach is the subject of the present chapter. In Hamilton-Jacobi theory, the central equation capturing the dynamics of the mechanical system is the Hamilton-Jacobi equation which is a first-order, non-linear partial differential equation. Remarkably and contrary to the other formulations of classical mechanics, the entire multi-dimensional dynamics is described by a single equation. Even for relatively simple mechanical systems, the corresponding Hamilton-Jacobi equation can be hard or even impossible to solve analytically. However, its virtue lies in the fact that it offers a useful, alternative way of identifying conserved quantities even in cases where the Hamilton-Jacobi equation itself cannot be solved directly. In addition, Hamilton-Jacobi theory has played an important historical role in the development of quantum mechanics, since the Hamilton-Jacobi equation can be viewed as a precursor to the Schr\"odinger equation~\cite{BW,Goldstein,Sakurai}.}

\section*{}
For simplicity we consider Hamiltonian systems $(T^*M,H)$ of cotangent type\index{Hamiltonian!system!of cotangent type}, that is for which the phase space is the cotangent bundle 
$T^*M$ of a (connected) manifold with canonical symplectic form $\o = \o_{\textnormal{can}}$. As we have shown in \chref{Hamiltonian_systems}, this includes the 
systems obtained from Lagrangian mechanical systems with a nice Lagrangian via Legendre transform. 
By Darboux's theorem every Hamiltonian system can be locally identified with an open subset of a Hamiltonian system of 
cotangent type.
 
We will now describe a method for the solution of Hamilton's equation\index{Hamilton's equation} which is based on the following simple idea.
Consider a function $S$ on the base manifold $M$. It gives rise to a $1$-form $dS : M \ra T^*M$ and for every curve 
$\g : I \ra M$ we can consider the curve
\[ \tilde{\g} : I \ra T^*M, \quad t\mapsto dS_{\g (t)}.\]
We would like to know when $\tilde{\g}$ is an integral curve of the Hamiltonian vector field $X_H$. Since $H$ is an integral of motion, a necessary condition is
that $H\circ \tilde{\g}$ is constant. Projecting Hamilton's equation $\tilde{\g}' = X_H(\tilde{\g})$ to $M$ we obtain
\be \label{1storderEq} \g' = d\pi X_H(\tilde{\g} ) .\ee
We will prove in the next theorem that, conversely, this equation (for $n=\dim M$ functions) is sufficient to solve Hamilton's equation (for $2n$ functions) if we 
assume that $H$ is constant  on the section $dS$ of $T^*M$.  Notice that in canonical coordinates $(q^i,p_i)$ associated with local coordinates 
$(x^i)$ in $M$ 
the  equation \re{1storderEq} corresponds to the system 
\[ \dot{q}^i = \frac{\p H}{\p p_i}(\tilde{\g}),\quad i=1,\ldots,n,\]
where $q^i(t)=q^i(\tilde{\g}(t)) = x^i(\g (t))$.  This is exactly half (that is, $n$ out of $2n$) of Hamilton's equations. 

\begin{Th} \label{HJThm} Let $(T^*M,H)$ be a Hamiltonian system of cotangent type and $S\in C^\infty (M)$. 
Then the following are equivalent. 
\begin{enumerate} 
\item For all solutions $\g$ of  \re{1storderEq}, the curve $\tilde{\g} = dS_\g$ is a solution of Hamilton's equation. 
\item The Hamiltonian vector field $X_H$ is tangent to the image of $dS: M \ra T^*M$.  
\item The function $H \circ dS: M \ra \bR$ is constant.
\end{enumerate} 
\end{Th} 
To prove this theorem we will use the following fundamental lemma.
\bl \label{fundL}Let $\a$ be a smooth $1$-form on $M$. We consider it as a smooth map
\[ \varphi_\a : M \ra T^*M.\]
Then the following holds: 
\begin{itemize}
\item[(i)]
The pull back of the Liouville form $\l$ under this map is given by 
\[ \varphi_\a^*\l = \a.\] 
\item[(ii)]$\,$ 
The pull-back of the canonical symplectic form $\o$  is 
\[ \varphi_\a^*\omega=d\a. \] 
\end{itemize} 
\el
\pf (i) For $v\in TM$, we compute
\[   (\varphi_\a^*\l) (v) = \l  (d\varphi_\a v) = \a (d\pi d\varphi_\a v) = \a (v). \]
Part (ii) follows from (i):
\[ \varphi_\a^*\omega = \varphi_\a^*d\l = d \varphi_\a^*\l = d\a .\] 
\epf 
 \bd An immersion $\varphi : N \ra M$ of an $n$-dimensional manifold $N$ into a $2n$-dimensional symplectic manifold $(M,\o )$ is called 
  \emph{Lagrangian}\index{Lagrangian!immersion} if it satisfies $\varphi^*\o =0$. 
  \ed 
Lemma \ref{fundL} (ii) immediately implies:
\bc The embedding $\varphi_\a : M \ra T^*M$ defined by a smooth 
$1$-form $\a$ on a manifold $M$ is Lagrangian if and only if $\a$ is closed. 
\ec 

\bd Let $(V,\o )$ be a symplectic vector space. A subspace $U\subset V$ is called \emph{Lagrangian}\index{Lagrangian!subspace}
if $U$ coincides with 
\[ U^\perp = \{ v\in V \mid \o (v,u)=0\quad\mbox{for all}\quad u\in U \} .\] 
\ed 
It is easy to prove the following proposition (see \appref{exercises}, \excref{25}). 
\bp An immersion $\varphi : N \ra M$ of an $n$-dimensional manifold $N$ 
into a $2n$-dimensional symplectic manifold $(M,\o )$ is Lagrangian if and only if 
its tangent spaces $d\varphi T_xN\subset T_{\varphi (x)}M$ are Lagrangian for all $x\in N$. 
\ep 
 \pf (of Theorem \ref{HJThm}) Let us first show that 1.\ and 2.\ are equivalent. 
We first assume 1.\ and show that $X_H(\varphi_{dS}(x))$ is tangent to the 
Lagrangian submanifold $N=\varphi_{dS}(M)\subset T^*M$ for all $x\in M$. 
Let $\g : I=(-\e ,\e )\ra M$, $\e >0$,  be a solution of \re{1storderEq} with initial condition
$\g (0) = x$. Then the 
 curve $\tilde{\g} = \varphi_{dS}\circ \g$ is a solution to Hamilton's equation and lies in $N$.  Therefore,  
 \[ X_H(\tilde{\g}) = \tilde{\g}' \in TN\]
and, in particular,  $X_H(\varphi_{dS}(x))=X_H(\tilde{\g}(0)) = \tilde{\g}'(0) \in TN$. So we have proven that
1.\ implies 2. To show the converse, we observe that given a solution $\g: I \ra M$ of \re{1storderEq} and $t\in I$, 
the vector $\tilde{\g}'(t)\in T_{\tilde{\g}(t)}N$ is the unique tangent vector in $T_{\tilde{\g}(t)}N$ which 
projects to $\g'(t)=d\pi X_H(\tilde{\g}(t))$. Since the vector $X_H(\tilde{\g}(t))\in T_{\tilde{\g}(t)}T^*M$ projects to the 
same vector in the base, we see that $\tilde{\g}'(t)=X_H(\tilde{\g}(t))$ if and only if $X_H(\tilde{\g}(t))\in TN$. 

Next we prove the equivalence of 2.\ and 3. Since $N\subset T^*M$ is Lagrangian, the vector field 
$X_H$ is tangent to $N$ if and only if at all points $u\in N$ we have 
\[ X_H(u) \in (T_uN)^\perp \iff dHT_uN=0 \iff d(H\circ \varphi_{dS})|_{\pi (u)}=0.\]

  \epf 
 The partial differential equation
 \[ H \circ dS = \textnormal{const}\]
 of part 3 of Theorem \ref{HJThm} is called the \emph{Hamilton-Jacobi equation}\index{Hamilton-Jacobi equation}. 
 We can summarize part of Theorem \ref{HJThm} as follows. 
If $S$ is a solution to the Hamilton-Jacobi equation then every  integral curve of $X_H$ passing through a point of the 
Lagrangian submanifold $N=\varphi_{dS}(M)\subset T^*M$ is fully 
contained in $N$ and these integral curves are obtained by solving a system 
of $n$ ordinary differential equations (rather than $2n$). 
  
Next we will show that given not only one solution, but a smooth family of solutions 
of the Hamilton-Jacobi equation depending 
on $n$ parameters we can (at least locally) completely solve Hamilton's equations of motion, provided the family is 
non-degenerate in a certain sense. The non-degeneracy assumption will ensure that the family 
cannot be locally reduced to a family depending on fewer parameters.  
\bd\deflabel{phi_S} Let $(T^*M,H)$ be a Hamiltonian system of cotangent type\index{Hamiltonian!system!of cotangent type} and $n=\dim M$. 
A \emph{family} of solutions of the Hamilton-Jacobi equation is a smooth function
\[ S : M\times U\ra \bR,\quad (x,u) \mapsto S(x,u),\]  
where $U\subset \bR^n$ is an open subset, such that 
for all $u\in U$ the function $x\mapsto S^u(x) := S(x,u)$ on $M$ is a solution of the Hamilton-Jacobi equation. 
It is called \emph{non-degenerate} if the map
\[\eqlabel{phi_S} \Phi_S: M\times U \ra T^*M,\quad (x,u) \mapsto dS^u|_x\in T^*_xM\subset T^*M\]
is of maximal rank. 
\ed 
 Let $(S^u)_{u\in U}$ be a smooth family of solutions of the Hamilton-Jacobi equation\index{Hamilton-Jacobi equation}. 
Then, for all $u\in U$, the image of $dS^u: M \ra T^*M$ is a Lagrangian submanifold $N_u$.
The smooth map $\Phi_S$ maps each submanifold $M \times \{ u\} \subset M\times U$ diffeomorphically to the submanifold $N_u \subset T^*M$. 
Moreover, it is fiber preserving in the sense that it maps the fibers $\{ x\} \times U$, $\; x\in M$, of the trivial projection $M\times U \ra M$ into 
the fibers $T^*_xM$ of the cotangent bundle.
The next proposition is left as an exercise (see \appref{exercises}, \excref{26}). 
\bp Let $(T^*M,H)$ be a Hamiltonian system of cotangent type\index{Hamiltonian!system!of cotangent type}, $n=\dim M$ and let $S: M \times U\ra \bR$ be a smooth $n$-parameter family of solutions of the Hamilton-Jacobi equation. Then the following conditions are equivalent.
\item[(i)] The family is non-degenerate. 
\item[(ii)]
$\Phi_S: M \times U \ra T^*M$ is a local diffeomorphism.
\item[(iii)] For all $x\in M$, 
\[ \Phi_S|_{\{ x\} \times U}: \{ x\} \times U \cong U \ra T_x^*M, \quad u\mapsto dS^u|_x,\]
is a local diffeomorphism.    
\item[(iv)] For all $x\in M$, 
\[ \Phi_S|_{\{ x\} \times U}: \{ x\} \times U \cong U \ra T_x^*M,\]
is of maximal rank.
\item[(v)] For all $(x,u)\in M\times U$ the $n\times n$-matrix 
\[ \left( \frac{\p^2 S(x,u)}{\p x^i \p u^j}   \right)  \]
is invertible, where $(x^i)$ are local coordinates in a neighborhood of $x\in M$ and 
$(u^i)$ are (for instance)  standard coordinates in $U\subset \bR^n$. 
\ep 
\bl \label{pullbackL}Let $(T^*M,H)$ be a Hamiltonian system of cotangent type\index{Hamiltonian!system!of cotangent type}, $n=\dim M$, and $S: M \times U\ra \bR$ a smooth non-degenerate $n$-parameter family of solutions of the Hamilton-Jacobi equation\index{Hamilton-Jacobi equation}. Then the pull back of the canonical symplectic form by the local diffeomorphism $\Phi_S : M\times U \ra T^*M$ is given by 
\[ \Phi_S^*\o = \sum \frac{\p^2 S}{\p x^i\p u^j}dx^i\wedge du^j,\]
with respect to any local coordinate system $(x^1,\ldots ,x^n)$ on $M$, where $(u^1,\ldots ,u^n)$ are standard coordinates 
on $U\subset \bR^n$.   
\el 

\pf
Let us denote by $(q^i,p_i)$ the local coordinates on $T^*M$ associated with the local coordinates $(x^i)$ on $M$. 
Then  
\[ q^i\circ \Phi_S = x^i,\quad p_i \circ \Phi_S = \frac{\p S}{\p x^i}\]
and, hence, 
\[ \Phi_S^*\l = \sum \frac{\p S}{\p x^i}dx^i \implies \Phi_S^*\o = d\Phi_S^*\l  =   \sum \frac{\p^2 S}{\p x^i\p u^j}dx^i\wedge du^j.\]
\epf 
\bt  Let $(T^*M,H)$ be a Hamiltonian system of cotangent type\index{Hamiltonian!system!of cotangent type}, where $M$ is connected, $n=\dim M$ and $S: M \times U\ra \bR$ a smooth non-degenerate $n$-parameter family of solutions of the Hamilton-Jacobi equation\index{Hamilton-Jacobi equation}. Then the motions of the Hamiltonian system $(T^*M,H)$ contained in the open subset
$\Phi_S(M\times U) \subset T^*M$ are of the form $\Phi_S \circ \g$, where $\g : I \ra M\times U$ is a motion
of the Hamiltonian system 
\[ (M\times U,\; \tilde{\o}:=\phi_S^*\o ,\; \tilde{H} :=H \circ \Phi_S).\] 
The equations of motion of the latter system can be solved by passing to local 
coordinates of the form 
\[ \left(\tilde{q}^i:=u^i, \tilde{p}_i:= -\frac{\p S}{\p u^i}\right).\]
In such coordinates we have 
\[ \tilde{\o} = \sum d\tilde{q}^i\wedge d \tilde{p}_i,\quad \frac{\p \tilde{H}}{\p \tilde{p}_i}=0.\]
The function $\tilde{H}= \tilde{H}(\tilde{q}^1,\ldots ,\tilde{q}^n)$ depends only on the $\tilde{q}^i$ and not on $\tilde{p}_i$, and 
the solutions are given by 
\[ \tilde{q}^i(t) = a^i,\quad \tilde{p}_i(t) = -\int \frac{\p \tilde{H}}{\p \tilde{q}^i}(a^1,\ldots ,a^n)dt,\]
where $a^i$ are arbitrary real constants.    
\et 

\pf 
Since $\Phi_S^*\o = \tilde{\o}$ and $\phi_S^*H=\tilde{H}$ it is clear that $\Phi_S$ maps motions 
of $(M\times U,  \tilde{\o}, \tilde{H})$ to motions of $(T^*M, H)$. In this way
we obtain all motions lying in the image of $\Phi_S$.  Due to the non-degeneracy of the matrix
$\left(\frac{\p^2 S(x,u)}{\p x^i \p u^j}\right)$, the functions $(\tilde{q}^i, \tilde{p}_i)$ form a system of local
coordinates, when restricted to an appropriate neighborhood of any given point of $M\times U$. 
We compute 
\[  \sum d\tilde{q}^i\wedge d \tilde{p}_i = \sum du^i \wedge d \left( -\frac{\p S}{\p u^i} \right) = \sum \frac{\p^2 S}{\p x^i\p u^j}dx^i\wedge du^j.\]
By Lemma \ref{pullbackL}, this is $\tilde{\o}$. In virtue of the  Hamilton-Jacobi equation, for every $u\in U$,  
the function $x\mapsto \tilde{H}(x,u)=H\circ \Phi_S(x,u) = H\circ dS^u|_x$ is constant, that is, it depends only on $u$. 
In the local coordinates $(\tilde{q}^i, \tilde{p}_i)$ this means that $\tilde{H}=\tilde{H}(\tilde{q})$ is a function solely of the $\tilde{q}^i$. 
So Hamilton's equations reduce to 
\[ \dot{\tilde{q}}^i = 0,\quad \dot{\tilde{p}}_i=-\frac{\p \tilde{H}}{\p \tilde{q}^i}(\tilde{q}),\]
which are trivially solved as indicated.
\epf 
Next we discuss an example, cf.\ \cite[Ch.\ 9, Sec.\ 47]{A} and references therein. Consider a particle of 
unit mass moving in the Euclidean plane
under the gravitational potential generated by two equal masses placed at 
distance $2c>0$, say at the points $f_1=(c,0)$ and $f_2=(-c,0)$.  
This problem can be conveniently studied by considering the distances 
$r_1$ and $r_2$ to the points $f_1$ and $f_2$, respectively. The gravitational potential 
is 
\[ V= -\frac{k}{r_1}- \frac{k}{r_2}\] 
for some constant $k>0$. 
Away from the 
Cartesian coordinate axes one can use the globally defined functions 
\[ \xi := r_1+r_2,\quad \eta := r_1-r_2,\] 
as coordinates. The level sets of $\xi$ and $|\eta|$ form a \emph{confocal} system of ellipses  and hyperbolas, that is a system with common focal points $f_1, f_2$. 
From the geometry of conic sections we know that the tangent line at a point $q$ of any ellipse or hyperbola bisects the angle between the two lines connecting $q$ with the foci. This implies that the ellipses and hyperbolas of the 
confocal system intersect orthogonally.  Therefore the Euclidean metric is a linear combination
of $d\xi^2$ and $d\eta^2$ (with functions as coefficients). 
\bp\proplabel{4.11} In the coordinates  $(\xi, \eta )$ the Euclidean metric takes the form 
\[  g= \frac{\xi^2-\eta^2}{4(\xi^2 -4c^2)}d\xi^2 +  \frac{\xi^2-\eta^2}{4(4c^2-\eta^2)}d\eta^2.\]
\ep 
\pf It is sufficient to check the formula on the ellipses and hyperbolas of the confocal system. 
The ellipse can be parametrized as $\g (t) = (a\cos t, b\sin t)$, where $a^2-b^2=c^2$. 
Then, along the ellipse,  
\begin{eqnarray*} r_1^2  &=& (a\cos  t -c)^2 + b^2\sin^2t= a^2\cos^2t -2ac\cos t + c^2 +b^2\sin^2 t\\
&=& c^2\cos^2t -2ac\cos t + a^2 = (a- c\cos t)^2
\end{eqnarray*}
and, thus,  
\[ r_1 = a-c\cos t,\quad r_2 = a+ c\cos t,\quad \xi = 2a,\quad \eta = -2c\cos  t.\]
This implies $4c^2-\eta^2 = 4c^2\sin^2 t=\dot{\eta}^2$ and, hence, 
\[ g(\g',\g') =  \frac{\xi^2-\eta^2}{4(4c^2-\eta^2)}\dot{\eta}^2= \frac14 (\xi^2-\eta^2)= a^2-c^2 \cos^2 t = a^2 \sin^2 t + b^2 \cos^2 t.\]
The result coincides with $g_{\textnormal{can}}(\g' ,\g')$, the square of the Euclidian length of $\g'= (-a\sin t , b \cos t)$. The analogous calculation for the hyperbola
is left to the reader (see \appref{exercises}, \excref{32}). This proves that $g_{\textnormal{can}} = g$ on the domain of definition of the coordinates $(\xi, \eta )$. 
\epf 
\bc In the coordinates $(\xi , \eta)$, the Hamiltonian of a particle of 
unit mass moving in the Euclidean plane
under the gravitational potential generated by two equal masses placed at $f_1=(c,0)$ and $f_2=(-c,0)$
has the form 
\be \label{HEqu}H = 2p_\xi^2 \frac{\xi^2 -4c^2}{\xi^2-\eta^2} + 2p_\eta^2\frac{4c^2-\eta^2}{\xi^2-\eta^2} - \frac{4k\xi}{\xi^2 -\eta^2}.\ee
\ec 
\pf
For any local coordinate system $(x^1,x^2)$ we have 
\[ H(q,p)= \frac12\sum g^{ij}(q)p_ip_j +V(q),\]
where $(q^i,p_i)$ are the 
corresponding local coordinates on the cotangent bundle of $\bR^2$. 
Inverting the diagonal matrix representing the metric in the coordinates $(\xi , \eta )$ and dividing by $2$
yields the kinetic term in \re{HEqu}. The potential term is  
\[ V=  -\frac{k}{r_1}- \frac{k}{r_2}= -\frac{k(r_2+r_1)}{r_1r_2}= -\frac{4k\xi }{\xi^2-\eta^2}.\] 
\epf 
In the coordinates $(\xi , \eta )$, the Hamilton-Jacobi equation $H\left(q,\frac{\p S}{\p q}\right)=C=\textnormal{const}$ takes the form
\[ 2\left( \frac{\p S}{\p \xi}\right)^2 (\xi^2 -4c^2) + 2\left( \frac{\p S}{\p \eta}\right)^2 (4c^2-\eta^2) -4k\xi = C (\xi^2 -\eta^2) .\] 
The variables $(\xi ,p_\xi)$ and $(\eta , p_\eta )$ can be separated as follows. 
We look for solutions of the system 
\begin{eqnarray*}
&&2\left( \frac{\p S}{\p \xi}\right)^2 (\xi^2 -4c^2)  -4k\xi - c_2\xi^2 = c_1,\\
&&2\left( \frac{\p S}{\p \eta}\right)^2 (4c^2-\eta^2)  + c_2\eta^2 = -c_1,
\end{eqnarray*} 
where $c_1, c_2$ are constants. 
Solving each equation for the partial derivatives $\p S/\p \xi$ and $\p S/\p \eta$, respectively, and integrating yields
the following two-parameter family of solutions of the Hamilton-Jacobi equation
\[ S(\xi, \eta , c_1,c_2) =  \int \sqrt{\frac{c_1+c_2\xi^2 +4k\xi }{2(\xi^2 -4c^2)}}d\xi +\int \sqrt{\frac{-c_1-c_2\eta^2}{2(4c^2-\eta^2)}}d\eta.\]
\begin{remark} Hamilton-Jacobi theory can be extended to the case of time-dependent 
Hamiltonians $H(q,p,t)$. The corresponding generalization of the Hamilton-Jacobi equation has the form
\be \label{HJT} \frac{\p S}{\p t} + H\left( q,\frac{\p S}{\p q},t\right) =0,\ee
where $S=S(q,t)$ is now allowed to depend on time $t$. The choice of the letter $S$ is related to 
the action functional of a Lagrangian mechanical system as defined in Definition \ref{actionDef}. 
Let us very briefly explain this relation. Let $(M,\mathcal{L})$ be a Lagrangian mechanical system with possibly time-dependent
Lagrangian.  Fix a point $x_0\in M$ and a time $t_0$. Suppose that for $(x,t)\in M\times \bR$ in a suitable domain 
there exists a unique motion $s\mapsto \g(s)$ (inside a suitable domain in $M$) such that $\g (t_0) = x_0$ and $\g (t) = x$. 
Then we can define
\[ S_{(x_0,\,t_0)}(x,t) := \int_{t_0}^t \mathcal{L}(\g'(s),s)ds.\]
Under appropriate assumptions, it can be shown that $S_{(x_0,t_0)}$ is a smooth family of solutions of the time-dependent Hamilton-Jacobi equation\index{Hamilton-Jacobi equation!Time-dependent} \re{HJT}, see 
\cite[Ch.\ 9, Sec.\ 46]{A}. 
\end{remark}\clearpage{}%
\clearpage{}%

\chapter{Classical field theory}
\chlabel{fieldthy} 

\abstract*{A second cornerstone of classical physics besides point-particle mechanics is field theory. Classical field theory is essentially an infinite collection of mechanical systems (one at each point in space) and hence can be viewed as an infinite-dimensional generalization of classical mechanics. More precisely, solutions of classical mechanical systems are smooth curves $t\mapsto\gamma(t)$ from $\bR$ to $M$. In classical field theory, curves from $\bR$ are replaced by maps from a higher-dimensional source manifold. In this more general framework we also allow for Lagrangians with explicit time dependence. Another key feature of classical field theory is its manifest incorporation of the laws of Einstein's theory of relativity.
This chapter begins with definitions and properties of the central objects, namely fields, Lagrangians, action functionals, and the field-theoretic version of the Euler-Lagrange equations. Modern covariant field theory is customarily formulated in the language of jet bundles, which is also utilized and thus introduced here. We study symmetries and conservation laws of classical field theories in the second part of this chapter, which culminates in the field-theoretic version of Noether's theorem. The penultimate section is devoted to a thorough presentation, from a mathematical perspective, of some prominent examples of classical field theories, such as sigma models, Yang-Mills theory, and Einstein's theory of gravity. A key ingredient of matter-coupled Einstein gravity is the energy-momentum tensor, which is studied in detail in the final section. This chapter is largely based on Ref.~\cite{Ol}.}

\abstract{A second cornerstone of classical physics besides point-particle mechanics is field theory. Classical field theory is essentially an infinite collection of mechanical systems (one at each point in space) and hence can be viewed as an infinite-dimensional generalization of classical mechanics. More precisely, solutions of classical mechanical systems are smooth curves $t\mapsto\gamma(t)$ from $\bR$ to $M$. In classical field theory, curves from $\bR$ are replaced by maps from a higher-dimensional source manifold. In this more general framework we also allow for Lagrangians with explicit time dependence. Another key feature of classical field theory is its manifest incorporation of the laws of Einstein's theory of relativity.
This chapter begins with definitions and properties of the central objects, namely fields, Lagrangians, action functionals, and the field-theoretic version of the Euler-Lagrange equations. Modern covariant field theory is customarily formulated in the language of jet bundles, which is also utilized and thus introduced here. We study symmetries and conservation laws of classical field theories in the second part of this chapter, which culminates in the field-theoretic version of Noether's theorem. The penultimate section is devoted to a thorough presentation, from a mathematical perspective, of some prominent examples of classical field theories, such as sigma models, Yang-Mills theory, and Einstein's theory of gravity. A key ingredient of matter-coupled Einstein gravity is the energy-momentum tensor, which is studied in detail in the final section. This chapter is largely based on Ref.~\cite{Ol}.}

\section{The Lagrangian, the action and the Euler-Lagrange equations}
The \emph{fields}\index{Field|see{Classical field theory}} of a \emph{classical field theory}\index{Classical field theory} are basically smooth maps $f: \mathfrak{S}\ra \mathcal{T}$  
from a source manifold\index{Manifold!Source} $\mathfrak{S}$\nomenclature[aSfrak]{$\mathfrak{S}$}{source manifold (possibly with boundary)} (possibly with boundary) 
to a target manifold\index{Manifold!Target} $\mathcal{T}$\nomenclature[aTcal]{$\mathcal{T}$}{target manifold}.  The theory is defined by an action functional\index{Action} of the form
\nomenclature[aSf]{$S{[}f{]}$}{action functional of a classical field theory}
\be\eqlabel{class_ft_action}
 S{[}f{]} = \int_{\mathfrak{S}} \mathcal{L}(j^k(f))dvol, 
\ee
where $dvol$\nomenclature[advol]{$dvol$}{volume element on $\mathfrak{S}$} is a fixed volume element\index{Volume!element|see{form}}\index{Volume!form} on $\mathfrak{S}$ and  the \emph{Lagrangian}\index{Lagrangian} $\mathcal{L}$  is a smooth function on a certain 
bundle $\mathrm{Jet}^k( \mathfrak{S},\mathcal{T})$\nomenclature[aJetkST]{$\mathrm{Jet}^k( \mathfrak{S},\mathcal{T})$}{jet bundle over $\mathfrak{S}$} over $\mathfrak{S}$. The elements of  the fiber 
$\mathrm{Jet}^k_x( \mathfrak{S},\mathcal{T})$ over $x\in \mathfrak{S}$ are defined as equivalence classes of smooth maps $\mathfrak{S}\ra \mathcal{T}$, where 
two maps are equivalent if their Taylor expansions with respect to some local coordinate system coincide up to order $k$ at $x$. This condition does not
depend on the particular choice of local coordinates. The equivalence class of $f$ with respect to the above relation is denoted by $j^k_x(f)\in \mathrm{Jet}^k_x( \mathfrak{S},\mathcal{T})$ and is called    the \emph{$k$-th order jet}\index{Jet} of $f$ at $x$. The map 
\nomenclature[ajkf]{$j^k(f)$}{smooth section of the jet bundle $\mathrm{Jet}^k( \mathfrak{S},\mathcal{T})$}
\[ 
 j^k(f) :  \mathfrak{S}\ra  \mathrm{Jet}^k( \mathfrak{S},\mathcal{T}),\quad x\mapsto j^k_x(f) ,
\]
is a smooth section of the \emph{jet bundle}  $\mathrm{Jet}^k( \mathfrak{S},\mathcal{T})$. 

In many physically interesting cases the source manifold $\mathfrak{S}$ is interpreted as a space-time\index{Space-time} of a certain dimension. 
The Lagrangian mechanical systems discussed so far can be considered as classical field theories of order $k=1$ with one-dimensional space-time.  
The source  $\mathfrak{S} =I$ is an interval and the target  manifold $\mathcal{T}$ is the configuration space of the mechanical system. 

For simplicity, we will assume for the moment that 
$\mathfrak{S}$ is either a bounded domain $\O \subset \bR^n$ with smooth boundary or $\O=\bR^n$, and that $\mathcal{T} = \bR^m$. (Recall that 
a domain is a connected open set.) 
The basic ideas can be explained with almost no loss of generality in this setting, cf.\ \cite{Ol}. 
For the convergence of the action integral one needs to require certain boundary conditions on $f$, depending on the particular problem under consideration. 

Using standard coordinates $(x^1,\ldots ,x^n)$ on $\O\subset \bR^n$ and $(y^1,\ldots ,y^m)$ on $\bR^m$ we can trivialize 
the  bundle $\mathrm{Jet}^k(\O, \bR^m)$ and so identify it with $\Omega \times V$, where $V=\mathrm{Jet}^k_0(\bR^n , \bR^m)$ 
is the vector space consisting of vector-valued Taylor polynomials of order $k$. It has natural global 
coordinates denoted by $u^a_I$, where $a = 1,\ldots ,m$ and $I=(i_1,\ldots ,i_\ell )\in \{ 1,\ldots ,n\}^\ell$ runs through all unordered multi-indices\index{Multi-index}
of length $\ell  \le k$. So multi-indices which differ only by a permutation are not distinguished. The dimension
of $V$ is $m \binom{n+k}{k}$.
For any smooth map $f : \O \ra \bR^m$ and multi-index $I=(i_1,\ldots ,i_\ell )$ of length $\ell \le k$ we have 
\[ u^a_I(j^k(f))= \partial_I f^a= \partial_{i_1}\cdots \partial_{i_\ell}f^a ,\]
where $\p_i=\p_{x^i}$ and $f^a = y^a \circ f$, $a=1,\ldots ,m$. So the $k$-th jet of $f$  at some point $x\in \O$ is simply given $x$ together with the partial derivatives up to order $k$ of the components $f^a$ of $f$ at $x$.

\begin{Ex} Consider the case $n=3$, $m=1$ and $k=2$. With obvious notational simplifications,  the natural coordinates of $V=\mathrm{Jet}^2_0(\bR^n , \bR)$ are 
\[ u, u_1, u_2, u_3, u_{11},u_{12}, u_{13}, u_{22}, u_{23}, u_{33}\] 
and $\dim V= 10=\binom{5}{2}$.
\end{Ex}
In order to  derive the equations of  motions of a classical field theory it is useful to 
introduce the 
\emph{total derivative}\index{Total derivative}\index{Derivative!Total} $D_i=D_{x^i}$ which is defined by
\[ D_i=\p_i +\sum_I u^a_{I,i}\frac{\p}{\p {u^a_I}},\]
such that, by the chain rule,  
\[ \p_i (\mathcal{L}(j^kf))= (\p_i\mathcal{L})(j^kf) +\sum_I (\p_{I,i}f^a) \frac{\p \mathcal{L}}{\p {u^a_I}}(j^kf)= (D_i\mathcal{L})(j^kf). \]
For a multi-index $I=(i_1,\ldots ,i_\ell )$ of length $|I| := \ell$ we define
\[ D_I =D_{i_1}\cdots D_{i_\ell}\]
and 
\[ (-D)_I=(-D_{i_1})\cdots (-D_{i_\ell}) = (-1)^\ell D_I.\]
\bt \label{ELFTThm} Let $\mathcal{L}$ be a smooth function on $\mathrm{Jet}^k(\O, \bR^m)$ and consider the corresponding action
functional (with respect to the canonical volume form). 
A smooth map $f: \O\ra \bR^m$ is a critical point of the action 
under smooth variations with compact support in $\O$ if and only if it has finite action and is a solution of the following system of 
partial differential equations of order $\le k$: 
\be\label{ELeq}  \a_a := \sum_{|I|\le k} (-D)_I\frac{\p \mathcal{L}}{\p u^a_I}(j^kf)=0,\quad a=1,\ldots , m. \ee  
\et
\bd The equations \re{ELeq}  are called the \emph{Euler-Lagrange equations}\index{Euler-Lagrange!equations} or \emph{equations of motion}
of the classical field theory defined by the Lagrangian $\mathcal{L}$. A \emph{solution}\index{Solution!of the classical field theory} of the classical field theory 
is, by definition, a solution of its equations of motion (irrespective of whether it has finite action or not).  
\ed
In order to compare to the Euler-Lagrange equations in mechanics we observe 
that 
\[  \sum_{|I|\le k} (-D)_I\frac{\p}{\p u^a_I} = \frac{\p}{\p u^a} + \sum_{1\le |I|\le k}(-D)_I\frac{\p}{\p u^a_I} .\]
So for a first order Lagrangian we obtain
\[ \frac{\p}{\p u^a} - \sum_{i}D_i\frac{\p}{\p u^a_i}.\]
In the case $n=1$ these are precisely the Euler-Lagrange equations of classical mechanics if we denote 
$u^a=q^a$, $u^a_1=\hat{q}^a$, $x^1=t$, and $D_1= \frac{d}{dt}$. 
\pf (of Theorem \ref{ELFTThm}) A smooth map $f$ with finite action is a critical point of the action if and only if for all 
variations $h$ with compact support we have 
\[ \left. \frac{d}{d\e}\right|_{\e =0}S[f+\e h]=0.\] 
We compute 
\begin{eqnarray*}
\left. \frac{d}{d\e }\right|_{\e =0} \mathcal{L}(j^k(f+\e h)) &=&\left. \frac{d}{d\e }\right|_{\e =0} \mathcal{L}(j^k(f)+\e j^k(h))  =
\sum \frac{\mathcal{\p L}}{\p u^a_I}(j^kf) u_I^a (j^k(h))
\\
&=&\sum \frac{\mathcal{\p L}}{\p u^a_I}(j^kf) \p_I h^a .
\end{eqnarray*} 
Applying the divergence theorem to $\p\O$ if $\p\O\neq \emptyset$, or to the boundary of an open ball containing the support of $h$ if $\O = \bR^n$, 
we obtain 
\begin{eqnarray*}\left. \frac{d}{d\e}\right|_{\e =0}S[f+\e h]&=&  \sum \int_\O  \frac{\mathcal{\p L}}{\p u^a_I}(j^kf) \p_I h^advol = \sum \int_\O (-\p)_I\left( \frac{\mathcal{\p L}}{\p u^a_I}(j^kf) \right) h^advol\\ 
&=& \sum \int_\O \left( (-D)_I \frac{\mathcal{\p L}}{\p u^a_I} \right) (j^kf) h^advol = \sum \int_\O \a_ah^advol.\end{eqnarray*}  
This vanishes for all $h$ if and only if $\a_a=0$ for all $a=1,\ldots ,n$.  
\epf 
The one-form $\a= \sum \a_ady^a$ along $f$ is called the \emph{Euler-Lagrange} one-form\index{Euler-Lagrange!one-form}. It generalizes the 
one-form which we encountered in Lagrangian mechanics. We will also consider the \emph{Euler-Lagrange}
operators\index{Euler-Lagrange!operator}
\nomenclature[aEa]{$E_a$}{component of Euler-Lagrange operator}
\be\eqlabel{ELop}
 E_a :=  \sum_{|I|\le k} (-D)_I\frac{\p}{\p u^a_I},\quad a=1,\ldots ,n,
\ee
which are differential operators acting on smooth functions on the manifold $\mathrm{Jet}^k(\O , \bR^m)$, such as 
$\mathcal{L}$.  The operator $E_a$ is related to the function $\a_a\in C^\infty (\O )$ by
\[ E_a(\mathcal{L})(j^kf) = \a_a.\]

It is clear that the Lagrangian of a classical field theory determines the equations of motion and, hence, the solutions of the theory. 
Adding a constant to the Lagrangian does not change the equations of motion. More generally, we will show that adding 
a \emph{total divergence} does not alter the equations of motion. 
\bd A \emph{total divergence}\index{Total divergence} is a function $f$ on $\mathrm{Jet}^k(\O , \bR^m)$ of the form
\nomenclature[aDivP]{$\mathrm{Div}\, P$}{total divergence}
\[ 
 f = \mathrm{Div}\, P := \sum_{i=1}^n D_iP^i,
\]
where $P=(P^1,\ldots ,P^n)$ is a  smooth vector-valued function on $\mathrm{Jet}^\ell (\O , \bR^m)$ for some $\ell \ge k$. 
\ed 
\bp\proplabel{euler_op_vanish_on_L} Let $\mathcal{L} = \mathrm{Div}\, P$ be a total divergence. Then the Euler-Lagrange operators $E_a$ vanish
on $\mathcal{L}$:
\[ E_a(\mathcal{L})=0.\]   
\ep 
\pf Let $f,h : \O \ra \bR^m$ be smooth maps and assume that $h$ has compact support in $\O$. 
Let $\O'$ be a domain with smooth boundary, such that $\O'$  contains the support of $h$ and is relatively compact 
in $\O$. Then, as in the proof of Theorem \ref{ELFTThm}, we have that 
\begin{eqnarray*}  \sum \int_{\O} E_a(\mathcal{L}(j^kf))h^a dvol &=& \sum \int_{\O'} E_a(\mathcal{L}(j^kf))h^a dvol\\ 
&=& \sum \int_{\O'} \left( (-D)_I \frac{\mathcal{\p L}}{\p u^a_I} \right) (j^kf) h^advol\\
&=&  \sum \int_{\O'}  \frac{\mathcal{\p L}}{\p u^a_I}(j^kf) \p_I h^advol\\
&=&   \left. \frac{d}{d\e}\right|_{\e =0}\int_{\O'} \mathcal{L}(j^k(f+\e h))dvol.\end{eqnarray*} 
By the divergence theorem,  
\begin{eqnarray*} \int_{\O'} \mathcal{L}(j^k(f+\e h))dvol &=&\int_{\O'} \mathrm{Div}\, P(j^\ell (f+\e h))dvol\\
&=& \int_{\p \O'} \langle P(j^\ell f),\nu \rangle dvol_{{\p \O'}},
\end{eqnarray*} 
where $\nu$ denotes the outer unit normal and  $dvol_{{\p \O'}}$ the induced volume form of $\p \O'\subset \bR^n$. Since the 
result does not depend on $\e$, we conclude that $E_a(\mathcal{L}(j^kf)))=0$ for all smooth maps $f:\O \ra \bR^m$. This proves
that $E_a(\mathcal{L})=0$.
\epf  
\bt Let $\mathcal{L}_1, \mathcal{L}_2\in C^\infty (\mathrm{Jet}^k(\bR^n , \bR^m))$ be two Lagrangians of order $\le k$ 
defined on $\bR^n$. Then  $\mathcal{L}_1$ and $\mathcal{L}_2$ 
have the same Euler-Lagrange equations if and only if $\mathcal{L}_1-\mathcal{L}_2$  is a total divergence\index{Total divergence}. 
\et 
\pf By the previous proposition,  we already know that $\mathcal{L}_1$ and $\mathcal{L}_2$ 
have the same Euler-Lagrange equations if $\mathcal{L}_1-\mathcal{L}_2$
is a total divergence. Therefore it suffices to show that a Lagrangian $\mathcal{L}\in C^\infty (\mathrm{Jet}^k(\bR^n , \bR^m))$ such that $E_a(\mathcal{L})=0$ for all
$a$ (that is a Lagrangian with trivial equations of motion) is necessarily a total divergence.  For every $f\in C^\infty (\bR^n, \bR^m)$ 
we have 
\be \label{LjkfEq}
\mathcal{L}(j^kf) = \int_{0}^1  \frac{d}{d\e} \mathcal{L}(\e j^kf)d\e - \mathcal{L}(s_0),
\ee 
where $s_0$ stands for the zero section 
of  the vector bundle $\mathrm{Jet}^k(\bR^n , \bR^m) \ra \bR^n$, that is  the $k$-th order jet
of the constant map $\bR^n \ra \bR^m$, $x\mapsto 0$. Clearly the function $\mathcal{L}(s_0)\in C^\infty (\bR^n)$ does not depend on $f$ and  
can be written in the form $\mathcal{L}(s_0(x)) =\p_1 F_1(x)$ for some function $F_1\in C^\infty (\bR^n )$. This shows that it is a total divergence $\sum D_iF_i$, since we
can simply put  $F_i=0$ for $2\le i\le n$. It remains to show that the integral on the right-hand side of \re{LjkfEq} is also
a total divergence. By the chain rule we have  
\[ \frac{d}{d\e} \mathcal{L}(\e j^kf) = \sum \p_I f^a\frac{\p \mathcal{L}}{\p u^a_I}(\e j^kf).\]
Performing partial integrations for each term $\ell =|I|\le k$ brings this to the form
\be \label{FaDEqu} \sum  f^a(-D)_I\frac{\p \mathcal{L}}{\p u^a_I}(\e j^kf) + (\mathrm{Div} P_{\e })(j^{2k}f),\ee
where $P_\e= (P^1_\e ,\ldots ,P^n_\e )$ is a vector-valued function on
$\mathrm{Jet}^{2k-1}(\bR^n,\bR^m)$ depending smoothly on all the variables, including the parameter $\e$. (Notice that $\mathrm{Div} P_{\e }$ is thus
a function on $\mathrm{Jet}^{2k}(\bR^n,\bR^m)$ depending smoothly on all the variables, including the parameter $\e$.) 
The first term in \re{FaDEqu} vanishes by the assumption $E_a(\mathcal{L})= \sum (-D)_I\frac{\p \mathcal{L}}{\p u^a_I}=0$. 
Thus,
\[ \int_{0}^1  \frac{d}{d\e} \mathcal{L}(\e j^kf)d\e = \left(  \mathrm{Div}\, \int_0^1P_{\e }d\e\right)(j^{2k}f)\]
is a total divergence. \epf

\section{Automorphisms and conservation laws}
We begin by observing that for every pair of smooth manifolds $M$, $N$ the group $\mathrm{Diff}(M)\times \mathrm{Diff}(N)$ acts naturally on 
$C^\infty (M,N)$. In fact, given a group element $g= (\varphi, \psi )$ and a smooth map $f:M\ra N$ we have 
\[ g\cdot f = \psi \circ f \circ \varphi^{-1}.\] 
This action induces an action on the jet bundle $\mathrm{Jet}^k(M,N)$:
\[ g \cdot (j^k_xf) := j^k_{\varphi (x)} (g\cdot f),\quad x\in M.\]

\bd Let $M,N $ be smooth manifolds, $dvol$ a volume form\index{Volume!form} on $M$, $n=\dim M$ and $\mathcal{L} \in C^\infty (\mathrm{Jet}^k(M,N))$ a Lagrangian.  An element $g=(\varphi ,\psi) \in \mathrm{Diff}^+(M) \times \mathrm{Diff} (N)$ is called an \emph{automorphism}\index{Automorphism} of the Lagrangian $n$-form $\mathcal{L}dvol$  if for all
$f\in C^\infty (M,N)$, $x\in M$: 
\be \label{AutLdvolEq} \mathcal{L}(g\cdot j^k_{x}f) (\varphi^*dvol)_x = \mathcal{L}(j^k_xf)dvol_x.\ee
Here $\mathrm{Diff}^+(M) \subset \mathrm{Diff}(M)$\nomenclature[aDiff+M]{$\mathrm{Diff}^+(M)$}{subgroup of orientation preserving diffeomorphisms of $M$} denotes the subgroup of orientation preserving diffeomorphisms of 
$M$.  
\ed 
Pulling back the $n$-forms in equation \re{AutLdvolEq} by $\varphi^{-1}$ we obtain the equivalent equation
\be  \label{AutLdvolEq2} \mathcal{L}(g\cdot j^k_{\varphi^{-1}(x)}f) dvol_x = \mathcal{L}(j^k_{\varphi^{-1}(x)}f)((\varphi^{-1})^*dvol)_x.\ee 
\bp If $g\in \mathrm{Diff}^+(M) \times \mathrm{Diff} (N)$ is an automorphism of $\mathcal{L}dvol$, then 
\[ S{[}g\cdot f{]}=\int_M   \mathcal{L}(j^k(g\cdot f)) dvol = S{[}f{]}=\int_M \mathcal{L}(j^kf)dvol\]
for all $f\in C^\infty (M,N)$. 
\ep 
\pf 
This follows from 
\begin{eqnarray*} \left[  (\varphi^{-1})^*\left( \mathcal{L}(j^kf)dvol \right) \right]_x &=& \mathcal{L}(j^k_{\varphi^{-1}(x)}f)((\varphi^{-1})^*dvol)_{x} \stackrel{\re{AutLdvolEq}}{=} \mathcal{L}(g\cdot j^k_{\varphi^{-1}(x)}f) dvol_x\\ 
&=& 
\mathcal{L}( j^k_x (g\cdot f) )dvol_x. 
\end{eqnarray*}
\epf 
The calculation in this proof shows that $g=(\varphi ,\psi) \in \mathrm{Diff}^+(M) \times \mathrm{Diff} (N)$ is an automorphism of $\mathcal{L}dvol$
if and only if for all $f\in C^\infty (M,N)$:
\[  (\varphi^{-1})^*\left( \mathcal{L}(j^kf)dvol \right) = \mathcal{L}( j^k (g\cdot f) )dvol,  \]
or, equivalently,
\[ \varphi^* \left( \mathcal{L}( j^k (g\cdot f) )dvol\right) =  \mathcal{L}(j^kf)dvol .\]
\bd\label{def_cons_law} A \emph{conservation law}\index{Conservation!law} for a Lagrangian $\mathcal{L} \in C^\infty (\mathrm{Jet}^k(\bR^n,\bR^m))$ is a total divergence\index{Total divergence}
$\mathrm{Div}\, P$ which vanishes on all solutions of the Euler-Lagrange equations of $\mathcal{L}$. 
\ed 
\begin{Th}[Noether]  \label{NThm}\index{Noether!theorem} Consider a classical field theory defined by a Lagrangian $\mathcal{L} \in C^\infty (\mathrm{Jet}^k(\bR^n,\bR^m))$ and 
denote by $dvol$ the standard volume form\index{Volume!form} of $\bR^n$. With  every local one-parameter group of local automorphisms of $\mathcal{L}dvol$  one can associate a conservation law of the form $\mathrm{Div}\, P=\sum Q^aE_a(\mathcal{L})$, 
where $Q^a= Y^a-\sum u_i^aX^i$ is determined by the vector field $\sum_{i=1}^n X^i\frac{\p}{\p x^i}+ \sum_{a=1}^mY^a\frac{\p}{\p y^a}$ 
generating the 
local one-parameter group. 
\end{Th}
For the proof of Noether's theorem we will use a series of lemmas. Let $X, Y$ be smooth vector fields on $\bR^n$ and $\bR^m$, respectively. 
Then the flow of $Z=X+Y$ on $\bR^n \times \bR^m$ induces a flow on $\mathrm{Jet}^k(\bR^n,\bR^m)$ and we denote the 
corresponding vector field on $\mathrm{Jet}^k(\bR^n,\bR^m)$  by $\mathrm{pr}^{(k)}Z$. 
\bd The vector field $\mathrm{pr}^{(k)}Z$\nomenclature[aprkZ]{$\mathrm{pr}^{(k)}Z$}{$k$-th prolongation of $Z$} is called the $k$-th \emph{prolongation}\index{Prolongation} of $Z$. 
\ed   
\bl \label{infL} Consider a Lagrangian $\mathcal{L} \in C^\infty (\mathrm{Jet}^k(\bR^n,\bR^m))$ and the standard volume form $dvol$ on $\bR^n$. Let $X, Y$ be smooth vector fields on $\bR^n$ and $\bR^m$, respectively. 
Then $Z=X+Y$ is an infinitesimal automorphism\index{Automorphism!Infinitesimal} of $\mathcal{L}dvol$ if and only if 
\be (\mathrm{pr}^{(k)}Z)(\mathcal{L}) + \mathcal{L}\mathrm{div} X=0. \ee
\el 
\pf  Recall that the divergence $\mathrm{div}X$ of a smooth vector field $X$ on $\bR^n$ is characterized by the 
equation 
\[ L_Xdvol = \mathrm{div}Xdvol,\]
where $L_X$ denotes the Lie derivative. The differential equation characterizing infinitesimal 
automorphisms is obtained by differentiating the equation \re{AutLdvolEq} with respect to $t$ after evaluation on a one-parameter group
$g_t=(\varphi_t,\psi_t)\in \mathrm{Diff}^+(\bR^n) \times \mathrm{Diff} (\bR^m)$. The derivative of the left-hand side is
\[ \left. \frac{d}{dt}\right|_{t=0}  \mathcal{L}(g_t\cdot j^k_{x}f) (\varphi_t^*dvol)_x = (\mathrm{pr}^{(k)}Z)(\mathcal{L})|_{j^k_{x}f}dvol_x + \mathcal{L}(j^k_{x}f)(\mathrm{div} X)_xdvol_x,\]
whereas the right-hand side does not depend on $t$.  
\epf
\bl \label{prolongL} Let $X, Y$ be smooth vector fields on $\bR^n$ and $\bR^m$, respectively. Then the $k$-th 
prolongation of $Z=X+Y$ is given by 
\be \label{prolongEq}  \mathrm{pr}^{(k)}Z = Z + \sum_{1\le |J|\le k}\sum_a Y_J^a\frac{\p}{\p u^a_J},\ee
where 
\[ Y_J^a=D_JQ^a+\sum u_{J,i}^aX^i,\quad Q^a= Y^a-\sum u_i^aX^i.\] 
\el
\pf Since $X$ and $Y$ commute, the flow $\varphi_t^Z$ of $Z=X+Y$ decomposes as 
the composition 
$\varphi_t^Z=\varphi_t^X\circ \varphi_t^Y$ of the flows of its summands. This implies 
that $\tilde{\varphi}_t^Z=\tilde{\varphi}_t^X\circ \tilde{\varphi}_t^Y$, where 
$\tilde{\varphi}_t^X, \tilde{\varphi}_t^Y, \tilde{\varphi}_t^Z$ denote the induced flows on $\mathrm{Jet}^k(\bR^n,\bR^m)$. 
Differentiation with respect to $t$ yields that
\[ \mathrm{pr}^{(k)}Z = \mathrm{pr}^{(k)}X + \mathrm{pr}^{(k)}Y.\] 
Therefore, it suffices to check the formula \re{prolongEq} in the special cases $X=0$ and $Y=0$.  

We first consider the case $X=0$. Then $Q^a=Y^a$ and $Y_J^a=D_JY^a$. So, what we have to show
is 
\be \label{prolYEq} \mathrm{pr}^{(k)}Y = Y + \sum_{1\le |J|\le k}\sum_a D_J Y^a\frac{\p}{\p u^a_J}.\ee
Let us denote by $\psi_t$ the flow of $Y$ on $\bR^m$ and by $g_t=(\mathrm{Id},\psi_t)$ the corresponding local one-parameter group of local diffeomorphisms of $\bR^n\times \bR^m$. The induced action on $\mathrm{Jet}^k(\bR^n,\bR^m)$ is given by 
\[ g_t \cdot j^k_xf = j^k_x(\psi_t\circ f) = \left(x,\psi_t (f(x)), \left\{ \p_J (\psi_t^a\circ f)(x))\right\}_{1\le |J|\le k,\; a=1,\ldots, m}\right)\]
for all $f\in C^\infty (\bR^n,\bR^m)$ and $x \in \bR^n$, where $\psi_t^a = y^a\circ \psi_t$. Differentiating this equation with respect to $t$ 
yields 
\begin{eqnarray*} \mathrm{pr}^{(k)}Y |_{j_x^kf}&=& \left(0,Y_{f(x)},\{ \p_J(Y^a\circ f)(x)\}_{1\le |J|\le k,\; a=1,\ldots, m}\vphantom{j^k_xf}\right)\\
&=&\left(0,Y_{f(x)},\{ (D_JY^a)(j^k_xf)\}_{1\le |J|\le k,\; a=1,\ldots, m}\right).\end{eqnarray*} 
This is precisely the right-hand side of \re{prolYEq}.  

Next we assume $Y=0$, in which case $Q^a=-\sum u_i^aX^i$. 
Firstly, we consider the case $k=1$ as a warm up. 
Now 
\[ Y_j^a=D_jQ^a+\sum u_{i,j}^aX^i = -\sum u_i^a\p_j X^i.\]
So, what we have to show is
\be \label{prolXEq} \mathrm{pr}^{(1)}X = X - \sum u^a_i\p_j X^i\frac{\p}{\p u^a_j}.\ee
Let us denote by $\varphi_t$ the flow of $X$ and by $g_t=(\varphi_t,\mathrm{Id})$  the corresponding local one-parameter group of local diffeomorphisms of $\bR^n\times \bR^m$. It acts on  $\mathrm{Jet}^1(\bR^n,\bR^m)$ by 
\[ g_t \cdot j^k_xf= j^k_{\varphi_t (x)}(f\circ \varphi_t^{-1}) = (\varphi_t(x),f(x),\{ \p_j (f^a\circ \varphi_t^{-1})|_{\varphi_t (x)}\}_{j,a} ).\]
Differentiation with respect to $t$ yields 
\[  \mathrm{pr}^{(1)}X|_{j^k_xf} = \left. \frac{d}{dt}\right|_{t=0} g_t \cdot j^k_xf=\left( X_x,0, 
\left\{ \left. \frac{d}{dt}\right|_{t=0}\p_j (f^a\circ \varphi_t^{-1})|_{\varphi_t (x)} \right\}\right).\]
We compute 
\[  d (f\circ \varphi_t^{-1})|_{\varphi_t (x)} = df \circ d (\varphi^{-1}_t)|_{\varphi_t(x)}= \left[ (\varphi_t^{-1})^*df\right]_{\varphi_t (x)}.\]
In order to differentiate this with respect to $t$, we consider the time-dependent matrix-valued function 
$F_t(x):=F(t,x) := (\varphi_t^{-1})^*df|_x$. So, what we have to differentiate is $F(t,\varphi_t(x))$. Its total time-derivative is
\begin{eqnarray*} \left. \frac{d}{dt}\right|_{t=0} F(t,\varphi_t(x)) &=& \frac{\p F}{\p t}(0,x) + dF_0X_x = -L_Xdf|_x + (\mathrm{Hess}\, f)(X_x,\cdot )\\
&=& -d(dfX)|_x+ (\mathrm{Hess}\, f)(X_x,\cdot )= -df\circ dX|_x.\end{eqnarray*}
This shows that 
\begin{eqnarray*}  \left. \frac{d}{dt}\right|_{t=0}\left( \p_j (f^a\circ \varphi_t^{-1})|_{\varphi_t (x)}\right) &=& -df^a\circ dX|_x\p_j = -\sum\p_if^a \p_j X^i(x)\\
&=&- \left( \sum u^a_i\p_jX^i\right) (j^1_xf),\end{eqnarray*}
proving \re{prolXEq} in the case $k=1$.

Let us finally consider the case of general $k\ge 1$. We first claim that for every multi-index of length $\le k$ we have 
\be \label{helpEq} (D_JQ^a)(j^{k+1}f)= -\p_JX(f^a).\ee
In fact, 
\[ Q^a(j^kf)=-\sum (u_i^aX^i)(j^kf)= -\sum \p_if^aX^i=-X(f^a)\]
implies  \re{helpEq}. As a consequence, 
\[ Y^a_J (j^{k+1}f)= \left(D_JQ^a +\sum u^a_{J,i}X^i\right)(j^{k+1}f) =  -\p_JX(f^a)+\sum (\p_{J,i}f^a)X^i.\]
So, what we have to show is 
\be \label{ClaimEq} \left. \frac{d}{dt}\right|_{t=0}\left( \p_J (f^a\circ \varphi_t^{-1})|_{\varphi_t (x)}\right)= -\p_JX(f^a)|_x+\sum (\p_{J,i}f^a)X^i|_x.\ee
To calculate the left-hand side we put $F_t(x):=F(t,x) :=  \p_J (f^a\circ \varphi_t^{-1})_x$. Then we see
that 
\begin{eqnarray*}  \left. \frac{d}{dt}\right|_{t=0}\left( \p_J (f^a\circ \varphi_t^{-1})|_{\varphi_t (x)}\right) &=&  \left. \frac{d}{dt}\right|_{t=0}F(t,\varphi_t(x)) =
\frac{\p F}{\p t}(0,x) +dF_0X_x\\
&=& -\p_JX(f^a)|_x + d (\p_Jf^a)X|_x.\end{eqnarray*}
This coincides with the right-hand side  of \re{ClaimEq}.
\epf 
Now we can prove Noether's theorem.   
\pf (of Theorem \ref{NThm}) We have to show that $\sum Q^aE_a(\mathcal{L})$ is a total divergence. 
By partial integration we have  
\[
\sum Q^aE_a(\mathcal{L}) = \sum Q^a(-D)_J\frac{\p \mathcal{L}}{\p u^a_J} = \sum (D_JQ^a)\frac{\p \mathcal{L}}{\p u^a_J} + \mathrm{Div}\, V 
\]
for some vector valued function $V=(V^1,\ldots, V^n)$ on $\mathrm{Jet}^{2k-1}(\bR^n,\bR^m)$. 
It suffices to show that $\sum (D_JQ^a)\frac{\p \mathcal{L}}{\p u^a_J}$ is a total divergence. Recall that we defined $Y^a_J= D_JQ^a+\sum u^a_{J,i}X^i$ for 
$1\le |J| \le k$. The formula holds also for $|J|=0$: 
\[
 Y^a= Q^a+\sum u^a_iX^i . 
\]
Using this, Lemma \ref{prolongL},  and Lemma \ref{infL} we have 
\begin{eqnarray*} \sum (D_JQ^a)\frac{\p \mathcal{L}}{\p u^a_J} &=& \sum (Y^a_J -\sum u^a_{J,i}X^i) \frac{\p \mathcal{L}}{\p u^a_J} = (-X+ \mathrm{pr}^{(k)}Z) \mathcal{L} 
-\sum u^a_{J,i}X^i\frac{\p \mathcal{L}}{\p u^a_J}\\
&=& -\sum X^iD_i\mathcal{L} +(\mathrm{pr}^{(k)}Z) \mathcal{L} = -\sum X^iD_i\mathcal{L} -\mathcal{L} \mathrm{div}\, X\\ 
&=& -\sum D_i(\mathcal{L}X^i),\end{eqnarray*}
which is a total divergence. 
\epf  
In the proof of Theorem~\ref{NThm} we have shown that $\sum Q^aE_a(\mathcal{L})$ is the total divergence of the
vector-valued function with components $P^i=V^i-\mathcal{L}X^i$, where $X=\sum X^i\p_i$ is the projection of the infinitesimal 
automorphism\index{Automorphism!Infinitesimal} $Z$ onto the source manifold $\bR^n$ and $V^i$ is a function on $\mathrm{Jet}^{2k-1}(\bR^n,\bR^m)$ obtained by partial integration. 
This means that $P=(P^1,\ldots ,P^n)$ can be explicitly computed by performing the partial integrations. In the case $k=1$ 
we obtain the following result. 
\bc \label{1storderCor} Let $\mathcal{L} \in C^\infty (\mathrm{Jet}^1(\bR^n,\bR^m))$ be a first order Lagrangian. Then Noether's conservation
law associated with an infinitesimal automorphism\index{Automorphism!Infinitesimal} $Z=X+Y$ as in Theorem  \ref{NThm} is the 
total divergence of the vector-valued function $P=(P^1,\ldots ,P^n)$ on $\mathrm{Jet}^1(\bR^n,\bR^m)$ given by
\[ P^i = -\sum Q^a\frac{\p \mathcal{L}}{\p u^a_i} -\mathcal{L}X^i,\quad Q^a= Y^a-\sum u_j^aX^j.\] 
\ec 
\pf It suffices to observe that $V^i:= -\sum Q^a\frac{\p \mathcal{L}}{\p u^a_i}$ satisfies 
\[  -\sum Q^aD_i\frac{\p \mathcal{L}}{\p u^a_i} = \sum (D_iQ^a)\frac{\p \mathcal{L}}{\p u^a_i} + \sum D_i V^i.\]
\epf 
As another corollary we obtain the generalization to time-dependent Lagrangian mechanical systems\index{Noether!theorem!for time-dependent systems} on $\bR^m$ of Noether's theorem (Theorem \ref{NoetherThmMech}),  
which concerned time-independent mechanical systems. 
\bc \label{LMSNoetherL}Let $\mathcal{L}=\mathcal{L}(t,q^1,\ldots, q^m,\hat{q}^1,\ldots ,\hat{q}^m)$ be a time-dependent Lagrangian mechanical system on $\bR^m$. Then 
every infinitesimal automorphism\index{Automorphism!Infinitesimal} $Z=X+Y$, $X=X^1\p_t\in \mathfrak{X}(\bR)$, $Y=\sum Y^a\p_{q^a}\in \mathfrak{X}(\bR^m)$, of $\mathcal{L}dt$ gives rise to an integral of motion 
\[ f=-P^1=\sum Q^a\frac{\p \mathcal{L}}{\p \hat{q}^a} +\mathcal{L}X^1, \quad Q^a= Y^a-\sum \hat{q}^aX^1.\]  
\ec 
\begin{Ex}Notice that in the case $X=0$, an infinitesimal automorphism\index{Automorphism!Infinitesimal} $Y\in \mathfrak{X}(\bR^m)$ of $\mathcal{L}dt$ 
is the same as 
an infinitesimal automorphism\index{Automorphism!Infinitesimal} of $\mathcal{L}$ and the integral of motion takes the familiar form of Theorem \ref{NoetherThmMech}:
\[ f= \sum Y^a\frac{\p \mathcal{L}}{\p \hat{q}^a} = d\mathcal{L}Y^{\textnormal{ver}}.\] 
The prolongation of  a vector field $Y\in \mathfrak{X}(\bR^m)$ given in \re{prolYEq} simplifies  in the considered case $k=1=n$ as 
\[ \mathrm{pr}^{(1)}Y = Y +\sum D_tY^a\frac{\p}{\p \hat{q}^a}\]
and thus $Y$ is an infinitesimal automorphism\index{Automorphism!Infinitesimal} if and only if
\[  Y(\mathcal{L}) +\sum D_tY^a\frac{\p \mathcal{L}}{\p \hat{q}^a}=0.\]
Evaluating this along the velocity vector field $\g'$ of a smooth curve $\g : I \ra \bR^m$, 
we obtain
\[  \sum Y^a(t)\frac{\p \mathcal{L} (\g' (t))}{\p q^a} +\sum \dot{Y}^a(t)\frac{\p \mathcal{L}(\g'(t))}{\p \hat{q}^a}=0,\]
where $Y^a(t) := Y^a(\g(t))$. 
So $Y$ is an infinitesimal automorphism\index{Automorphism!Infinitesimal} if and only if the latter equation holds for all $\g$. 
\end{Ex}
Next we state Noether's theorem for time-dependent Lagrangian mechanical systems in its general form replacing $\bR^m$ by an arbitrary smooth manifold. 
\bt\index{Noether!theorem!for time-dependent systems} Let $\mathcal{L}\in C^\infty(\bR \times TM)$ be a time-dependent Lagrangian mechanical system on a smooth manifold $M$. Then 
every infinitesimal automorphism\index{Automorphism!Infinitesimal} $Z=X+Y$, $X=X^1\p_t\in \mathfrak{X}(\bR)$, $Y\in \mathfrak{X}(M)$, of $\mathcal{L}dt$ gives 
rise to an integral of motion 
\be \label{IMEqu} f=Y^{\textnormal{ver}}(\mathcal{L}) + (\mathcal{L}-\xi (\mathcal{L}))X^1,\ee
where $\xi\in \mathfrak{X}(TM)$ denotes as usual the vertical vector field generated by scalar multiplication $(s,v)\mapsto e^sv$ in the fibers of $TM$ and $Y^{\textnormal{ver}}$ denotes the vertical lift of $Y$. 
\et 

\pf Let $(U,\varphi)$ be a local chart of $M$. According to  Corollary \ref{LMSNoetherL}, we have the following integral of motion 
on $TU$:  
\[ f_{(U,\varphi )}=\sum Q^a\frac{\p \mathcal{L}}{\p \hat{q}^a} +\mathcal{L}X^1, \quad Q^a= Y^a-\sum \hat{q}^aX^1,\]  
where $Y^a$, $a=1,\ldots , m$, $m=\dim M$, are the components of $Y$ with respect to the 
local chart $(q^1,\ldots ,q^m)=\varphi$ and $(t,q^1,\ldots ,q^m,\hat{q}^1,\ldots ,\hat{q}^m)$  are the corresponding 
local coordinates of $\bR\times TM=\mathrm{Jet}^{1}(\bR,M)$, defined on $\bR \times TU=\mathrm{Jet}^{1}(\bR,U)$. 
Since $\sum Y^a\frac{\p}{\p \hat{q}^a} =Y^{\textnormal{ver}}|_U$ and $\sum \hat{q}^a\frac{\p}{\p \hat{q}^a}=\xi|_U$, we see that 
$f_{(U,\varphi)}$ is the restriction of the globally defined (and manifestly\footnote{Note that in the definition of $f$, we did not use any coordinates.} coordinate independent) function $f$. This shows that $f$ is an integral of motion. 
\epf 
\begin{Ex} Consider the special case when $Y=0$. The expression for the prolongation of $X=X^1\p_t\in \mathfrak{\bR}$  on $\mathrm{Jet}^1(\bR,M)$ 
in local coordinates $(t,q,\hat{q})$ follows 
immediately from  \re{prolXEq}:
\be \mathrm{pr}^{(1)}X = X - \sum \hat{q}^a\p_tX^1\frac{\p}{\p \hat{q}^a}.\ee
The vector field $\mathrm{pr}^{(1)}X$ is in fact coordinate independent and obviously coincides with $X -\p_tX^1 \xi$. 
We conclude that $X$ is an infinitesimal automorphism\index{Automorphism!Infinitesimal} of $\mathcal{L}dt$ if and only if 
\[  0=X(\mathcal{L}) -\p_tX^1 \xi (\mathcal{L})+\mathcal{L}\p_tX^1=X(\mathcal{L}) +(\mathcal{L}-\xi (\mathcal{L}))\p_tX^1.\]
The right-hand side vanishes, in particular, if $\mathcal{L}$ is time-independent and $X=\p_t$. The corresponding integral of motion \re{IMEqu}
is precisely 
\[ \mathcal{L}-\xi (\mathcal{L})=-E,\]
the energy, up to the factor $-1$.   So we see that the energy 
is the integral of motion associated with the invariance of the 
Lagrangian under translations in time.\index{Time!translations}
\end{Ex} 
The notion of an infinitesimal automorphism of a Lagrangian $n$-form can be generalized as follows. 
\bd Let $X$ and $Y$ be smooth vector fields on $\bR^n$ and $\bR^m$, respectively, $\mathcal{L}\in C^\infty (\mathrm{Jet}^k(\bR^n,\bR^m))$ 
a Lagrangian and $dvol$ the standard volume form\index{Volume!form} of $\bR^n$. We say that $Z=X+Y$ is an \emph{infinitesimal automorphism}\index{Automorphism!Infinitesimal}  
of $\mathcal{L}dvol$ \emph{up to a divergence} if
\[ (\mathrm{pr}^{(k)}Z)(\mathcal{L}) + \mathcal{L}\mathrm{div} X\]
is a total divergence. 
\ed
Noether's theorem can be easily generalized as follows (see \appref{exercises}, \excref{41}). 
\begin{Th}[Noether] \label{uptodivThm}\index{Noether!theorem} Under the above assumptions, let $Z=X+Y$ be an  infinitesimal automorphism 
of $\mathcal{L}dvol$ up to a divergence.  Then 
\[ \sum Q^aE_a(\mathcal{L}),\quad \mbox{defined by}\quad  Q^a= Y^a-\sum u_i^aX^i,\] 
is a total divergence, 
where $X^i$  and $Y^a$ are the components of $X\in  \mathfrak{X}(\bR^n)$ and
$Y\in \mathfrak{X}(\bR^m)$. 
\end{Th}

\section{Why are conservation laws called conservation laws?}
Consider a  classical field theory defined by a Lagrangian $n$-form 
$\mathcal{L}dvol$, where $\mathcal{L} \in C^\infty (\mathrm{Jet}^k(\bR^n,\bR^m))$ and $dvol$ is the 
standard volume form\index{Volume!form} on $\bR^n$. We denote the standard coordinates of $\bR^n$ by 
$(x^0,\ldots x^{n-1})$ and think of $t=x^0$ as the time-coordinate and of $\vec{x}=(x^1,\ldots ,x^{n-1})$ 
as the spatial position. Suppose that we are given a conservation law
$\mathrm{Div}\, P$, $P\in C^\infty (\mathrm{Jet}^{\ell }(\bR^n,\bR^m),\bR^n)$, $\ell \ge k$. We denote by 
\[ J(t,\vec{x}) := P(j^{\ell}_xf), \]
the evaluation of $P$ on a solution $f\in C^\infty (\bR^n,\bR^m)$. 

\bd  The vector-valued  function 
\nomenclature[aJ]{$J$}{(Noether) current}\nomenclature[aJ0]{$J^0$}{charge density}\nomenclature[aJvec]{$\vec{J}$}{flux density}
\[ 
 J=(J^0,\vec{J}) : \bR^n \ra \bR^n=\bR \times \bR^{n-1}
\]
is called the \emph{current}\index{Current}, the function $J^0$ is called the \emph{charge density}\index{Charge!density}, 
$\vec{J}$ is called the \emph{flux density}\index{Flux density} and the spatial integral 
\be \label{QEqu} Q(t) := \int_{\bR^{n-1}}J^0(t,\vec{x})dx^1\wedge \cdots \wedge dx^{n-1}\ee\nomenclature[aQ]{$Q$}{(Noether) charge}
is called the \emph{charge}.

If $P$ is associated with an infinitesimal automorphism $Z$ of $\mathcal{L}dvol$, as described in 
Theorem \ref{NThm} (or, more generally, in Theorem \ref{uptodivThm}),  then   
$J$ is called the \emph{Noether current}\index{Noether!current}\index{Conserved current|see{Noether current}} and $Q$ is called the \emph{Noether charge}\index{Noether!charge} associated with $-Z$.   \ed 
The sign is included here such that the notation of Theorem \ref{NThm} is consistent with the usual 
conventions for the Noether charge in the physics literature. With this notation, the 
Noether current of an infinitesimal automorphism $Z=X+Y$ of a first order Lagrangian $\mathcal{L}dvol$ on $\bR^n$ is given by
\be \label{NcurrentEq} J^i = \sum_a \left(Y^a\circ f  -\sum_j \p_jf^aX^j\right)\frac{\p \mathcal{L}}{\p u^a_i}(j^1f) +\mathcal{L}(j^1f)X^i,\ee
see Corollary \ref{1storderCor}. 
Note that in the case $n=1$, the current reduces to the charge density, 
$J=J^0=Q$, and as we already observed,  the equation $\mathrm{Div}\, P=0$
reduces to the statement that $P$ is an integral of motion, that is to $Q'(t)=0$ (for all solutions $f$). This is generalized 
in the next theorem. 
\bt \label{conservedThm} Suppose that the current falls off sufficiently fast at infinity, in the sense that for every 
bounded interval $I$ 
there exists a Lebesgue integrable function $F: \bR^{n-1}\ra [0,\infty]$  such that 
\begin{eqnarray*} |J^0(t,\vec{x})|  &\le& F(\vec{x}),\\
\left| \frac{\p J^0}{\p t}(t,\vec{x})\right| &\le& F(\vec{x}),
\end{eqnarray*}
for all $(t,\vec{x})\in I\times \bR^{n-1}$ and 
\[ \lim_{r\ra \infty} \int_{\p B_r(0)} \langle \vec{J},\nu \rangle dvol_{\p B_r(0)} =0,\]
where $\nu$ denotes the outer unit normal of the sphere $\p B_r(0)\subset \bR^{n-1}$. 
Then the  charge is conserved, that is $t\mapsto Q(t)$ is constant.
\et  
\pf In virtue of the first inequality, we see that the total charge 
\re{QEqu} is finite by Lebesgue's theorem. The second inequality then shows that 
$Q$ is differentiable and
\[ Q'(t) = \int_{\bR^{n-1}}\frac{\p J^0}{\p t}(t,\vec{x})d^{n-1}x, \]
where we have abbreviated $d^{n-1}x=dx^1\wedge \cdots \wedge dx^{n-1}$. 
Since 
\[ \mathrm{div}\, J= \p_tJ^0+\mathrm{div}_{\bR^{n-1}}\vec{J}=0,\] 
we can rewrite this as 
\begin{eqnarray*} Q'(t) &=&  -\int_{\bR^{n-1}}\mathrm{div}_{\bR^{n-1}}\, \vec{J} (t,\vec{x})d^{n-1}x = -\lim_{r\ra \infty} \int_{B_r(0)} \mathrm{div}_{\bR^{n-1}}\, \vec{J} (t,\vec{x})d^{n-1}x\\ 
&=&-\lim_{r\ra \infty} \int_{\p B_r(0)} \langle \vec{J},\nu \rangle dvol_{\p B_r(0)} ,
\end{eqnarray*}
and the resulting limit is zero by the last assumption of the theorem. 
\epf

\section{Examples of field theories} 
Here we discuss some examples of classical field theories. The examples play important roles in active research on contemporary theories of high-energy particle physics, such as the standard model of particle physics and string theory.

\subsection{Sigma models}\index{Sigma model}
Let $(M,g)$ and $(N,h)$ be pseudo-Riemannian manifolds of dimension $m$ and $n$, respectively.
We assume that $(N,h)$ is oriented and denote its volume form by $dvol_h$. 
The most natural\footnote{This is to be understood in the sense that it generalizes the kinetic term in the standard Lagrangian~\eqref*{cm_std_lagr} of classical mechanics.} first order Lagrangian $n$-form  $\mathcal{L}dvol_h$  for maps $f\in C^\infty (N,M)$ is 
given by 
\be \label{SigmamodelEq}  \mathcal{L} (j^1f) =  \frac12 \langle df, df \rangle, \ee
where $\langle \cdot ,\cdot \rangle = \langle \cdot ,\cdot \rangle_{h,g}$ is the 
(possibly indefinite) fiber-wise scalar product on $T^*N\otimes f^*TM$ induced by $h$ and $g$. 
In local coordinates $(x^i)$ in a neighborhood of  $x\in N$ and $(y^a)$ in a neighborhood of $f(x)\in M$ we have 
\[ df_x = \sum \frac{\p f^a(x)}{\p x^i}dx^i|_x\ot \left. \frac{\p}{\p y^a}\right|_{f(x)}\]
and 
\[  \langle df_x, df_x \rangle = \sum  g_{ab}(f(x)) h^{ij}(x)\frac{\p f^a(x)}{\p x^i}\frac{\p f^b(x)}{\p x^j},\]
where $(h^{ij})$ denotes the inverse matrix of $(h_{ij})$. 
\bt The Euler-Lagrange equations for the Lagrangian $n$-form $\mathcal{L}dvol_h$ given by 
\re{SigmamodelEq}  are equivalent to   
\be \eqlabel{harmmapEq} \tau (f) := \mathrm{tr}_h \n df =0,\ee
where $\n$ is the connection in the vector bundle $T^*N\otimes f^*TM$ induced by the Levi-Civita
connections in $TN$ and $TM$. 
\et
\pf  See \appref{exercises}, \excref{40}. 
\epf 
\bd A smooth map $f: N\ra M$ which is a solution of \eqref{harmmapEq} is called a \emph{harmonic map}.\index{Harmonic!map}  
The vector field $\tau (f)$\nomenclature[gtau]{$\tau (f)$}{tension of $f$} along $f$ is called the \emph{tension}\index{Tension} of $f$. 
\ed
\begin{Ex} (Harmonic functions)\index{Harmonic!function} Consider the case when the target manifold $(M,g)$ is simply the Euclidean line $\bR$.
Then $f^*TM$ can be identified with the trivial line bundle over $N$ and 
\nomenclature[sDelta]{$\Delta$}{Laplace operator}
\[ 
 \tau (f) = \Delta f := \mathrm{div}\, \mathrm{grad}\, f
\]
is given by the  \emph{Laplace operator}\index{Laplace operator} $\Delta$ associated with the 
pseudo-Riemannian metric $h$ on the source manifold $N$.  The solutions
of the equation $\Delta f=0$ are called \emph{harmonic functions}. Recall that 
\[ \mathrm{grad}\, f = h^{-1}df \]
and 
\[ \mathrm{div}\, X = \mathrm{tr}\, \n X\]
for every vector field $X$ on $(N,h)$.  So indeed 
\[  \mathrm{div}\, \mathrm{grad}\, f = \mathrm{tr}\, \n (h^{-1}df)   =  \mathrm{tr}_h \n df = \tau (f).\] 
Here we have used that $\n_v (h^{-1} df) = h^{-1}\n_vdf$ for every $v\in TN$. Note that the 
pseudo-Riemannian Laplace equation $\Delta f=0$ is linear, whereas the harmonic map equation 
\eqref*{harmmapEq} is in general non-linear, since the connection $\n$ depends on $f$. 
Observe that in the special case when $(N,h)$ is a pseudo-Euclidean vector space we can write the metric 
as $h=\sum \e_i (dx^i)^2$, where $\e_i\in \{\pm 1\}$, and $\Delta f = \sum \e_i \p_i^2$. 
\end{Ex}
\begin{Ex} \label{lsmEx} Generalizing the previous example, we consider the case when the target manifold $(M,g)$ is a pseudo-Euclidean vector space. Then the 
components $\tau^a(f)$ of the tension $\tau (f)$ with respect to an affine coordinate system on the target manifold
$M$ are given by 
\[ \tau^a (f) = \Delta f^a.\]
So a smooth map $f$ from a pseudo-Riemannian manifold to a pseudo-Euclidean vector space is harmonic if and only if its components $f^a$ are harmonic functions. 
\end{Ex}
In the physics literature the Lagrangian \re{SigmamodelEq} is called a \emph{sigma-model} and in the case of 
Example \ref{lsmEx}  it is called a \emph{linear sigma model}\index{Sigma model!Linear}, since the equations of motion are linear. 
The components of the map
$f: N \ra M$ are considered as scalar fields. We can enlarge the class of sigma-model Lagrangians by 
including a potential: 
\[ \mathcal{L}(j^1f)=\frac12 \langle df , df \rangle -V(f),\]
 where $V\in C^\infty (M)$.
\begin{Ex} (Geodesics as harmonic maps)\index{Geodesic!as harmonic map} Consider the case when the source manifold $(N,h)$ is the Euclidean line $\bR$. Then 
a harmonic map $f: \bR \ra (M,g)$ is the same as a geodesic. 
\end{Ex}
\bp The Lie group $\mathrm{Isom}^+(N,h)\times \mathrm{Isom}(M,g)$\nomenclature[aIsom+Nh]{$\mathrm{Isom}^+(N,h)$}{subgroup of orientation preserving isometries of the metric $h$ on $N$} consists of automorphisms of the sigma-model Lagrangian
$n$-form 
$\mathcal{L}dvol_h$ defined by \re{SigmamodelEq}. 
\ep 
\pf This follows from the fact that the group $\mathrm{Isom}^+(N,h)$ of orientation preserving isometries of $(N,h)$ preserves the 
metric volume form $dvol_h$ and that $\mathrm{Isom}(N,h)\times \mathrm{Isom}(M,g)$ preserves 
the Lagrangian $\mathcal{L}$. 
\epf

\subsection{Pure Yang-Mills theory}
\label{pureYMsec}
Let $E$ be a real or complex vector bundle of rank $k$ over an oriented pseudo-Riemannian manifold $(M,g)$ with a \emph{reduction of the 
structure group} of $E$ to some compact\footnote{Recall that a compact subgroup of a Lie group is automatically a Lie subgroup.} subgroup $G\subset \GL{k,\bK }$, where $\bK =\bR$ or $\bC$\nomenclature[aK]{$\bK$}{field (in this book, $\bK =\bR$ or $\bC$)}. Such a \emph{$G$-reduction}\index{$G$!-reduction} in the vector bundle $E$ is defined 
as a principal $G$-subbundle $\mathcal{F}_G(E) \subset \mathcal{F}(E)$ of the bundle of frames of $E$. The elements 
of $\mathcal{F}_G(E)$ are called \emph{$G$-frames}\index{$G$!-frame}. 
\begin{Ex} An $\mathrm{O}(k)$-reduction in a real vector bundle $E$ is equivalent to a (positive definite fiber) metric $h$ in $E$. The corresponding
principal $\mathrm{O}(k)$-subbundle of the frame bundle  $\mathcal{F}(E)$\nomenclature[aFcalE]{$\mathcal{F}(E)$}{frame bundle of $E$} is the bundle of orthonormal frames in the metric vector bundle $(E,h)$.  
A $\U{k}$-reduction in a complex vector bundle $E$ is equivalent to a Hermitian metric $h$ in $E$. The corresponding
principal $\mathrm{U}(k)$-subbundle of the frame bundle  $\mathcal{F}(E)$  is the bundle of unitary frames in the Hermitian vector bundle $(E,h)$.  
\end{Ex}
Note that since $G$ is compact every $G$-reduction induces a metric $h$ in $E$ if $\bK=\bR$ and a 
Hermitian metric in $E$ if $\bK = \bC$. 
A connection $\n$ in a vector bundle $E$ endowed with a $G$-reduction $\mathcal{F}_G(E)$ is called 
a \emph{$G$-connection}\index{$G$!-connection} if the parallel transport with respect to $\n$ maps $G$-frames to $G$-frames.  
We denote the curvature of $\n$ by
\nomenclature[aFnabla]{$F^\n$}{curvature of a connection $\n$}
\[ 
 F=F^\n \in \G (\Lambda^2T^*M \ot {\gg}(E)),
\]
where $\gg (E_x)$ denotes the Lie algebra of the subgroup $G(E_x)\subset \GL {E_x}$ of elements preserving the set of $G$-frames in $E_x$, $x\in M$. 
For convenience, we will write $\so (E)$, $\su (E)$, $\ggl (E)$ etc.\ rather than $\so (k,\bR )(E)$, $\su (k)(E)$, $\ggl (k,\bK )(E)$ etc. 
\begin{Ex} An $\mathrm{O}(k)$-connection in a metric vector bundle $(E,h)$ of rank $k$ is the same as a metric connection. 
A $\U {k}$-connection in a Hermitian  vector bundle $(E,h)$ of rank $k$ is the same as a Hermitian connection. 
\end{Ex}
Given a vector  bundle $E$ with structure group $G$ over a pseudo-Riemannian manifold $(M,g)$, the space of fields of \emph{pure Yang-Mills theory}\index{Yang-Mills!theory} is 
the affine  space $\mathcal{A}_G(E)$ of $G$-connections in $E$. By choosing a reference connection $\n^0$ we can write 
$\n = \n^0 + \Phi$, where $\Phi  : M \ra T^*M\ot \gg (E)$ is a smooth section. In this way we can identify the affine space $\mathcal{A}_G(E)$
with the vector space $\G (T^*M\ot \gg (E))$. Using this identification, one can define jets of $\n$ by considering jets of the map $\Phi = \n -\n^0$. 
The Yang-Mills Lagrangian\index{Yang-Mills!lagrangian} is 
\be \label{YMLagrEq}  \mathcal{L} (j^1(\n ) ) = -\frac12 \langle F,F\rangle, \ee
where $\langle \cdot ,\cdot \rangle$ is the (fiber-wise) scalar product on $\Lambda^2 T^*M \ot \gg(E)$ obtained as the 
tensor product  of the scalar product ${\langle \cdot ,\cdot \rangle}_\Lambda$ on $\Lambda^2 T^*M$ induced by 
$g$ and the  scalar product $\langle A, B \rangle_\gg =  - \mathrm{tr}\, (AB)
$ on $\gg (E)$. Notice that $\langle A, A \rangle_\gg = -\mathrm{tr}\, A^2 = \mathrm{tr}\, A A^\dagger \ge 0$, where
$A^\dagger$ denotes the adjoint of $A$ with respect to the $G$-invariant metric $h$. In particular, $\langle \cdot , \cdot \rangle_\gg$ is 
real valued, irrespective of whether $E$ is a real or a complex vector bundle. The scalar product ${\langle \cdot ,\cdot \rangle}_\Lambda$ is normalized such that $\a\wedge \b$ is of unit length if
$\a$, $\beta$ are orthonormal.  

Recall that every connection satisfies the \emph{Bianchi identity}\index{Bianchi identity} 
\be \label{BianchiEq} d^\n F =0,\ee
where $d^\n$\nomenclature[adnabla]{$d^\n$}{covariant exterior derivative} is the covariant exterior derivative acting on differential forms with values 
in $\mathrm{End}(E)\supset \gg (E)$.  
\bt \label{YMThm} The Euler-Lagrange equations\index{Euler-Lagrange!equations}\index{Yang-Mills!equations} for the Yang-Mills Lagrangian \re{YMLagrEq} are equivalent to
\be \label{YMEq} d^\n \ast F=0.\ee
\et
Before we begin the proof of this proposition let us first discuss an example.  

\begin{Ex} (Maxwell theory)\index{Maxwell theory} Let us consider the special case when $(M,g)=(\bR^4,dt^2-\sum_{\a=1}^3 (dx^\a)^2)$ is 
the four-dimensional Minkowski space\index{Minkowski!space} and $E$ is a Hermitian line bundle.  So the structure 
group is the Abelian group $G=\U {1}$. Identifying the Lie algebra $\gu (1) = \sqrt{-1}\bR \cong \bR$ with $\bR$, we can consider
$F$ as a real-valued $2$-form.  The Bianchi and Yang-Mills equations reduce to $dF=d\ast F=0$. So the differential form $F$ is closed and co-closed, which 
implies the second order equations $\ast d\ast dF=0$ and $d\ast d \ast F=0$\nomenclature[sast]{$\ast$}{Hodge star operator}. 
The spatial components $F_{\a \b}=F (\p_\a ,\p_\b )$, $\a ,\b \in \{ 1,2,3\}$, define a real-valued time-dependent 
$2$-form in $\bR^3$. Identifying $\bR^3\cong \Lambda^2(\bR^3)^*$ by means of contraction of a vector with the Euclidean volume 
form, this $2$-form defines a time-dependent vector field $\vec{B}$ in $\bR^3$. The remaining components 
$E^\a := F_{0\a }=F(\p_t ,\p_\a)$, $\alpha = 1,2,3$, define a time-dependent vector field $\vec{E}=\sum_\a E^\a \p_\a$ in $\bR^3$. The vector fields $\vec{E}$\nomenclature[aEvec]{$\vec{E}$}{electric field} and $\vec{B}$\nomenclature[aBvec]{$\vec{B}$}{magnetic field} can be interpreted as the electric and the magnetic field in Maxwell's theory of electromagnetism and the Yang-Mills equation 
reduces  to half of Maxwell's equations in the vacuum, that is  
\[ \mathrm{div}\, \vec{E} =0,\quad \mathrm{rot}\, \vec{B} = \frac{\p}{\p t} \vec{E},\]
whereas the Bianchi identity reduces to the other half of Maxwell's vacuum equations, that is  
\[ \mathrm{div}\, \vec{B} =0,\quad \mathrm{rot}\, \vec{E} = -\frac{\p}{\p t} \vec{B},\]
see \appref{exercises}, \excref{44}.  
\end{Ex}

In order to derive the Yang-Mills equation \re{YMEq} it is helpful to first rewrite the Yang-Mills Lagrangian as:
\be -\frac12 \langle F,F\rangle dvol_g = -\frac12 \langle F\wedge \ast F\rangle_\gg,\ee
where $\langle \cdot \wedge \cdot \rangle_\gg$ denotes the $\Lambda^nT^*M$-valued pairing of $\Lambda^\ell T^*M \ot \gg (E)$ with 
$\Lambda^{n-\ell }T^*M\ot \gg (E)$ obtained by 
combining the wedge product and the scalar product on $\gg (E)$. Here $n=\dim M$ and $\ell =2$, but in later calculations we will also need the case $\ell=1$.  
This is a consequence of the following lemma. 
\bl For every section $\a$ of  $\Lambda^2T^*M \ot {\gg}(E)$ we have 
\be \langle \a , \a \rangle dvol_g = \langle \a \wedge \ast \a \rangle_\gg .\ee 
\el 
\pf Let $(e_i)_{i=1,\ldots ,n}$ be a local frame of $TM$ and put $\a_{ij} := \a (e_i,e_j)$. 
Then we can write 
\[ \a = \frac12 \sum \a_{ij}e^{ij},\]
where $(e^i)$ is the dual local frame of $T^*M$ and we have abbreviated $e^i\wedge e^j =: e^{ij}$.  
Now we can compute 
\begin{eqnarray*} 
\langle \a \wedge \ast \a \rangle_\gg &=& \frac14 \sum \langle \a_{ij},\a_{k\ell}\rangle_\gg e^{ij} \wedge \ast  e^{k\ell} \\ &=&  
\frac14 \sum \langle \a_{ij},\a_{k\ell}\rangle_\gg \langle e^{ij} , e^{k\ell}\rangle_\Lambda dvol_g 
\\ &=&  \langle \a , \a \rangle dvol_g.
\end{eqnarray*}
\epf 

\bp Let $(e_i)$ be any local frame in $TM$ and $(e^i)$ the dual frame in $T^*M$. Then  
the Yang-Mills Lagrangian is locally given by 
\[ -\frac12 \langle F,F\rangle = \frac14 \sum \mathrm{tr}\, (F_{ij }F^{ij }),\]
where $F_{ij} = F(e_i ,e_j)$, $F^{ij }= \sum g^{ii'}g^{jj'}F_{i' j'}$, 
$g^{ij}=g^{-1}(e^i,e^j)$. 
\ep 
\pf Choosing an orthonormal frame $(e_i)$, the calculation in the proof of the previous lemma shows that 
\begin{eqnarray*}  -\frac12 \langle F,F\rangle &=& -\frac18  \sum \langle F_{ij},F_{k\ell}\rangle_\gg \langle e^{ij} , e^{k\ell}\rangle_\Lambda\\
 &=& -\frac14 
\sum \langle F_{ij},F_{ij}\rangle_\gg \langle e^{ij} , e^{ij}\rangle_\Lambda = -\frac14 
\sum \langle F_{ij},F^{ij}\rangle_\gg = \frac14 \sum \mathrm{tr}\, (F_{ij}F^{ij}).
\end{eqnarray*} 
It is easy to check that the 
expression $\sum F_{ij}F^{ij}$ is independent of the choice of frame.  
This proves that the formula $-\frac12 \langle F,F\rangle= \frac14 \sum \mathrm{tr}\, (F_{ij}F^{ij})$ holds in every local frame. 
\epf 

As a corollary we obtain the usual expression for the Yang-Mills Lagrangian\index{Yang-Mills!lagrangian} in the physics literature. 
\bc In local coordinates $(x^\mu)$, $\mu =1,\ldots ,n$, the Yang-Mills Lagrangian on $(M,g)$ is given by 
\[ -\frac12 \langle F,F\rangle = \frac14 \sum \mathrm{tr}\, (F_{\mu \nu }F^{\mu \nu }),\]
where $F_{\mu \nu} = F(\p_\mu ,\p_\nu )$, $F^{\mu \nu }= \sum g^{\mu \mu'}g^{\nu \nu'}F_{\mu' \nu'}$, 
$g^{\mu \nu}=g^{-1}(dx^\mu,dx^\nu)$. 
\ec 
\bl \label{curvLemma} Let $\n$ be a $G$-connection in $E$ and 
\[ \a\in \O^1(\gg (E)) = \G (T^*M\ot \gg (E))\] 
a one-form with values in $\gg (E)$. Then 
\[ F^{\n + \a} = F^\n + d^\n \a + \a \wedge \a ,\]
where $(\a \wedge \a) (X,Y) := \a (X)\a (Y) -\a (Y) \a (X) = [\a (X), \a (Y)]$ for all $X, Y\in \mathfrak{X}(M)$.  
\el 
\pf See \appref{exercises}, \excref{45}.
\epf 
\bc Let $\n$ be a $G$-connection in $E$, $\a\in \O^1(\gg (E))$, and consider the curvature $F_\e$ of the one-parameter family 
of $G$-connections $\n^\e = \n +\e \a$. Then 
\[ \left. \frac{d}{d\e }\right|_{\e =0}  \langle F_\e \wedge \ast F_\e \rangle_\gg = 2\langle d^\n \a \wedge \ast F\rangle_\gg ,\] 
where $F=F_0$.
\ec 
\pf Since $F_\e= F + \e d^\n \a +  \e^2 \a \wedge \a$, we have 
\[ \frac12 \left. \frac{d}{d\e }\right|_{\e =0}  \langle F_\e \wedge \ast F_\e \rangle_\gg = \frac12 \left. \frac{d}{d\e }\right|_{\e=0} \langle F_\e , F_\e \rangle dvol_g=  \langle d^\n \a , F \rangle dvol_g.  \] 
\epf 
\pf (of Theorem \ref{YMThm}) Using the preceding notation we compute   
\begin{eqnarray*}   -\left. \frac{d}{d\e }\right|_{\e=0}  \mathcal{L} (j^1(\n^\e ))dvol_g &=&
\langle d^\n \a \wedge \ast F \rangle_\gg = -\tr (d^\n \a \wedge \ast F)\\ 
&=& -\tr \left( d^\n  (\a \wedge \ast F) +  \a \wedge d^\n \ast F \right)\\ 
&=&  
-\tr d^\n  (\a \wedge \ast F)  + \langle \a \wedge d^\n \ast F \rangle_\gg.\end{eqnarray*}
Since 
\[ \tr d^\n  (\a \wedge \ast F) = d  \tr (\a \wedge \ast F) ,\]
we conclude that for variations $\a$ with compact support we have 
\[ -\int_M \left. \frac{d}{d\e }\right|_{\e=0}  \mathcal{L} (j^1(\n^\e ))dvol_g = \int_M \langle \a \wedge d^\n \ast F \rangle_\gg dvol_g.\]
This proves that the equations of motion of pure Yang-Mills theory are indeed $d^\n \ast F =0$. 
\epf

\subsection*{Gauge transformations}
Given a vector bundle $E\ra M$,  
we denote by $\mathrm{Aut} (E)$ its group of automorphisms\index{Automorphism!Group of}. 
If the vector bundle is equipped with a $G$-reduction $\mathcal{F}_G(E)\subset \mathcal{F}(E)$, then 
the subgroup of $\mathrm{Aut} (E)$ consisting of those automorphisms which preserve the reduction
will be denoted by $\mathrm{Aut}_G(E)$. It coincides with the group of sections of 
the group bundle $G(E)\subset \GL{E}$.  
\bd  $\mathrm{Aut}_G(E)$\nomenclature[aAutrmGE]{$\mathrm{Aut}_G(E)$}{gauge group of the vector bundle $E$ with $G$-reduction $\mathcal{F}_G(E)$} is called the \emph{gauge group}\index{Gauge!group} of the vector bundle $E$ with $G$-reduction $\mathcal{F}_G(E)$. 
The elements of $\mathrm{Aut}_G(E)$ are called \emph{gauge transformations}\index{Gauge!transformations}. 
\ed 
Gauge transformations  play a similar role  in the study of vector bundles to diffeomorphisms 
in the study of manifolds. 
The group $\mathrm{Aut} (E)$ acts $C^\infty (M)$-linearly on the space of smooth sections $\G (E)$ by  
\[  \mathrm{Aut} (E)\times \G (E) \ra \G (E),\quad  (\varphi , s ) \ra \varphi s.\]
\bp The action of $\mathrm{Aut}_G(E)$ on $\G (E)$ induces an affine action on
the space $\mathcal{A}_G(E)$ of $G$-connections. 
\ep 
 
 \pf The transformation of  $\mathcal{A}_G(E)$ induced by $\varphi \in \mathrm{Aut}_G(E)$ is 
 \be \label{nablaprimeEq}  \n \mapsto \n' = \n^\varphi := \varphi \circ \n \circ \varphi^{-1}.\ee
 To see that it is affine, we apply $\n'$ to a section $s\in \G (E)$:
 \[ \n' s = \varphi (\n \varphi^{-1}s) = \varphi \n (\varphi^{-1}) s + \n s. \]
 So 
 \be \label{shiftEqu} \n' = \n + \varphi \n (\varphi^{-1}) \ee 
 is related to $\n$ by a translation. 
 \epf 
 \bp Let $\n$ be a connection in a vector bundle $E\ra M$ and $\varphi \in \mathrm{Aut} (E)$ an automorphism. 
 Then the  curvature $F'$ of the gauge transformed connection 
 $\n' = \n^\varphi$ is related to the curvature $F$ of $\n$ by the natural $C^\infty (M)$-linear action of  $\mathrm{Aut} (E)$
 on $\O^2(\mathrm{End}(E))$: 
 \[ F' = \varphi \circ F \circ \varphi^{-1}.\]
 \ep 
 \pf This is a straightforward consequence of \re{nablaprimeEq} and the definition of the curvature. 
 \epf 
 \bc \label{YMgaugeinvCor}
 The Yang-Mills Lagrangian \re{YMLagrEq} is invariant under $\mathrm{Aut} (E)$. 
 \ec 
 \pf This follows from the fact that the trace form $(A,B) \mapsto \tr (AB)$ on $\ggl (E)$, which was used
 in the definition of $\langle F ,F\rangle$,  is invariant under conjugation. 
 \epf 
 
This clarifies the role of gauge transformations in Yang-Mills theory. For that reason, Yang-Mills theory is also known as \emph{gauge theory}\index{Gauge!theory}. In particular in the physics literature, often a further distinction is made between Abelian and non-Abelian gauge theories depending on the property of the respective gauge group. In this context, only the latter is customarily referred to as Yang-Mills theory. This additional distinction is used not only for historical reasons but also because the corresponding physical theories differ significantly in both cases~\cite{J,Ko,M}.
 
 \begin{remark} 
Let us denote by $\mathcal{A}_G^{YM}(E)\subset \mathcal{A}_G(E)$\nomenclature[aAcalGYME]{$\mathcal{A}_G^{YM}(E)$}{subset of Yang-Mills $G$-connections} the subset consisting of \emph{Yang-Mills connections}\index{Yang-Mills!connection}, that is 
solutions to the Yang-Mills
equations. It follows from Corollary \ref{YMgaugeinvCor} that the gauge group $\mathrm{Aut}_G (E)$ acts on
$\mathcal{A}_G^{YM}(E)$. The quotient 
\be \label{moduliYM} \mathcal{A}_G^{YM}(E)/\mathrm{Aut}_G (E)\ee
is the set of equivalence classes of Yang-Mills connections. 
Notice that the space of connections as well as the group of
gauge transformations are of infinite dimension. Nevertheless 
it is possible to prove that under suitable extra assumptions 
(on the signature of the metric $g$, the connections, and gauge transformations), 
quotients similar to \re{moduliYM} are sometimes finite-dimensional smooth manifolds
encoding differential topological information about the smooth manifold $M$.  
These problems are studied in mathematical work on gauge theory.  A particularly 
successful instance is \emph{\mbox{Donaldson} theory}\index{Donaldson theory} (see, for example, ref.~\cite{DK}), which is concerned with
four-dimensional smooth manifolds $M$, and restricts attention to the class
of self-dual or anti-self-dual connections (with respect to a Riemannian metric $g$), that is connections satisfying
the \emph{self-duality equation}\index{Self-duality equation|see{Donaldson theory}} 
\[ \ast F = F\]
or the \emph{anti-self-duality equation}\index{Anti-self-duality equation|see{Donaldson theory}}  
\[ \ast F = -F.\]
In virtue of the Bianchi identity, each of these equations implies the Yang-Mills equation: 
\[ d^\n \ast F = \pm d^\n F =0.\]
\end{remark}

\subsection*{The connection one-form} 
Let $\n$ be a connection in a vector bundle $E\ra M$ of rank $k$. 
Given a  local frame $f=(f_1,\ldots ,f_k)$ of $E$, defined 
on some open subset $U\subset M$, there exists a 
system of one-forms $A^i_j$, $i,j\in \{ 1,\ldots ,k\}$, such that 
\be \label{conn-oneformEqu} \n_X f_i = \sum A^j_i(X)f_j\ee
for all vector fields $X$ on $U\subset M$. The matrix-valued 
one-form 
\nomenclature[aA]{$A$}{connection one-form}
\[ 
 A = A^{\n ,f} = (A^i_j)_{i,j\in \{1,\ldots ,k\}} 
\]
is called the \emph{connection one-form}\index{Connection one-form} 
of $\n$ with respect to the local frame $f$. The equation \re{conn-oneformEqu}
can be written in matrix-notation as 
\[ \n f = fA.\]
Solving it with respect to $A$ we obtain 
\[ A = f^{-1}\n f,\]
where $f^{-1}=(f_1^*,\ldots ,f_k^*)^\intercal$  is the column vector which is  the transposed of the 
dual frame $f^*=(f_1^*,\ldots ,f_n^*)$. The notation is consistent with the 
interpretation of a frame $b=(b_1,\ldots ,b_k)$ of $E$ at $x\in M$ (such as $b=f_x$) as an isomorphism 
\[ b : \bK^k \ra E_x,\quad v \ra bv=\sum v^ib_i,\]
where $bv$ is the product of the $E_x$-valued row vector $b=(b_1,\ldots ,b_k)$ with the column vector  
$v=(v^1,\ldots ,v^k)^\intercal$. The inverse of this map is given by 
\[ E_x \ra \bK^k,\quad w \mapsto (b_1^*(w),\ldots ,b_k^*(w))^\intercal.\]

\bp Let $\n$ be a connection and $f$ a local frame in a vector bundle $E$ over $\bK\in \{ \bR , \bC\}$. The connection form $A'  := A^{\n' , f}$ of a gauge transformed connection 
$\n' = \n^\varphi$, $\varphi \in \mathrm{Aut}(E)$, is related to the connection form $A= A^{\n , f}$ of $\n$ by
\be \label{gaugetrafoEqu}  A' = \psi A \psi^{-1} +\psi d(\psi^{-1}),\ee
where $\psi = f^{-1}\circ \varphi \circ f : U \ra \GL {k,\bK }$ is the matrix-valued function representing the automorphism $\varphi$ in the frame $f$. 
\ep 
\pf Using that $\varphi = f \circ \psi \circ f^{-1}$ we compute 
\begin{eqnarray*} \a  &:=& \n' -\n \stackrel{ \re{shiftEqu} }{=} \varphi \n (\varphi^{-1}) = f  \psi  f^{-1} \left[ (\n f )  \psi^{-1}  f^{-1} + f \n (\psi^{-1}  f^{-1})\right] \\
&=&
f \psi A \psi^{-1} f^{-1} + f\psi \n (\psi^{-1}  f^{-1}) = f \psi A \psi^{-1} f^{-1} + f\psi d(\psi^{-1}) f^{-1} + f\n (f^{-1}) \end{eqnarray*} 
and, hence, 
\[ A' -A = f^{-1} \a (f) = \psi A \psi^{-1} + \psi d(\psi^{-1}) + \underbrace{\n (f^{-1})f}_{=-f^{-1}\n f=-A}.\] 
\epf 
\bp With the notation of the previous proposition we have 
\[ A^{\n' ,f} = A^{\n , f'},\]
where $f' = \varphi^{-1}  \circ f= f\circ \psi^{-1}$.
\ep 

\pf 
From the formula $A^{\n , f}= f^{-1}\n f$ we can immediately deduce the general transformation rule 
\be A^{\varphi \circ \n \circ \varphi^{-1} , \varphi  \circ f } = A^{\n , f}.\ee 
Substituting $f'$ for $f$ in this formula, we obtain the claim.  
\epf 
The last proposition shows that a gauge transformation $\nabla \mapsto \nabla' = \n^\varphi= \varphi \circ \nabla \circ \varphi^{-1}$ by $\varphi$ 
has the same effect on the connection form as a change of frame $f\mapsto f' = \varphi^{-1} \circ f = f \circ \psi^{-1}$ by the 
inverse transformation $\varphi^{-1}$, or equivalently, by the corresponding change of basis $\psi^{-1}$ in $\bK^k$. 

Substituting a one-parameter group $\psi_s : U \ra \GL {k,\bK}$ into \re{gaugetrafoEqu}  and differentiating with respect to $s$,
we obtain the local expression for the action  on a connection $\n$ of an infinitesimal gauge transformation\index{Gauge!transformations!infinitesimal} represented by $\tau  : U \ra \ggl (k,\bK )$:
\[ A \mapsto [\tau ,A] -d\tau .\]
To apply Noether's theorem (Theorem~\ref{NThm}) and to compute the Noether current \re{NcurrentEq} corresponding to an infinitesimal gauge transformation, one can use the local trivialization $f$ of the bundle $E$ and a system of coordinates 
$(x^\mu)$, $\mu = 1,\ldots ,n$,  on $U\subset M$ to describe a connection $\n$ by 
its connection form $A= \sum A_\mu dx^\mu = \sum A^i_{j\mu}dx^\mu\ot e_j^*\ot e_i$, which we consider as a map 
\[ U \ra U \times ((\bR^n)^* \ot \ggl (k, \bK)),\quad x\mapsto (x,(A^i_{j\mu}(x))).\] 
Here $(e_1,\ldots , e_k)$ denotes the standard basis of $\bK^k$. 
Now one can compute the corresponding Noether current by the formula 
\[ J^\nu = \sum_a  Y^a(A)\frac{\p \mathcal{L}}{\p u^a_\nu }(j^1A),\]
where $a$ is an index numerating the elements of the basis $(dx^\mu\ot e_j^*\ot e_i)$ of 
$V:=(\bR^n)^* \ot \ggl (k,\bK)$ such that 
\[ (u^a (j^1A)) = (u^i_{j\mu}(j^1A))=(A^i_{j\mu})\quad\mbox{and}\quad (u^a_\nu (j^1A)) =   (u^i_{j\mu, \nu}(j^1A)) = (\p_\nu A^i_{j\mu}).\]  The 
components $(Y^a) = (Y^i_{j\mu })$ of the infinitesimal automorphism 
\[ \mathfrak{X}(U \times V)\ni Y = \sum Y^a\frac{\p}{\p u^a}: U\times V \ra V,\quad \]
\[  (x,A)\mapsto Y(x,A)= [\tau (x) , A] -d\tau (x)\] 
are given by
\[ Y^i_{j\mu }(x,A) = \sum \tau^i_{j'}(x)A^{j'}_{j \mu} - A^i_{j' \mu}\tau^{j'}_j(x)-\p_\mu \tau^i_j(x),\quad A = (A^i_{j\mu}).\]
In order to obtain an explicit expression for $J^\nu$ it suffices to calculate
the partial derivatives 
\[ \frac{\p \mathcal{L}}{\p u^a_\nu }(j^1A) = \frac{\p \mathcal{L}}{\p u^i_{j\mu,\nu }}(j^1A) =  (F^{\mu \nu})^j_i,\]
where  $((F^{\mu \nu})^i_j)_{i,j\in \{1,\ldots ,k\}}$ is the matrix representing the endomorphism $F^{\mu \nu}$ with respect to the local frame $f$. 
So we obtain:
\bp\proplabel{noether_curr_infinit_gauge_trafo} The Noether current associated with an infinitesimal gauge transformation\index{Gauge!transformations!infinitesimal} represented by $\tau  : U \ra \ggl (k,\bK )$ is given by
\[ J^\mu = \tr \, \sum F^{\mu \nu}([\tau , A_\nu] -\p_\nu \tau )  .\]
\ep

\subsection{The Einstein-Hilbert Lagrangian}\seclabel{EHlagr}
In theories of gravity the space-time metric is no longer fixed 
but is a \emph{dynamical} field of the theory, that is subject to equations of motion. 
The most important Lagrangian for a pseudo-Riemannian metric $g$ on a smooth manifold $M$ 
is the \emph{Einstein-Hilbert Lagrangian}\index{Einstein-Hilbert Lagrangian} $n$-form
\be\eqlabel{EH_lagr}
 \mathcal{L}(j^2g) = \frac{1}{16\pi \k}scal\,  dvol_g \; ,
\ee
where $scal$\nomenclature[ascal]{$scal$}{scalar curvature of $(M,g)$} is the scalar curvature\index{Scalar curvature} of $(M,g)$ and $\k$\nomenclature[gkappa]{$\k$}{gravitational coupling constant} is the gravitational coupling constant\index{Gravitational!coupling constant}. 
It describes pure gravity, that is, gravity in the absence of matter fields. More complicated theories of gravity can be obtained by
including matter fields described, for instance, by adding to the Einstein-Hilbert term a sigma-model Lagrangian (for a map from $M$ to some target manifold) 
or a Yang-Mills Lagrangian (for a connection in some vector bundle $E\ra M$).   Notice that the fields of the resulting theory are coupled through the common dependence 
on the space-time metric, in the sense that the resulting equations of motion generally form a coupled system of partial differential equations.
\bt \label{EHEoms} The Euler-Lagrange equations\index{Euler-Lagrange!equations} for the Einstein-Hilbert Lagrangian are equivalent to the 
\emph{Einstein vacuum equation}\index{Einstein!vacuum equation}:\nomenclature[aRic]{$Ric$}{Ricci curvature tensor}
\be \label{EinsteinEq} Ric -\frac12 scal\cdot g=0,\ee
which is equivalent to $Ric=0$ if $n=\dim M \ge 3$ and is satisfied for every pseudo-Riemannian metric $g$ if $n\le 2$. 
\et 

\pf Let $g_\e$ be a smooth family of pseudo-Riemannian metrics on $M$ such that  
the symmetric tensor field 
\be \label{hEq} h :=\left. \frac{\p }{\p \e}\right|_{\e=0} g_\e\ee
has compact support. In order to compute the derivative of $\mathcal{L}(j^2g_\e)$ with respect to $\e$,  we 
first observe that 
\be \label{detEqu}  \left. \frac{\p }{\p \e}\right|_{\e=0} dvol_{g_\e} =\frac12 \tr (g^{-1}h)dvol_g,\ee
where $g=g_0$.  The derivative of the curvature tensor $R_\e$ of the Levi-Civita 
connection $\n^\e$ of the metric $g_\e$ is expressed in terms of 
 the tensor field 
\[  \a := \left. \frac{\p }{\p \e}\right|_{\e=0}  \n^\e =  \left. \frac{\p }{\p \e}\right|_{\e=0} (\n^\e -\n )\in \O^1 (\End TM) \]
by  
\[  \left. \frac{\p }{\p \e}\right|_{\e=0}R_\e = d^\n \a,\quad \n = \n^0,\]
as follows easily from Lemma~\ref{curvLemma}. As a consequence, the 
Ricci curvature $Ric_\e$ of $g_\e$ can be computed as follows:
\[  \left. \frac{\p }{\p \e}\right|_{\e=0} Ric_{\e}(X,Y) =  \left. \frac{\p }{\p \e}\right|_{\e=0} \tr (Z\mapsto R_\e (Z,X)Y) =  \tr (Z\mapsto d^\n \a (Z,X)Y),\]
where $X,Y,Z\in \mathfrak{X}(M)$. Omitting $Z$ we can write this in terms of a local frame $(e_i)$ of $TM$ as 
\be \label{RicEq} \left. \frac{\p }{\p \e}\right|_{\e=0} Ric_{\e}(X,Y) = \sum e^i\left[ (\n_{e_i}\a )_XY-  (\n_X \a)_{e_i}Y)\right], \ee
where  $\a_X = \a(X)$ denotes the endomorphism field obtained by evaluating the one-form $\a$ on the vector field $X$. 
We have used that $d^\n \a (X,Y)Z = (\n_X\a)_YZ-(\n_Y\a)_XZ$ for all $X,Y,Z\in \mathfrak{X}(M)$. 
For the scalar curvature $scal_\e$ of $g_\e$ we obtain
\begin{multline}\label{scalEqu}  \left. \frac{\p }{\p \e}\right|_{\e=0} scal_{\e}  = \left. \frac{\p }{\p \e}\right|_{\e=0}( \tr \, g_\e^{-1}Ric_{\e} )=
 -\tr \, (g^{-1} h g^{-1} Ric) \\ 
 + \underbrace{ \tr\, \left(g^{-1} \left. \frac{\p }{\p \e}\right|_{\e=0} Ric_{\e}\right) }_{=:f},
 \end{multline}
where $Ric$ denotes the Ricci curvature\index{Ricci curvature} of $g$. If we can show that $f$ is a total divergence,  then \re{detEqu} and \re{scalEqu}   imply that 
\[ \int_M   \left. \frac{\p }{\p \e}\right|_{\e=0} scal_\e \, dvol_{g_\e} =  -\int_M \tr \, \left(g^{-1}h\left(g^{-1}Ric -\frac{scal}{2} \mathrm{Id}\right)\right)dvol_g. \] 
From this formula we see that the equations of motion are equivalent to \re{EinsteinEq}. Taking the trace of $g^{-1}Ric -\frac{scal}{2} \mathrm{Id}$, 
we see that every solution has $scal=0$ if $n\neq 2$, which implies $Ric=0$ by \re{EinsteinEq}. To compute $f$ at some point 
$p\in M$ we can assume that $\n_{e_i}e_j=0$ at $p$. Then from \re{RicEq} we compute at $p$: 
\begin{eqnarray*} f &=& \sum g^{jk}e^i\left[ (\n_{e_i}\a )_{e_j}e_k-  (\n_{e_j} \a)_{e_i}e_k)\right]\\
&=&  \sum e^i \n_{e_i}(g^{jk}\a _{e_j}e_k)-  \sum g^{jk}e_k\n_{e_j} (e^ i\circ \a_{e_i}) , 
\end{eqnarray*}
where $g^{jk} = g^{-1}(e^j,e^k)$. The first sum is the divergence of the vector field $\tr_g \a$ obtained by contraction of the 
first two factors of $T^*M\ot T^*M\ot TM $ with the help of the metric. The other sum is the divergence of the vector field 
$v = g^{-1} \sum e^i \circ \a_{e_i}$ 
obtained by contraction of the first and last factors by duality and identification of the remaining factor $T^*M$ with $TM$ using the metric: 
\[ \sum g^{jk}e_k\n_{e_j} (e^ i\circ \a_{e_i})= \sum g^{jk}e_k\n_{e_j} gv=\sum g^{jk}g(\n_{e_j} v,e_k) = \sum e^j \n_{e_j} v = \mathrm{div}\, v.\]
So we have proven that $f$ is the divergence of the vector field $\tr_g \a-v$, which does not depend on the particular frame. 
\epf 
Let $(M,g,or)$ be an oriented pseudo-Riemannian manifold and denote by $dvol_{g,or}$ its metric volume form\index{Volume!form} with 
respect to the orientation $or$. 
In local coordinates $(x^1,\ldots ,x^n)$ on some open set $U\subset M$, we have 
\be\eqlabel{dvolg}
 dvol_{g,or}|_U = \varepsilon \sqrt{|\det (g_{ij})|}dx^1\wedge \cdots \wedge dx^n,\quad \varepsilon \in \{ 1, -1\} \; ,
\ee
if $dx^1\wedge \cdots \wedge dx^n\in \varepsilon \cdot or$, where $g_{ij}=g(\p_i,\p_j)$.  

\bp For all  $\varphi\in \mathrm{Diff} (M)$, we have 
\[ scal_{\varphi^* g}dvol_{\varphi^*g,\varphi^*or} =  \varphi^*(scal_g dvol_{g,or}) .\] 
In particular, the Einstein-Hilbert Lagrangian $n$-form $scal_gdvol_g$, where $dvol_g= dvol_{g,or}$, is invariant under all orientation 
preserving diffeomorphisms of $M$: 
\[ scal_{\varphi^* g}dvol_{\varphi^*g} =  \varphi^*(scal_g dvol_{g}) .\] 
\ep 
\pf  
 For all diffeomorphisms $\varphi$ of $M$ we have 
\[ dvol_{\varphi^*g,\varphi^*or} = \varphi^*dvol_{g,or}\quad \mbox{and}\quad scal_{\varphi^* g} = \varphi^*scal_g.\]
\epf
The Einstein-Hilbert Lagrangian can be generalized by considering instead 
\be \label{cosmoEq}\mathcal{L} (j^2g) = \frac{1}{16\pi \k}(scal\, +2 \Lambda ) dvol_g, \ee  
where $\Lambda$\nomenclature[gLambda]{$\Lambda$}{cosmological constant} is a constant known as the \emph{cosmological constant}\index{Cosmological constant}. 
The proof of Theorem~\ref{EHEoms} does also show the following.  
\bt The equations of motion of the Einstein-Hilbert Lagrangian with cosmological constant \re{cosmoEq} are equivalent to 
\be \label{EinsteinEqcosmo} Ric -\frac12 scal\cdot g=\Lambda g.\ee
This equation is equivalent to the system  
\[ \Lambda = \frac{2-n}{2n}scal\quad \mbox{and}\quad Ric = \frac{scal}{n} g.\]   
\et 
Notice that if $n\le 2$, then necessarily $\Lambda=0$ and every metric is again a solution. 
If $n\ge 3$, then one can show (exercise) that if $g$ is an \emph{Einstein metric}\index{Einstein!metric}, that is 
\[ Ric = f g\] 
for some function $f$, then 
$f=  \frac{scal}{n}= const$.  Thus $g$ solves \re{EinsteinEqcosmo} for $\Lambda =  \frac{2-n}{2n}scal$. In other words, 
the solutions of \re{EinsteinEqcosmo} for $n\ge 3$ are precisely the Einstein metrics such that 
$scal = \frac{2n}{2-n}\Lambda$. Observe that given an Einstein metric, the equation  $scal = \frac{2n}{2-n}\Lambda$ can be always 
solved by rescaling the Einstein metric by a positive constant provided that $scal$ has the same sign
as $\Lambda$. So it is sufficient to distinguish only 3 cases: $\Lambda =0$, $\Lambda >0$ and $\Lambda <0$.

To end this section we explain now how to obtain the Einstein metrics as solutions of a variational problem 
without introducing a cosmological constant. For that we consider the Einstein-Hilbert Lagrangian but
restrict to variations $g_\e$ with compact support  (that is \re{hEq} has compact support) that are volume preserving, that is 
$\left. \frac{\p }{\p \e}\right|_{\e=0} dvol_{g_\e} =0$. 
\bt The equations of motion for the Einstein-Hilbert Lagrangian under volume preserving variations
with compact support  are equivalent to 
\be \label{tracefreeEq} Ric^0 =0,\ee
where $Ric^0= Ric -\frac{scal}{n}g$ denotes the trace-free part of $Ric$. 
A pseudo-Riemannian metric $g$ is a solution of \re{tracefreeEq} if and only if $Ric = fg$ for some 
function $f$. 
\et  
\pf
Owing to \re{detEqu}, we obtain \re{tracefreeEq} by projecting \re{EinsteinEqcosmo} to its trace-free part.
\epf

\section{The energy-momentum tensor}
In the previous subsection, we discussed pure gravity, that is, gravity in the absence of matter fields. The key ingredient for coupling matter fields to Einstein gravity is the energy-momentum tensor~\cite{LL,MTW,W}. It can be introduced by reconsidering two important conservation laws of classical mechanics in the context of classical field theory.
Recall that the reason for conservation of energy\index{Conservation!of energy} 
in classical mechanics is the invariance of the Lagrangian under 
time-translations\index{Time!translations}.  Similarly, invariance under spatial translations\index{Spatial translations}
implies conservation of momentum\index{Conservation!of momentum}.

As a concrete simple example to illustrate at least some of the features of matter-coupled Einstein gravity, let us start by considering a first order field theory  for maps
$f : \mathfrak{S} \ra \mathcal{T}$ from some source manifold $\mathfrak{S}$  to some
target manifold $\mathcal{T}$.  Since our considerations will be local, we may restrict, by a choice of local coordinates, to 
$\mathfrak{S} = \bR^n$, $\mathcal{T}=\bR^m$, and we take the standard volume form on $\bR^n$. 
Then we can consider translations $X= \sum \vc^\mu \p_\mu$ in the source manifold $\mathfrak{S} = \bR^n$, where
$\vc^\mu$ are constants. \pagebreak[3] The vector field $X$ is an infinitesimal automorphism of a Lagrangian $\mathcal{L} \in C^\infty (\mathrm{Jet}^1(\bR^n,\bR^m))$
if and only if 
\[ \mathcal{L}(j^1_xf) = \mathcal{L}(x,f(x), \p f(x)),\quad \p f (x) := (\p_\mu f (x)),\] 
does not explicitly depend on $x=(x^\mu)\in \bR^n$. This follows from the fact that 
the prolongation of $X= \sum \vc^\mu \p_\mu\in \mathfrak{X}(\bR^n)$ is given by 
$\mathrm{pr}^{(1)}(X) = X\in \mathfrak{X}(\mathrm{Jet}^1(\bR^n,\bR^m))$. This can be seen either from 
Lemma \ref{prolongL} or by observing that the action of a translation by a vector $\vc = (\vc^\mu)\in \bR^n$ 
as a diffeomorphism on $\mathrm{Jet}^1(\bR^n,\bR^m)$ is simply
\[ (x,f(x), \p f (x)) \mapsto (x+\vc,f(x),\p f(x)).\]
The Noether current \re{NcurrentEq} corresponding to the infinitesimal translation $-\vc$ is given by
\[ J^\mu = \sum \p_\nu f^a \vc^\nu \frac{\p \mathcal{L}}{\p u^a_\mu}(j^1f) -\mathcal{L}(j^1f)\vc^\mu.\]
We can write it as $J^\mu = \sum \vc^\nu T^\mu_\nu$, where\nomenclature[aTmunu]{$T^\mu_\nu$}{component of the energy-momentum tensor}
\be\label{EMT} T^\mu_\nu := \p_\nu f^a \frac{\p \mathcal{L}}{\p u^a_\mu}(j^1f) - \mathcal{L}(j^1f)\d^\mu_\nu\ee
are the components of the \emph{energy-momentum tensor}\index{Energy-momentum tensor} or \emph{stress-energy tensor}\index{Stress-energy tensor|see{Energy-momentum tensor}}. (Caveat: We will redefine the energy-momentum tensor below.)
As a consequence of Noether's theorem\index{Noether!theorem} (Theorem~\ref{NThm}) we have the following result. 
\bp\label{propTdivfree} If $\mathcal{L} \in C^\infty (\mathrm{Jet}^1(\bR^n,\bR^m))$ is invariant under translations in $\bR^n$,
then the energy-momentum tensor is divergence-free (when evaluated on solutions of the equations of motions), that is
\be\eqlabel{Tdivfree} \sum  \frac{\p}{\p x^\mu}T^\mu_\nu =0.\ee
\ep 

\begin{Ex} Consider the linear sigma-model\index{Sigma-model!Linear} 
\[ \mathcal{L}(j^1f) = \frac12 \sum g_{ab}(f) \p_\mu f^a\p^\mu f^b -V(f),\quad f : \bR^n\ra \bR^m,\] 
with potential $V\in C^\infty (\bR^m)$, where greek indices are raised with the inverse of the constant metric $h= \sum h_{\mu \nu}dx^\mu dx^\nu$, for example $\p^\mu = \sum h^{\mu\nu} \p_\nu$ and $(h^{\mu\nu})$ being the matrix inverse to $(h_{\mu\nu})$. 
The corresponding energy-momentum tensor, written as  $(0,2)$-tensor field, is symmetric and given by 
\[  T_{\mu \nu} = \sum g_{ab}(f)\p_\mu f^a\p_\nu f^b - \mathcal{L}(j^1f) h_{\mu \nu}.\]
In index-free notation the right-hand side reads
\[ f^*g - \mathcal{L}h.\]
When $h=dt^2-\sum_{\a =1}^{n-1} (dx^\a)^2$ is the Minkowski metric\index{Minkowski!metric}, then  
the \emph{energy density}\index{Energy!density} (i.e.\ the charge density associated with the infinitesimal automorphism\index{Automorphism!Infinitesimal} $-\p_t$)
\[ T^{00}=T^0_0= \frac12 g(\p_tf, \p_t f) + \underbrace{\frac12 \sum_{\a =1}^{n-1}g(\p_\a f ,\p^\a f) +V(f)}_{=:\,\tilde{V}(f,\p_1f,\ldots, \p_{n-1}f)},\] 
resembles the energy in mechanics (and coincides with it if $n=1$), if we consider $\tilde{V}$ as potential energy. 
The \emph{non-gravitational energy}\index{Energy!Non-gravitational} is in this situation defined as the spatial integral of the energy density: 
\be\eqlabel{EMEnergy} E(t) = \int_{\bR^{n-1}} T^{00}(t,x^1,\ldots,x^{n-1}) d^{n-1}x.\ee
It is constant under appropriate boundary conditions at spatial infinity (as in Theorem~\ref{conservedThm}). 
Similarly, one defines the \emph{momentum density}\index{Momentum!density} as the time-dependent spatial vector field 
\[ \sum_{\a =1}^{n-1} T^{\a 0} \p_\a\] 
on $\bR^{n-1}$ and the \emph{momentum} as the  spatial integral 
\be\eqlabel{EMMomentum} \vec{P}(t) = \sum_{\a=1}^{n-1}P^\a(t) \p_\a,\quad P^\a(t) = \int_{\bR^{n-1}} T^{\a 0}(t,x^1,\ldots,x^{n-1}) d^{n-1}x.\ee
Notice that the momentum density  $(T^{\a 0})_{\a =1,\ldots ,n-1}$ is the flux density associated with the infinitesimal automorphism $-\p_t$.
Its components coincide up to sign with the charge densities $T^0_\a = -T^{\a 0}$ associated with the spatial translations\index{Spatial translations} $-\p_\a$, $\a = 1,\ldots , n-1$. 
\end{Ex} 

It is important to remark that the physical notions of energy and momentum as defined above are only valid in the Newtonian limit of Einstein gravity, that is, for quasi-static matter systems in asymptotically flat space-times (see also the discussion in~\cite[Sect.\ 19.3]{MTW}). In general, the physical notions of energy and momentum become rather subtle concepts in Einstein's theory of general relativity. In fact, no general definitions of energy and momentum valid for arbitrary space-time metrics exist. Only for special classes of metrics there exist well-defined expressions such as, for example, ADM energy, Komar energy, Bondi energy and Hawking energy. For further details regarding this subtle aspect of Einstein gravity, the reader is referred to~\cite{LL,MTW,W}.

\begin{Ex} Consider the Yang-Mills Lagrangian\index{Yang-Mills!lagrangian} $\mathcal{L}(j^1A ) = \frac14 \sum \tr\, (F_{\mu \nu }F^{\mu \nu})$.
Differentiating this Lagrangian with respect to $\p_\mu A_\rho$ (more precisely, with respect to $u^i_{j\rho ,\mu}$ in the notation of Section 
\ref{pureYMsec}) we obtain the (base-point dependent) linear map
\[ \mathfrak{g} \to \mathbb{K}, \quad B \mapsto  \tr\, (B F^{\mu \rho }).\] 
Therefore  from \re{EMT} we obtain that the corresponding energy-momentum tensor\index{Energy-momentum tensor} is given by 
\[   T^\mu_\nu (\textnormal{old}) = \sum \tr\,  \left( (\p_\nu A_\rho )F^{\mu \rho}\right) - \mathcal{L}(j^1A )\d^\mu_\nu.\] 
(We are working in a global trivialization of the vector bundle, such that the endomorphisms $F_{\mu \nu}$ can be 
considered as matrices.) 
At least in the case of Abelian gauge groups (the reader may try to extend this to the case of non-Abelian gauge groups), 
this tensor differs from the tensor 
\be  \label{symEMT} T^\mu_\nu (\textnormal{new})=\sum \tr\,  (F_{\nu \rho} F^{\mu \rho}) - \mathcal{L}(j^1 A )\d^\mu_\nu,\ee 
which is symmetric with respect to the metric $h$, 
by a tensor with vanishing divergence. 
In fact, for Abelian gauge groups $F_{\mu \nu } = \p_\mu A_\nu -\p_\nu A_\mu$ and it suffices to subtract 
\[ \sum \tr\,  \left( (\p_\rho A_\nu )F^{\mu \rho}\right)  = \sum \p_\rho  \tr\, (A_\nu F^{\mu \rho})\]
from $T^\mu_\nu (\textnormal{old})$.
The last equation holds because of the Yang-Mills equation, which for Abelian
gauge groups reduces to $\sum \p_\rho F^{\mu \rho} =0$. 
The tensor \re{symEMT}, which differs from the original energy-momentum tensor by symmetrization and is still divergence-free, will from now on be called the \emph{energy-momentum tensor}\index{Energy-momentum tensor!Re-defined} and its components will be denoted by $T^\mu_\nu$, rather than $T^\mu_\nu (\textnormal{new})$. Notice that the newly defined energy-momentum tensor is not only coordinate independent (as the previously defined  energy-momentum tensor) but, contrary to the previously defined one, is always invariant under gauge transformations. 
\end{Ex}

In the presence of gravity, the space-time metric $g$ (formerly denoted $h$) is considered as a dynamical field of the theory. The definition of the action functional~\eqref*{class_ft_action} is then modified to read
\[
 S{[}f, g{]} = \int_M L(j^k(f), j^\ell(g)) \, dvol_g \; , 
\]
where $dvol_g = \sqrt{|\det (g)|}\, dx^1\wedge \cdots \wedge dx^n$ in local coordinates $(x^1,\ldots ,x^n)$ on some open set $U\subset M$ (cf.~\eqref*{dvolg}). For physical reasons, it is important to distinguish between the gravitational Lagrangian\footnote{We remark that in some texts the definition of the Lagrangian in gravity theories differs by a factor of $\sqrt{|\det (g)|}$, namely $\mathcal{L} = \sqrt{|\det (g)|}\, L$. This will be used below.} $L_{\textnormal{GR}}$ and the matter Lagrangian $L_{\textnormal{matter}}$. The former depends only on $g$ and at least in ordinary Einstein gravity (as opposed to modified gravity theories) it is taken to be the Einstein-Hilbert Lagrangian $L_{\textnormal{EH}} = \frac{1}{16\pi \k} scal$ (cf.~\eqref*{EH_lagr}). The matter Lagrangian not only depends on $g$, but also on other fields through their $k$-th order jets. In total, $L = L_{\textnormal{GR}} + L_{\textnormal{matter}}$.

One can check that in the above examples the (symmetric) energy-momentum tensor\index{Energy-momentum tensor!Symmetric|see{Re-defined}}
can be obtained from the formula
\be\eqlabel{EMtensordef}
 T_{\mu \nu} := \frac{2}{\sqrt{|\det (g)|}} \frac{\d \mathcal{L}_{\textnormal{matter}}}{\d g^{\mu \nu}} = 2 \frac{\d L_{\textnormal{matter}}}{\d g^{\mu \nu}} + g_{\mu\nu} L_{\textnormal{matter}} \; ,
\ee
where $\frac{\d \mathcal{L}}{\d g^{\mu \nu}} :=  E_{g^{\mu \nu}} (\mathcal{L})$, and 
\be\eqlabel{EMT3_ELop}
 E_{g^{\mu \nu}} = \frac{\p }{\p g^{\mu \nu}}-  \sum \p_\lambda  \frac{\p }{\p {[} \p_\lambda g^{\mu \nu}{]}} \pm \cdots 
\ee 
denotes the Euler-Lagrange operator associated with the field $g^{\mu \nu}$ as defined in~\eqref*{ELop}. Here we are denoting the coordinates on the jet space ($u^{\mu \nu}, u^{\mu \nu}_{\lambda}, \ldots$) by the same symbols as their evaluation ($g^{\mu \nu}$, $\p_\nu g^{\mu\nu}, \ldots$) on $j^k(g^{-1})$, 
as customary in the literature. 
\Eqref{EMtensordef} is also often used as the definition of the (matter) energy-momentum tensor in the 
physics literature. With this definition, the energy-momentum tensor is always symmetric.

\bp\label{PropEEq} Let $\mathcal{L}= \mathcal{L}_{\textnormal{EH}} + \mathcal{L}_{\textnormal{matter}}$ be a gravity theory described as the sum of the Einstein-Hilbert 
Lagrangian for a pseudo-Riemannian metric $g$ and a Lagrangian\index{Matter Lagrangian}\index{Lagrangian!Matter}  $\mathcal{L}_{\textnormal{matter}}$ which depends on $g$ and on other fields through their 
$k$-th order jets. Then the equations of motion corresponding to the variation of the metric take the form:
\be \label{EEq} 
\frac{1}{8 \pi \k} \left(Ric_g - \frac12 scal_g \, g\right) = - T,\ee 
where $T=\sum T_{\mu \nu} dx^\mu dx^\nu$ is the energy-momentum tensor of $\mathcal{L}_{\textnormal{matter}}$ defined in~\eqref*{EMtensordef}. 
\ep 
\pf The Euler-Lagrange operator associated with $g^{\mu \nu}$ is  $\frac12 T$ for the matter Lagrangian and 
$\frac{1}{16 \pi \k} \left(Ric_g - \frac12 scal_g \, g\right)$ for the 
Einstein-Hilbert term.  For the latter statement we are using  the calculation in the proof of Theorem \ref{EHEoms} and taking into account that 
the variation of the inverse metric is related to the variation of the metric, $h=  \left. \frac{\p }{\p \e}\right|_{\e=0} g_\e$, by 
\[ \left. \frac{\p }{\p \e}\right|_{\e=0} g^{-1}_\e = -g^{-1} h g^{-1}.\]
\epf 
Notice that the minus sign on the right-hand side of the general \emph{Einstein equation}~\re{EEq} is usually absorbed by a sign change in the 
definition of the energy-momentum tensor $T$.

The full set of equations of motions corresponding to $\mathcal{L}= \mathcal{L}_{\textnormal{EH}} + \mathcal{L}_{\textnormal{matter}}$ is the Einstein equation \re{EEq} together with the matter equations of motions coming from varying $\mathcal{L}$ with respect to the matter fields. Due to the coupling to gravity, the matter equations of motions in general also contain terms involving the metric, thereby turning the full set of equations of motions into a system of coupled equations.

\bc For every solution of the Einstein equation\index{Einstein!equation} \re{EEq}, the energy-momentum tensor\index{Energy-momentum tensor} \eqref*{EMtensordef} of the matter Lagrangian\index{Matter Lagrangian}\index{Lagrangian!Matter}  
$\mathcal{L}_{\textnormal{matter}}$ satisfies the covariant divergence-free condition (cf. Proposition~\ref{propTdivfree}), namely 
\be\eqlabel{Tdivfree_gen}
 \sum \nabla_\mu T^{\mu\nu} = 0 \; , 
\ee
where $T^{\mu\nu} = \sum g^{\mu\rho} g^{\nu\sigma} T_{\rho\sigma}$ and $\nabla_\mu$ are the components of the Levi-Civita connection of $g$. 
\ec 

\pf This follows from the equations of motion provided in Proposition~\ref{PropEEq} and the fact that the \emph{Einstein tensor}\index{Einstein!tensor}  $Ric_g - \frac12 scal_g \, g$ is di\-ver\-gence-free, see \cite[Ch.\ 12, Lemma 2]{O}.
\epf 

Unlike \eqref{Tdivfree}, \eqref{Tdivfree_gen} in general does not correspond to a conservation law in the sense of Definition~\ref{def_cons_law}. The analog of the quantities defined in \eqrangeref{EMEnergy}{EMMomentum}, that is
\[ \int_\Sigma \sum \sqrt{|\det (g)|} T^{\mu\nu} d\sigma_\nu \; , \]
where the integration is performed over a spacelike hypersurface $\Sigma$, is conserved only if $\sum \p_\mu (\sqrt{|\det (g)|} T^{\mu\nu}) = 0$ holds rather than \eqref{Tdivfree_gen} (see, for example,~\cite[Ch.\ 11, Section 101]{LL} for further details).
To find a conservation law nevertheless, one needs to also take into account the contribution of the gravitational field (and its derivatives) to the total energy-momentum. This can be achieved by generalizing the definition~\eqref*{EMtensordef} to 
\be\label{EMT3} 
 T^{\textnormal{eff}}_{\mu \nu} := \frac{2}{\sqrt{|\det (g)|}} \frac{\d \mathcal{L}}{\d g^{\mu \nu}} = \frac{2}{\sqrt{|\det (g)|}} E_{g^{\mu \nu}} (\mathcal{L}) \; .
\ee
An analysis along the lines of~\cite[Ch.\ 11, Section 101]{LL} shows the following analog of Proposition~\ref{propTdivfree} in the presence of gravity:
\be\eqlabel{Tdivfree_gen1}
 \sum \p_\mu T^{\textnormal{eff}} {}^{\mu\nu} = \sum \p_\mu (T^{\mu\nu} + t^{\mu\nu}) = 0 \; , 
\ee
where $T^{\mu\nu} = \sum g^{\mu\rho} g^{\nu\sigma} T_{\rho\sigma}$ is the same as in~\eqref*{EMtensordef} after raising the indices, and $t^{\mu\nu}$ is called the \emph{energy-momentum pseudo-tensor}\index{Energy-momentum pseudo-tensor} of the gravitational field. The quantity $t^{\mu\nu}$ is a coordinate-dependent object that cannot be interpreted as the components of a tensor field. Note that \eqref{Tdivfree_gen1} is equivalent to \eqref{Tdivfree_gen}, with the difference being that the latter is manifestly covariant. From the above, we see that the distinction between energy-momentum carried by the gravitational field versus energy-momentum carried by the matter fields becomes a subtle issue in general relativity. For more details, the reader is referred to~\cite{LL,MTW,W}.
\clearpage{}%
\clearpage{}%

\appendix
\addtocontents{toc}{\protect\setcounter{tocdepth}{-1}} %
\chapter{Exercises}
\applabel{exercises} %

\section{Exercises for \texorpdfstring{\Chref{lagmech}}{Chapter~\ref{ch:lagmech}}}

\begin{enumerate}
\item\exclabel{1} Suppose that for every local chart $\varphi : U \rightarrow  \mathbb{R}^n$ on some smooth manifold $M$ we are given
a system of $n$ functions $\alpha_i^\varphi \in C^{\infty}(U)$, $i=1,\ldots ,n$.  Let us denote by $\mathcal{V}^i_\varphi\in C^{\infty}(U)$ 
the components of a smooth vector field $\mathcal{V}$
on $M$ with respect to the chart $\varphi$. Suppose that  for every  $\mathcal{V}$ and for every pair of charts $\varphi : U \rightarrow  \mathbb{R}^n$, $\tilde{\varphi} : \tilde{U} \ra \bR$  the functions $\sum  \alpha_i^\varphi \mathcal{V}^i_\varphi$ and $\sum  \alpha_i^{\tilde{\varphi}} \mathcal{V}^i_{\tilde{\varphi}}$ coincide on $U\cap \tilde{U}$. 
Show that there exists a smooth function $f_{\mathcal{V}}$ on $M$ and a smooth one-form $\a$ on $M$ such that $f_{\mathcal{V}}|_U = \sum  \alpha_i^\varphi \mathcal{V}^i_\varphi$  and  $\a|_U = \sum \a_i^\varphi dx^i$ for every chart
$\varphi =(x^1,\ldots ,x^n): U \rightarrow  \mathbb{R}^n$. Check that $\a (\mathcal{V})=f_\mathcal{V}$. 
\item Determine the equations of motion of a free particle in Euclidean space (see \exref{NewtonEx}) and find the general solution.
\item Determine the equations of motion for the harmonic oscillator (see \exref{NewtonEx}) and find the general solution.
\item Consider the Riemannian manifold $\tilde{M} = (\bR^{n+1}\setminus \{ 0\}, g_{\textnormal{can}})$, where $g_{\textnormal{can}}$ is the restriction of the Euclidean metric $\langle \cdot ,\cdot \rangle$ on $\bR^{n+1}$. Denote by $\G$ the cyclic group generated by the homothety $x\mapsto 2x$.
\begin{enumerate}
\item Show that the quotient $M=\tilde{M}/\G$ is diffeomorphic to $S^1 \times S^n$ and that the Riemannian metric 
\[ \tilde{g} := \frac{1}{r^2}g_{\textnormal{can}},\quad r(x) := \sqrt{\langle x,x\rangle},\quad x\in \tilde{M}, \]
induces a Riemannian metric $g$ on $M$. 
\item Determine all periodic motions for the Lagrangian mechanical system $(M,\mathcal{L})$, where $\mathcal{L}(v) = \frac12 g(v,v)$, $v\in TM$. 
\item Show that $(M,g)$ is locally conformally flat, that is for every $p\in M$ there exists an open neighborhood
$U\subset M$ and a positive function $f\in C^\infty(U)$ such that the Riemannian metric $f\cdot g|_U$ is flat. 
\item Does there exist a global function $f\in C^\infty (M)$ such that $fg$ is of constant (sectional) curvature?\\
{\it (Hint: You may use the classification of complete simply connected Riemannian manifolds of constant sectional curvature.
These manifolds are precisely the Euclidean spaces, the spheres and the hyperbolic spaces.)} 
\end{enumerate}
\item Let  $V$ be a smooth function on a Riemannian manifold $(M,g)$. Consider the Lagrangian
\be L(v) = \frac12 g(v,v) -V(\pi v),\quad v\in TM, \ee
where $\pi : TM \ra M$ is the canonical projection.  
Let $X$ be a Killing vector field on $(M,g)$ such that $X(V)=0$. 
Check by direct calculation (without using Noether's theorem) that the function $v\mapsto g(v,X(\pi v))$ on $TM$ is an integral of motion
of the Lagrangian mechanical system $(M,L)$. 
\item\exclabel{6} Show that the automorphism group of the Lagrangian mechanical system $(M,L)$ of the previous exercise is 
given by 
\[ \mathrm{Aut}(M,L) = \{ \varphi \in \mathrm{Isom}(M)| V\circ \varphi =V\} .\]
Deduce that the infinitesimal automorphisms of $(M,L)$ are precisely the Killing vector fields $X$ such that $X(V)=0$.
\item\exclabel{7} Let $\g : I \ra \bR^3 \setminus \{ 0\}$ be a smooth curve. Show that the following conditions are equivalent:  
\begin{enumerate}
\item $\g \times \g'=0$, where $\times$ denotes the cross product,  
\item $\g$ is a radial curve, that is $\g (I)$ is contained in the ray $\bR^{>0} v_0$ generated by some constant vector $v_0\in \bR^3 \setminus \{ 0\}$. Here $\bR^{>0}$ denotes the set of positive real numbers.
 \end{enumerate}
\item\exclabel{8} Consider a particle $\g : I \ra \bR^3 \setminus \{ 0\}$ of mass $m=1$ moving in Euclidean space under the influence of Newton's gravitational potential
$V=-\frac{M}{r}$, where units have been chosen such that the gravitational constant  $\k=1$.  
\begin{enumerate}
\item 
Determine the radial motions of the system. 
\item  Deduce the total fall time as a function of the initial radius if the initial velocity is zero. 
\end{enumerate}
\begin{it} Hints: Rather than calculating $r$ as a function of time $t$ you will notice that in general it is easier to calculate 
$t$ as a function of $r$. The function $r$ of $t$ is then implicitly determined as the inverse function and will not be calculated explicitly. 
The calculation of
$r\mapsto t(r)$ reduces to finding the primitive of a function. 
For that it might be helpful to calculate the derivative of the function
$F(x)=\sqrt{x+x^2} - \mathrm{arsinh}(\sqrt{x})$ for $x>0$.
\end{it}
\item\exclabel{9} (Conservation of angular momentum). Consider a particle $t\mapsto \g(t)\in \bR^3$ moving in Euclidean space 
according to Newton's law
\[  \frac{d}{dt} \vec{p}= F.\] 
Its \emph{angular momentum} at time $t$ is the vector
\[ \vec{L}(t) := \g (t) \times \vec{p}(t).\]
Show that $\vec{L}$ is subject to the equation 
\[  \frac{d}{dt} \vec{L} = \vec{M},\]
where  $\vec{M}:=\g \times F$ is the \emph{moment of force}. 
Deduce that the angular momentum 
is constant if the moment of force is zero. 
\item 
Let $V$ be a smooth function on a pseudo-Riemannian manifold $(M,g)$ and consider the Lagrangian
$\mathcal{L}(v)= \frac12 g(v,v) - V(\pi v)$, $v\in TM$. Assume that 
with respect to some coordinate system on $M$ the metric $g$ and the potential $V$ are both invariant under rotations 
in one of the coordinate planes, say in the $(x^1,x^2)$-plane. 
Show that there exists a corresponding integral of motion defined on the coordinate domain. 
How is it related with the notion of angular momentum? 
\item\exclabel{11} Show that a function on $\bR^3\setminus \{ 0\}$ is radial if and only if it is 
\emph{spherically symmetric}, that is invariant under $\SO 3$. Deduce that invariance under $\SO3$ is equivalent to invariance 
under $\mathrm{O}(3)$. 
\item\exclabel{12} Show that a vector field on $M=\bR^3\setminus \{ 0\}$ is radial if and only if it is 
\emph{spherically} \emph{symmetric}, that is invariant under the natural action of $\SO 3 \subset \mathrm{Diff}(M)$.  Deduce that invariance under $\SO3$ is equivalent to invariance 
under $\mathrm{O}(3)$. \\
\begin{it} Hint: Recall that the natural action of the diffeomorphism group $\mathrm{Diff}(M)$ on the vector space of vector fields $F : M \ra TM$ on 
a smooth manifold $M$ is given by
\[ F\mapsto F^{\varphi} := d\varphi \circ F \circ \varphi^{-1},\quad  \varphi \in  \mathrm{Diff}(M).\]
\end{it}
\item\exclabel{13} Let $\g : I \ra \bR^3\setminus \{ 0\}$ be the motion of a particle in a radial force field $F$ according to Newton's law $m\g'' = F(\g )$. 
Recall that the motion is planar due to the conservation of the angular momentum vector. 
Show that the area $A(t_0,t_1)$ swept out by the vector $\g$ during a time interval $[t_0,t_1]$ is given by $A(t_0,t_1)=\frac{L}{2}(t_1-t_0)$. 
\item\exclabel{14} Let $(M,g)$ be a pseudo-Riemannian manifold and $f$ a nowhere vanishing smooth function on $M$. Consider the
pseudo-Riemannian metric 
\[ g_N = g + fdu^2\]
on $N:= M\times \bR$, where $u$ is the coordinate on $\bR$. 
Show that geodesic equations for a curve $t\mapsto \g_N (t)= (\g (t) ,u(t)) \in N=M\times \bR$ can be separated 
into
\begin{enumerate}
\item 
 the equations of motion for $\g$ with respect to a Lagrangian of the form $\mathcal{L}(v) = \frac12 g(v,v) -V(\pi v)$, $v\in TM$,
where $V$ is a certain smooth function on $M$  related to $f$ and $\pi :TM \ra M$ is the canonical projection and
\item an ordinary differential equation for the function $t\ra u(t)$, which can be solved by integration once we know $t\mapsto \g (t)$. 
\end{enumerate}
\item Determine the radius $r$ as a function of the angle $\varphi$ for a motion of a particle of unit mass with non-zero angular momentum in Coulomb's electrostatic potential. Can you use the results about the motion in Newton's gravitational potential? What is the main difference? 
\item Let $V$ be a smooth function on a pseudo-Riemannian manifold, which is either bounded from above or from below. 
Show\footnote{See~\cite{Eisenhart} for results generalizing this exercise and also \excref{14}.} that there exists a function $f$ related to $V$ and a pseudo-Riemannian metric $g_N=g+fdu^2$ on $N=M\times \bR$, where $u$ is the coordinate on the 
$\bR$-factor,  such that the solutions of the Lagrangian mechanical system defined by $\mathcal{L}(v) = \frac12 g(v,v) -V(\pi v)$, $v\in TM$, correspond 
precisely to geodesics $t\mapsto \g_N (t) = (\g (t), u(t))$ in $(N,g_N)$ with a particular choice of the affine parameter $t$ and which satisfy $u'\neq 0$.
\item Let $V$ be a smooth function, which is bounded from below, on a complete Riemannian manifold $(M,g)$. 
Show that the solutions of the Lagrangian mechanical system 
defined by $\mathcal{L}(v) = \frac12 g(v,v) -V(\pi v)$, $v\in TM$, exist for all times. \\
{\it Hint: You may use the previous exercise to relate the problem to the completeness of a Riemannian manifold $(N,g_N)$ of the type $N=M\times \bR$, $g_N=g+fdu^2$. 
Recall that a Riemannian metric on a smooth manifold $M$ is called complete if every Cauchy sequence in $(M,g)$ converges and that this notion is equivalent to 
geodesic completeness by the Hopf-Rinow theorem.} 
\item Deduce from the previous exercise that for every smooth function $V$ on a compact Riemannian manifold the solutions of the Lagrangian mechanical system 
defined by $\mathcal{L}(v) = \frac12 g(v,v) -V(\pi v)$, $v\in TM$, exist for all times.

\section{Exercises for \texorpdfstring{\Chref{Hamiltonian_systems}}{Chapter~\ref{ch:Hamiltonian_systems}}}

\item\exclabel{19} Let $f$ be a smooth function on a symplectic manifold $M$. Compute the Hamiltonian vector field $X_f$
in a coordinate system $(q^1,\ldots ,q^n,p_1,\ldots ,p_n)$ defined on some open subset $U\subset M$ such that $\o|_U = \sum dp_i \wedge dq^i$.
\item\exclabel{20} Let $(M,\mathcal{L})$ be a Lagrangian mechanical system and denote by $\pi :TM \ra M$ the canonical projection. Show that the following conditions are equivalent:
\begin{enumerate}
\item $\mathcal{L} $ is non-degenerate.
\item $\phi_\mathcal{L}: TM \ra T^*M$ is a local diffeomorphism. 
\item For all $x\in M$, $\phi_{\mathcal{L}}|_{T_xM} : T_xM \ra T^*_xM$ is of maximal rank.
\item For all $v\in TM$, there exists a coordinate system $(x^i)$ defined on an open neighborhood $U$ of $\pi v$ such that the matrix $\left( \frac{\p^2 \mathcal{L}(v)}{\p \hat{q}^i\p \hat{q}^j}\right)$ is invertible, where $(q^1,\ldots ,q^n,\hat{q}^1,\ldots ,\hat{q}^n)$ are the corresponding coordinates on $TU$. 
\item For all $v\in TM$, and every coordinate system $(x^i)$ defined on an open neighborhood $U$ of $\pi v$ the matrix $\left( \frac{\p^2 \mathcal{L}(v)}{\p \hat{q}^i\p \hat{q}^j}\right)$ is  invertible. 
\end{enumerate}
\item\exclabel{21} Let $(M,\mathcal{L})$ be a Lagrangian mechanical system. Show that the following conditions are equivalent:
\begin{enumerate}
\item $\mathcal{L} $ is nice.
\item $\phi_\mathcal{L}: TM \ra T^*M$ is a diffeomorphism. 
\item $\mathcal{L} $ is non-degenerate and for all $x\in M$, $\phi_{\mathcal{L}}|_{T_xM} : T_xM \ra T^*_xM$ is a bijection. 
\item For all $x\in M$, $\phi_{\mathcal{L}}|_{T_xM} : T_xM \ra T^*_xM$ is a diffeomorphism. 
\end{enumerate}
\item\exclabel{22} Let $(M,g)$ be a pseudo-Riemannian manifold and denote by $\phi :TM\ra T^*M$ 
the isomorphism of vector bundles induced by $g$. Let $V$ be a smooth function on  
$M$ and consider the Lagrangian $\mathcal{L}(v)=\frac12 g(v,v)-V(\pi v)$, $v\in TM$.  Here $\pi : TM \ra M$ 
denotes the projection. Denote by $E\in C^\infty (TM)$ the energy and by $H=E\circ \phi^{-1}\in C^\infty (T^*M)$ the Hamiltonian. 
\begin{enumerate}
\item Show that if a curve $\tilde{\g} : I \ra T^*M$ is a motion of the Hamiltonian system $(T^*M,\o ,H)$, then 
the curve $\pi \circ \tilde{\g} : I \ra M$ is a motion of the Lagrangian system $(M,\mathcal{L})$. Here $\o$ denotes the canonical
symplectic form. 
\item Show that the map $\g \mapsto \phi \circ \g'$ from curves in $M$ to curves in $T^*M$ is inverse to the map $\tilde{\g} \mapsto \pi \circ \tilde{\g}$
from curves in $T^*M$ to curves in $M$ when restricted to solutions of the Euler-Lagrange equations and Hamilton's equations, respectively. 
Here $\pi :T^*M \ra M$ denotes the projection. 
\end{enumerate} 
\item\exclabel{23} State and prove a local result relating Lagrangian mechanical systems  
with non-degenerate Lagrangian to Hamiltonian systems, similar to the global result proven 
in \secref{legendre_trafo} for nice Lagrangians.  
\item\exclabel{24} Let $f$ be a smooth function on a finite-dimensional real vector space $V$ such that 
$\phi_f : V \ra V^*$ is a diffeomorphism and consider its Legendre transform $\tilde{f}\in C^\infty (V^*)$. 
Show that $\phi_{\tilde{f}} : V^* \ra V$ is also a diffeomorphism and that the Legendre transform
of $\tilde{f}$ is $f$. 
\item\exclabel{25} A subspace $U\subset V$ of a symplectic vector space $(V,\o )$ is called
\emph{isotropic} if $U \subset U^\perp$. Show that the maximal dimension of an isotropic subspace
is $n=\frac12 \dim V$ and that the isotropic subspaces $U$ of dimension $n$ are Lagrangian, that is $U=U^\perp$. Deduce 
that, with the definitions given in \chref{hjthy}, an immersed submanifold of a symplectic manifold is Lagrangian if and only if its tangent spaces are Lagrangian. 
\item\exclabel{26} Let $(T^*M,H)$ be a Hamiltonian system of cotangent type, $n=\dim M$, and let $S: M \times U\ra \bR$ be a smooth $n$-parameter family of solutions of the Hamilton-Jacobi equation. Show that the following conditions are equivalent.
\begin{enumerate}
\item[(i)] The family is non-degenerate. 
\item[(ii)] The map 
$\Phi_S: M \times U \ra T^*M$, defined in \defref{phi_S}, is a local diffeomorphism.
\item[(iii)] For all $x\in M$, 
\[ \Phi_S|_{\{ x\} \times U}: \{ x\} \times U \cong U \ra T_x^*M, \quad u\mapsto dS^u|_x,\]
is a local diffeomorphism.    
\item[(iv)] For all $x\in M$, 
\[ \Phi_S|_{\{ x\} \times U}: \{ x\} \times U \cong U \ra T_x^*M,\]
is of maximal rank.
\item[(v)] For all $(x,u)\in M\times U$ the $n\times n$-matrix 
\[ \left( \frac{\p^2 S(x,u)}{\p x^i \p u^j}   \right)  \]
is invertible, where $(x^i)$ are local coordinates in a neighborhood of $x\in M$ and 
$(u^i)$ are (for instance)  standard coordinates in $U\subset \bR^n$. 
\end{enumerate}
\item Let $(M,g)$ be a Riemannian manifold and let $x_0\in M$ be an equilibrium point of a given Lagrangian mechanical system $\mathcal{L}(v)=\frac{1}{2}g(v,v)-V(\pi v)$. Further assume that $x_0$ is a  local minimum and a non-degenerate critical point of $V$. The latter means that the Hessian of $V$ with respect to the Levi-Civita connection at $x_0$ is positive definite. Show that $x_0$ is a stable solution in the sense of Lyapunov stability, i.e. show that $\forall \varepsilon>0$ $\exists \delta>0$, such that for all solutions $\gamma:I\to M$, $0\in I$, of the corresponding Euler-Lagrange equations the following holds true:
\begin{align*}
&\sqrt{d_M(x_0,\gamma(0))^2+g_{\gamma(0)}(\dot\gamma(0),\dot\gamma(0))}<\delta\\
\Rightarrow\ & \sqrt{d_M(x_0,\gamma(t))^2+g_{\gamma(t)}(\dot\gamma(t),\dot\gamma(t))}<\varepsilon \qquad \forall t\in I \; ,
\end{align*}
where $d_M(\cdot,\cdot)$ denotes the metric on $M$ induced by $g$.\\
\textit{Hints: Use the Morse lemma to bring $V$ near $x_0$ to a certain form. Verify that for $K\subset M$ compact and contained in a chart domain, one can find $c>0$ and $C>0$, such that $c\langle \cdot,\cdot \rangle\leq g|_K(\cdot,\cdot)\leq C\langle\cdot,\cdot\rangle$, where $\langle \hat{q},\hat{q}\rangle=\sum \hat q_i^2$. The topology of $M$ coincides with the topology induced by $d_M$, see~\cite[p. 166]{KN}, in particular $V$ is continuous with respect to the metric $d_M$. Also, recall that the energy is an integral of motion, hence estimating $E(\dot{\gamma}(0))$ yields a global result for $\gamma$.}
\item
\begin{enumerate}
\item Consider a Lagrangian mechanical system in $\mathbb{R}^3$ with radial potential $\mathcal{L}(v)=\frac{1}{2}\langle v,v\rangle -V(r)$ and assume that $r(t)=r_0$ is the constant radial component of a motion with $|\vec L|>0$. Recall the corresponding expression for the energy 
\begin{equation*}
E=\frac{1}{2}\dot r^2 + V_{\textnormal{eff}}(r)
\end{equation*}
and show that such a solution is stable in the sense of the previous exercise if 
\begin{equation*}
\frac{d^2 V}{dr^2}(r_0)+\frac{3}{r_0}\frac{d V}{dr}(r_0)>0.
\end{equation*}

\item Consider $V(r)=-\frac{\alpha}{r^n}$, $\alpha>0$. Find all $n\in\mathbb{N}$ such that a stable motion $r(t)=r_0$ exists. Does the answer depend on $\alpha$? (Note that this question is related to Newtonian gravitation in $d=n+2$ dimensions. To show this, one needs to use a conservation law similar to the angular momentum which was used in dimension three.)

\item For which $b>0$ does the Lagrangian system corresponding to the potential $V(r)=-\frac{\alpha}{r}\exp{\left(-\frac{r}{b}\right)}$, $\alpha>0$, have stable motions with $r(t)=r_0$? Does the answer depend on $\alpha$?

\item For which $\alpha$ does the Lagrangian system corresponding to the potential $V(r)=\alpha\ln(r)$ have stable motions with $r(t)=r_0$?
\end{enumerate}
\item Let $q^1,\hdots,q^n,\hat{q}^1,\hdots,\hat{q}^n$ denote the standard coordinates on $T\bR^n$ . Consider the function $V:\bR\setminus\{0\}\to\bR$
defined by 
\[ V(x)=V_0\left( \left(\frac{x_0}{x}\right)^{12}-\left(\frac{x_0}{x}\right)^6\right),\] 
where $V_0, x_0\in \bR\setminus\{0\}$ are constants. The Lagrangian system
\begin{equation*}
\mathcal{L}=\frac{1}{2}\langle\hat{q},\hat{q}\rangle - \sum_{i=1}^{n-1} V(q^{i+1}-q^i)
\end{equation*}
describes small oscillations in a linear chain of $n$ atoms of unit mass. Note that the oscillations are assumed to be in direction of the chain.
\begin{enumerate}%
\item Determine the distance $a$ of two neighboring atoms in the equilibrium position.
\item Show that the Taylor expansion up to second order of $V$ at $a$ is given by
\begin{equation*}
\widetilde{V}(x)=-\frac{V_0}{4}+\frac{1}{2}k(x-a)^2
\end{equation*}
with $k=\frac{18}{\sqrt[3]{2}}\frac{V_0}{x_0^2}$.
\item Determine the equations of motion corresponding to the linearized problem
\begin{equation*}
\mathcal{L}_2=\frac{1}{2}\langle\hat{q},\hat{q}\rangle - \sum_{i=1}^{n-1} \widetilde{V}(q^{i+1}-q^i).
\end{equation*}
\end{enumerate}

\section{Exercises for \texorpdfstring{\Chref{hjthy}}{Chapter~\ref{ch:hjthy}}}

\item Consider a point $q$ on a conic section $C\subset \bR^2$ with focal points $f_1\neq f_2$.  
Show that the tangent line at $q$ bisects the angle formed by the two lines connecting $q$ with the focal points.  
\item\exclabel{32} Consider a hyperbola $H\subset \bR^2$ with focal points $f_1, f_2$ at distance $2c>0$. 
Recall that $H$ is defined by the equation $|r_1-r_2|=2a$, where  $r_1$ and $r_2$ are the distances to $f_1$ and $f_2$, and $a\in (0,c)$ is a constant.  
Show that the restriction 
of the Euclidean metric to $H$ is given by 
\[  \frac{\xi^2-4a^2}{4(\xi^2 -4c^2)}d\xi^2 ,  \]
where $\xi = r_1+r_2$. (This result is used in the proof of \propref{4.11}.)

\section{Exercises for \texorpdfstring{\Chref{fieldthy}}{Chapter~\ref{ch:fieldthy}}}

\item Consider the Lagrangian $\mathcal{L}(q,\hat{q})=\prod_{i=1}^n \hat{q}^i$ on $\bR^n$ and let $n\ge 2$. 
Show that the restriction of $\phi_\mathcal{L} : T\bR^n \ra T^*\bR^n$ to the open subset 
\[ U_\s =\{ (q,\hat{q})\in T\bR^n \mid \s \mathcal{L}> 0\}\subset T\bR^n,\]
where  $\s\in \{+,-\}$, 
is a diffeomorphism onto its image. Check that $\phi_\mathcal{L}(U_{+})=\phi_\mathcal{L}(U_{-})$ if $n$ is odd and that $\phi_\mathcal{L}(U_{+})$ and $\phi_\mathcal{L}(U_{-})$
are invariant under the antipodal map in the fibers of $T^*\bR^n$ if $n$ is even. 
 Determine the corresponding Hamiltonians on $\phi_\mathcal{L}(U_\s )$ and find the general solution of the corresponding Hamilton's equations.  
\item Show that 
\[ \dim \mathrm{Jet}^k_0(\bR^n, \bR ) = \binom{n+k}{k}.\] 
\item Show that for every smooth manifold there are canonical identifications
\begin{enumerate}
\item 
\[ \mathrm{Jet}^1(M,\bR ) = \bR \times T^*M,\]
\item
\[ \mathrm{Jet}^1(\bR ,M) = \bR \times TM.\]
\end{enumerate}
\item Let $P=(P^1,\ldots , P^n)$ be a 
smooth vector-valued function on $\mathrm{Jet}^k(\bR^n , \bR^m)$. 
Its total divergence $\mathrm{Div}\, P$ is a smooth function on $\mathrm{Jet}^{k+1}(\bR^n , \bR^m)$. 
We know from \propref{euler_op_vanish_on_L} that the Euler operators $E_a$, acting on functions on $\mathrm{Jet}^{k+1} (\bR^n , \bR^m)$, vanish on $\mathrm{Div}\, P$. 
Check this in the following cases by directly computing $E_a(\mathrm{Div}\, P)$:
\begin{enumerate}
\item $k=0$,
\item  $k=m=n=1$. 
\end{enumerate}
\item
\begin{enumerate}%
\item Show that the Euler-Lagrange equations of the free scalar field Lagrangian on the $n$-dimensional Minkowski space $(M,g)=(\bR^n,(dx^0)^2-\sum_{i=1}^{n-1} (dx^i)^2)$, given in index notation by
\begin{equation*}
\mathcal{L}=\frac{1}{2}\partial_\mu \varphi \partial^\mu \varphi -\frac{1}{2}m^2\varphi^2,
\end{equation*}
yield the Klein-Gordon equation $(\Box+m^2)\varphi=0$. 
\item Verify that in the above case $\Box=(-1)^{n-1}\star d \star d$, where $\star$ denotes the Hodge star-operator, 
defined by  
\[ \alpha \wedge \star \beta = \langle \alpha , \beta \rangle dvol, \]
where $\a,\b$ are differential forms on $n$-dimensional Minkowski space and $dvol$ is the metric volume form of the 
Minkowski metric $(dx^0)^2-\sum_{i=1}^{n-1} (dx^i)^2$ compatible with the 
standard orientation. 
\end{enumerate}

\item Consider the Lagrangian of the Kepler problem with Newton's potential in $\mathbb{R}^3$,
\begin{equation*}
\mathcal{L}(v)=\frac{1}{2}\langle v,v\rangle +\frac{M}{r}.
\end{equation*}
Show that each entry of the so-called Runge-Lenz vector (or Laplace-Runge-Lenz vector)
\begin{equation*}R:=
\left(\begin{matrix}
x_t\\
y_t\\
z_t
\end{matrix}
\right)
\times
\left(
\left(\begin{matrix}
x\\
y\\
z
\end{matrix}
\right)
\times
\left(\begin{matrix}
x_t\\
y_t\\
z_t
\end{matrix}
\right)
\right)
-\frac{M}{r}\left(\begin{matrix}
x\\
y\\
z
\end{matrix}
\right)
\end{equation*}
is an integral of motion. 

\item For a given $k\in\mathbb{N}$, determine the Euler-Lagrange equations for
\begin{equation*}
\mathcal{L}=\frac{1}{2}\sum_{|J|\leq k}u_J^2,\ \ \mathcal{L}\in C^\infty(\text{Jet}^k(\mathbb{R}^n,\mathbb{R})).
\end{equation*}
\item\exclabel{40} Let $(M,g)$ and $(N,h)$ be pseudo-Riemannian manifolds of dimension $m$ and $n$, respectively.
We assume that $(N,h)$ is oriented and denote its volume form by $dvol_h$. Determine the Euler-Lagrange equations 
for the Lagrangian $n$-form $\mathcal{L}dvol_h$ defined by  
\[ \mathcal{L} (j^1f) =  \frac12 \langle df, df \rangle,\quad f\in C^\infty (N,M), \]
where $\langle \cdot ,\cdot \rangle_x$ is the scalar product on $T^*_xN\otimes T_{f(x)}M$, $x\in N$, induced by $h$ and $g$. 
Show that the resulting equations are equivalent to
\[ \mathrm{tr}_h \n df =0,\]
where $\n$ is the connection on the vector bundle $T^*N\ot f^*TM\ra N$ induced by the Levi-Civita connections on $M$ and $N$.
\item\exclabel{41} Let $X$ and $Y$ be smooth vector fields on $\bR^n$ and $\bR^m$, respectively, $Z=X+Y$, $dvol$ the standard volume form of $\bR^n$, and $\mathcal{L}\in C^\infty (\mathrm{Jet}^k(\bR^n,\bR^m))$ a Lagrangian. Assume that 
\[ (\mathrm{pr}^{(k)}Z)(\mathcal{L}) + \mathcal{L}\mathrm{div} X\]
is a total divergence. Show that the same calculation as in the proof of Noether's theorem can be used to prove that 
\[ \sum Q^aE_a(\mathcal{L}),\quad \mbox{defined by}\quad  Q^a= Y^a-\sum u_i^aX^i,\] 
is a total divergence, 
where $X^i$  and $Y^a$ are the components of $X\in  \mathfrak{X}(\bR^n)$ and
$Y\in \mathfrak{X}(\bR^m)$, respectively. 
\item\exclabel{42} Let $(M,g)$ and $(N,h)$ be pseudo-Riemannian manifolds and $V\in C^\infty (N\times M)$. 
Consider the Lagrangian 
\[ \mathcal{L}(j^1f) = \frac12 \langle df,df\rangle - V(j^0f),\quad f\in C^\infty (N,M).\]
\begin{enumerate}
\item In the special case when $(N,h)$ is a pseudo-Euclidean vector space,  show that 
$\mathcal{L}dvol_h$ is invariant under the group of translations of the source manifold $N$ if $V\in C^\infty (M)\subset$ $C^\infty (N\times M)$ and compute the corresponding 
Noether currents. 
\item  In the special case when $(M,g)$ is a pseudo-Euclidean vector space,  show that 
$\mathcal{L}dvol_h$ is invariant under the group of translations of the target manifold $M$ if $V\in C^\infty (N)\subset$ $C^\infty (N\times M)$ and compute the corresponding 
Noether currents. 
\end{enumerate}
\item\exclabel{43} Let $\Omega\subset \bR^{n-1}$ be a bounded domain with smooth boundary. We consider  a Lagrangian 
$\mathcal{L}\in C^\infty (\mathrm{Jet}^k(U,\bR^m ))$ on the cylinder 
$U=\bR \times \Omega \subset \bR^n$ with standard coordinates $(x^0,x^1,\ldots, x^{n-1}) =: (t,\vec{x})$. 
Let $P : \mathrm{Jet}^\ell (\bR^n,\bR^m)\ra \bR^n$ be a smooth vector-valued function such that 
\[ \mathrm{Div}\, P|_{\mathrm{Jet}^{\ell +1} (U,\bR^m)}\] 
is a conservation law  of $\mathcal{L}dvol$, where $dvol$ is the standard volume 
form of $U$. 
Let $f\in C^\infty (U,\bR^m)$ be a solution of the Euler-Lagrange equations which extends smoothly to a neighborhood of the closure of $U$. 
We decompose the current $J=P(j^\ell f)=(J^0,\vec{J}) : U \ra \bR^n$ into the charge density $J^0$ and the 
flux density $\vec{J}$. 
Show that the time evolution of the charge
$Q(t)=\int_\O J^0(t,\vec{x})d^{n-1}x$ is given by 
\[ Q'(t) = -\int_{\p \O} \langle \vec{J},\nu \rangle dvol_{\p \O},\] 
where $\nu$ stands for the outer normal of $\p \Omega$ and $dvol_{\p \O}$ for the induced volume form
on $\p \Omega \subset \bR^{n-1}$. Conclude that the charge is constant if 
the flux density is tangent along the boundary of $\O$. 
\item\exclabel{44} Consider Yang-Mills theory on a Hermitian line bundle over Minkowski space $(M,g)=(\bR^4,dt^2-\sum_{\a=1}^3 (dx^\a)^2)$. Since 
the Lie algebra of $\U 1$ is one-di\-men\-sional, we can identify the 
curvature $F=F^\n$ with an ordinary real-valued $2$-form. 
Show that the Yang-Mills equation $d \ast F=0$ reduces to half of Maxwell's equations in the vacuum, that is  
\[ \mathrm{div}\, \vec{E} =0,\quad \mathrm{rot}\, \vec{B} = \frac{\p}{\p t} \vec{E},\]
whereas the Bianchi identity $d F=0$ reduces to the other half of Maxwell's vacuum equations, that is  
\[ \mathrm{div}\, \vec{B} =0,\quad \mathrm{rot}\, \vec{E} = -\frac{\p}{\p t} \vec{B},\]
where $\vec{E}=\sum E^\a \p_\a$ and $\vec{B}=\sum B^\a \p_\a$ are  the time-dependent vector fields on $\bR^3$ obtained by decomposing
$F$ into 
\[ E^\a  = F(\p_t ,\p_\a )\quad \mbox{and}\quad B^\a = -F(\p_\b ,\p_\g), \]
where $(\a , \b , \g)$ runs trough cyclic permutations of $\{ 1,2,3\}$. 
\item\exclabel{45} Let  $\n$ be a $G$-connection in a vector bundle $E$ and $\a$ 
a one-form with values in $\gg (E)$. Show that the curvature of the $G$-connection $\n + \a$
is given by 
\[ F^{\n + \a} = F^\n + d^\n \a + \a \wedge \a .\]
\item Express the Yang-Mills equations in local coordinates.
\item Check by direct calculation that the Noether current associated with an infinitesimal gauge transformation as stated in \propref{noether_curr_infinit_gauge_trafo} is conserved, that is divergence-free.
\item Prove that 
\[  \left. \frac{\p }{\p \e}\right|_{\e=0} dvol_{g_\e} =\tr (g^{-1}h)dvol_g,\]
for every smooth family of pseudo-Riemannian metrics $g_\e$, where 
$g=g_0$ and  $h= \left. \frac{\p }{\p \e}\right|_{\e=0}g_\e$. 
\end{enumerate}\clearpage{}%

\backmatter%
\clearpage{}%

\clearpage{}%
\printindex

\end{document}